\documentclass[11pt]{article}

\usepackage{graphicx}
\usepackage{amsmath,amsfonts,amssymb,bm}
\usepackage{mathptmx}
\usepackage{newtxtext}
\usepackage{newtxmath}
\usepackage{epstopdf}
\usepackage[scaled=.97]{helvet}
\usepackage{fancyhdr}
\usepackage{natbib}
\usepackage{url}
\usepackage{xcolor}
\usepackage{indentfirst}
\usepackage{multicol}
\usepackage{ifthen}
\usepackage{rotating}
\usepackage{appendix}
\usepackage{setspace}
\usepackage{lineno}
\usepackage{authblk}
\usepackage{hyperref}
\usepackage{tabularx}

\title{\bfseries Predicting Forecast Error for the HRRR Using LSTM Neural Networks:\\
A Comparative Study Using New York and Oklahoma State Mesonets}

\author[1]{D. Aaron Evans\thanks{Corresponding author: \href{mailto:aaevans@albany.edu}{aaevans@albany.edu}}}
\author[1]{Kara J. Sulia}
\author[2]{Nick P. Bassill}
\author[1]{Chris D. Thorncroft}
\author[3]{Jay C. Rothenberger}
\author[4]{Lauriana C. Gaudet}

\affil[1]{Atmospheric Sciences Research Center, University at Albany, SUNY, Albany, NY, USA}
\affil[2]{State Weather Risk Communication Center, University at Albany, SUNY, Albany, NY, USA}
\affil[3]{University of Oklahoma, Norman, OK, USA}
\affil[4]{The Weather Company, Andover, MA, USA}

\date{26 May 2026}

\begin{document}
\maketitle

%% ABSTRACT %%
\begin{abstract}
Long Short-Term Memory (LSTM) models are trained to predict forecast errors for the High-Resolution Rapid Refresh (HRRR) model using the New York State Mesonet and Oklahoma State Mesonet near-surface weather observations as ground truth. When evaluated using mean-absolute-error and percent improvement relative to the HRRR, LSTMs predict precipitation error most accurately, providing, on average, a 48\% improvement relative to the HRRR forecast, followed by wind error, providing, on average, a 15\% improvement, and then temperature error, providing, on average, a 25\% improvement. Precipitation errors exhibit an asymmetry, with overforecast precipitation detected more accurately than underforecast, while wind error predictions are consistent across over- and underforecast predictions. Temperature error predictions are relatively accurate but smoother, with respect to variance, than true observations. This paper describes an overview of LSTM performance with the expressed intent of providing forecasters with real-time predictions of forecast error at the point of use within the New York State and Oklahoma State Mesonets. In practice, the predicted errors can be used to adjust deterministic HRRR forecasts at the point of use, identify locations and variables with elevated uncertainty, and provide supplemental guidance for high-impact decision-making. This research demonstrates the potential of LSTM-based machine learning models to provide actionable, location-specific predictions of forecast error for high-resolution operational numerical weather prediction (NWP) systems. However, model performance is variable-dependent, and the approach relies on the availability of dense mesonet observations, which may limit applicability in data-sparse regions.\footnote{This manuscript is a preprint and has been submitted for peer review to the \textit{Weather and Forecasting} journal. The content is subject to change based on the outcome of the peer-review process and should not be considered final or definitive. Copyright in this Work may be transferred without further notice.}
\end{abstract}

\section{Introduction}
Numerical Weather Prediction (NWP) models are fundamental forecasting tools for operational organizations like the National Weather Service (NWS), as well as academic institutions and the private sector. To enhance the accuracy of operational models, researchers continually assess forecast biases and errors. Typically, understanding bias and error in NWP model output is accomplished using a suite of statistical verification methods and data analysis tools \citep{Casati2008, Ebert2013}. These methods are robust and insightful but require considerable computational resources and time \citep{Gilleland2013}. Furthermore, studies of forecast error and bias traditionally focus on a specific model version, climatological time period, or case study event \citep{Duda2023, Guan2017, Moskaitis2008}.  In the context of the High-Resolution Rapid Refresh (HRRR) model, prior studies, such as \citet{Gaudet2024}, have documented systematic biases and forecast errors in near-surface variables using a retrospective framework. 

This scrutiny often leads to post-hoc improvements through post-processing techniques that correct for biases or through refinements in the models’ computational frameworks and parameterizations, advancing overall model performance. However, the motivation behind the research herein is an ad-hoc improvement, which builds upon the prescient proposal by \citet{Gaudet2024} to equip end-users with the capability to predict both the magnitude and direction of forecast error in NWP models in real-time, at the point of use -- specifically, they advise that machine learning is best suited for this task.

Machine learning (ML) has become an increasingly prominent tool for bias correction in NWP and climate modeling. Early applications have focused on precipitation, \citet{li_2023} uses gradient boosting and neural networks to substantially reduce systematic biases in mountainous terrain by capturing nonlinear relationships with environmental variables observed through in-situ and remote sensing measurements. \citet{mouatadid2023adaptive} extended these approaches to temperature-correction and precipitation-correction of subseasonal forecasting, with ML-based adaptive-bias-correction frameworks leveraging dynamical forecasts and observations, which demonstrate large improvements in predictive skill relative to traditional statistical methods.  Emerging deep learning approaches further enable joint bias correction and downscaling, particularly for precipitation extremes and spatial structure. \citet{zhang2024} improves the downscaling of extreme weather events in low-resolution climate models by using a convolutional-LSTM, demonstrating that ML techniques can correct dynamical fields within multivariate frameworks that jointly adjust temperature, humidity, and wind patterns in climate and NWP models.

Despite these advances, relatively few studies have addressed the direct, real-time prediction of NWP forecast error at sub-mesoscale across multiple near-surface variables. There remains a need for methods that can capture temporal evolution, spatial heterogeneity, and region-specific dynamics while remaining computationally efficient for operational use and latency. This gap motivates the development of data-driven approaches to model the temporal and location-specific characteristics of forecast error across diverse meteorological regimes.

Long Short-Term Memory (LSTM) models are particularly well-suited for applications in atmospheric science: the ability of LSTMs to retain information over long time intervals, capture nonlinear dependencies, and process multivariate inputs makes them especially effective for forecasting tasks \citep{Sepp1997}. Google recently demonstrated that an LSTM architecture proved to be the most accurate and reliable approach to predict flooding \citep{Nearing2024}. Similarly, another study by \citet{Wang2022} employs a hybrid convolutional-LSTM and Empirical-Mode-Decomposition-LSTM approach to predict sea-level anomalies in the South China Sea up to 15 days in advance. While newer architectures, such as transformers \citep{Kucuk2024} and convolutional neural networks \cite[CNNs,][]{Lagerquist2020}, have gained traction in meteorological research, LSTMs remain a competitive choice.

Motivated by its balance of predictive skill, stability, and efficiency, we adopt an LSTM-based model for real-time prediction of HRRR forecast errors \citep{NCEP_HRRR, Dowell2022, James2022} using observations from the New York State Mesonet (NYSM) and the Oklahoma State Mesonet (OKSM). The primary contributions of this study include: (1) the development of a station-specific, data-driven approach for predicting forecast errors at the point of use, (2) a systematic evaluation of model performance across multiple meteorological variables, and (3) an analysis of how regional physical and dynamical processes influence predictive skill across geographically distinct domains.

Specifically, we address the following questions: how does LSTM predictive skill vary across distinct meteorological regimes? Are differences in forecast error prediction linked to region-specific physical and dynamical processes? And to what extent can a single modeling framework generalize across geographically and dynamically contrasting environments? These questions are examined through a comparative analysis of the New York and Oklahoma domains, whose contrasting geography and atmospheric dynamics provide a natural testbed for evaluating region-dependent forecast error behavior.

The remainder of this paper is organized as follows. The Data section describes the datasets and preprocessing methods. The Machine Learning Model section outlines the machine learning architectures and experimental design. The Results and Discussions section presents the results, including model performance and inter-domain comparisons, as well as potential dynamical mechanisms driving model performance. Finally, the Conclusions section summarizes the key findings and highlights directions for future work.

\section{Data}
\label{Data Section}
\subsection{Ground Truth Atmospheric Observations}
The LSTMs used in this study are trained on high-quality near-surface atmospheric observations from two statewide mesonet networks: the NYSM and the OKSM. These networks provide critical inputs for our proposed machine learning (ML) architecture, with rigorous data collection and quality assurance protocols.

\begin{table*}[ht]
\centering
\resizebox{\textwidth}{!}{%
\begin{tabular}{|l|l|l|}
\hline
\textbf{New York State Mesonet Features} & \textbf{Oklahoma State Mesonet Features} & \textbf{HRRR Model Features} \\
\hline\hline
Latitude & Latitude & —\\
Longitude & Longitude & —\\
Elevation & Elevation & —\\
2-Meter Temperature & 1.5-Meter Temperature & 2-Meter Temperature \\
9-Meter Temperature & 9-Meter Temperature & 2-Meter Specific Humidity \\
2-Meter Dew Point & 1.5-Meter Dew Point & 2-Meter Dew Point \\
2-Meter Relative Humidity & 1.5-Meter Relative Humidity & 2-Meter Relative Humidity \\
Solar Radiation & Solar Radiation & Downward SW Radiation \\
Atmospheric Pressure & Atmospheric Pressure & Downward LW Radiation \\
Mean Sea-Level Pressure & Mean Sea-Level Pressure & Mean Sea-Level Pressure \\
Mean 10-Meter Sonic Anemometer Wind Speed & Mean 10-Meter Anemometer Wind Speed & Total Wind Speed \\
10-Meter Sonic Anemometer Wind Speed & 10-Meter Anemometer Wind Speed & 10-Meter Wind U Component \\
Max 10-Meter Sonic Anemometer Wind Speed & Max 10-Meter Anemometer Wind Speed & 10-Meter Wind V Component \\
10-Meter Wind Direction & 10-Meter Wind Direction & 10-Meter Wind Direction \\
Total Hourly Precipitation & Total Hourly Precipitation & Total Hourly Precipitation \\
Snow Depth & — & Accumulated Snow \\
— & — & CAPE \\
— & — & Total Cloud Cover \\
— & — & 500-hPa Geopotential Height \\
\hline
\end{tabular}}
\caption{Combined list of NYSM, OKSM, and HRRR independent variables used as features in training the LSTMs.}
\label{tab:combined_features}
\end{table*}

\subsubsection{Network Overview and Comparison}
The NYSM, operational since 2018, comprises 127 weather stations\footnote{Lake Placid Station is excluded, as it was installed outside of the training period in May 2024.} across New York State, with an average spacing of 27 kilometers \citep[hereafter B20]{Brotzge2020}. The OKSM, which launched in 1994 as the first statewide environmental monitoring network in the United States, includes 118 active stations for our study period (January 2018 to December 2024), and has a spatial resolution of roughly 30 kilometers \citep{brock1995oklahoma, ziolkowska2017benefits}.

The OKSM served as a prototype for the NYSM, and many of its operational standards were adopted by the NYSM. Both networks are recognized for strict site selection criteria, precise sensor calibration, and robust quality control processes \citep[hereafter M07]{mcpherson2007statewide}. Both mesonets' data undergo automated and manual quality assurance processes in real time, as well as on a daily, weekly, monthly, and annual basis (B20, M07). Each observation is automatically assigned a quality flag: good, suspect, warning, or failure (B20, M07). The data used to train the ML models herein excludes data flagged with warning or failure. With respect to data availability, the NYSM maintains an average operational availability of approximately 97\% across the dataset, with station-level availability ranging from 99.9\% to 81.4\%. Similarly, the OKSM exhibits a slightly higher average availability of approximately 99\%, with values ranging from 99.9\% to 87.8\% across stations. These results highlight the overall robustness of both mesonet networks. However, they also underscore a limitation of the present modeling framework: the model is trained to rely on surface observations and therefore exhibits degraded performance in the presence of missing data.

\subsubsection{Data Pre-Processing}
Building on the pre-processing techniques developed by \citet{Gaudet2024}, the NYSM and OKSM observations are aligned with the temporal scale of the NWP model forecast. To align the temporal scale of the instantaneous observations, which are recorded every five minutes, with that of an NWP model forecast, the observations taken at the top of each hour are used as the true observed atmospheric conditions during training. There are two exceptions to this: total precipitation is accumulated over the hour, and wind speed is averaged over the hour. Mesonet observations were restricted to top-of-the-hour values to align with the temporal resolution of the HRRR model output. While this ensures temporal consistency between predictors and targets, aiding the model in faster convergence, it may limit the representation of sub-hourly variability, particularly for rapidly evolving processes such as convective precipitation and boundary-layer transitions. 

There are 16 meteorological variables used from the NYSM as features in training the LSTMs, whereas the OKSM has 15 features used in training, all of which are listed in Table \ref{tab:combined_features}. It should be noted, differences in sensor configurations and available variables between NYSM and OKSM may introduce minor inconsistencies in cross-domain comparisons. While efforts were made to align comparable variables, these differences are acknowledged as a potential source of uncertainty in comparing the two domains in this study.

\subsection{Numerical Weather Prediction Forecasts}
\subsubsection*{High-Resolution Rapid Refresh Forecast System}
The High-Resolution Rapid Refresh (HRRR) forecast system, developed by the National Oceanic and Atmospheric Administration (NOAA) in 2014 \citep{Dowell2022}, employs a cloud-resolving, convection-allowing implementation of the Advanced Research version of the Weather Research and Forecasting (WRF-ARW) model as its dynamical core \citep{NCEP_HRRR}. HRRR is optimal for short-range forecasting and is designed with a particular focus on the evolution of precipitating systems to aid with situational awareness \citep{Dowell2022}. HRRR uses a 3-kilometer Lambert Conformal Grid spanning the continental United States \citep{NCEP_HRRR} and is initialized every hour, providing hourly forecasts out to 18 hours. Although the HRRR is capable of longer (48-hour) forecasts with 00, 06, 12, 18 UTC initializations \citep{Dowell2022}, our research focuses on the first 18 hours, as this allows us to consistently analyze hourly initialization of LSTM performance.   

The HRRR’s fine spatial and temporal resolution, combined with advanced data assimilation techniques, as well as incorporating radar reflectivity, hybrid ensemble-variational assimilation of conventional weather observations, and cloud analysis for initializing stratiform cloud layers, makes it a critical tool for forecasters \citep{Dowell2022}. This reliance has driven significant development and improvement of the HRRR over the years. The LSTMs introduced herein are trained on three versions of the HRRR: HRRRv2 (1 January 2018 to 11 July 2018), HRRRv3 (12 July 2018 to 1 December 2020), and HRRRv4 (2 December 2020 to 31 December 2023). To assess the impact of training across multiple HRRR versions, a cross-validation experiment was conducted using a Friedman test with Nemenyi post-hoc analysis \citep{demsar2006statistical}. Results indicated no statistically significant degradation in model performance when training across versions, suggesting that the model learns relationships that are robust to version-specific differences. A detailed list of the meteorological variables from the HRRR used as features in training the LSTMs is provided in Table \ref{tab:combined_features}. 

\subsection{Geographic Information}
NWP models exhibit varying degrees of efficacy in parameterizing complex geographic factors such as aspect/slope, elevation, and land type. The challenge lies in simplifying these intricate land-atmosphere interactions into computational schemes that are efficient yet effective. Recent advancements in computational power have enabled the incorporation of more dynamic land-surface parameterization schemes into NWP models, which help more accurately capture the nuanced interactions between land surfaces and atmospheric processes \citep{Li2013}. However, due to the non-linear complexity of the earth system, NWP parameterization schemes still decidedly simplify land-atmosphere interactions to manage computational costs.

To enhance the predictive accuracy of the LSTMs, we developed a preprocessing scheme that incorporates information about the surrounding geography, including land-use/land-class (LULC), elevation, and aspect/slope for each NYSM station (see Appendix for maps of analyzed geographic variables). This approach was designed to allow the LSTM to gain insight into the intricate topography and heterogeneous LULC of New York State, which are critical components in understanding and predicting NWP forecast errors. Moreover, we have applied the same methodology to the OKSM, despite Oklahoma exhibiting comparatively less topographic variability and more homogeneous LULC than New York, while still containing notable terrain variation in its eastern regions. 

Geographical analysis begins with a buffer surrounding each NYSM, 12-km for LULC and 30-km for aspect/slope and elevation. Buffer sizes were selected to represent physically meaningful spatial scales of land–atmosphere interaction and were further evaluated using correlation analyses to identify the most informative scales for each geographic variable. Our findings indicate that selecting an appropriate buffer size is crucial. A buffer that is too small fails to capture a sufficient geographical scope to effectively model the representation of the surrounding area, while a buffer that is too large results in regional averages that may not accurately reflect local conditions. Buffer size was primarily determined using Pearson \citep{HahsVaughn2023}, Spearman-Rank, and Kendall-Rank \citep{Puth2015} correlation analyses, which examined the relationship between forecast error and the feature percentages at each NYSM station. In contrast, elevation statistics employed canonical correlation analysis, as it allows for a multivariate dataset to be compared against a target dataset \citep{Barnston1992} and provides a more comprehensive assessment of the topography surrounding a NYSM station, compared to LULC and aspect/slope, which are best analyzed by class.

Once the LULC, elevation, and aspect/slope data are collected for each NYSM and OKSM station, their respective geographic data are separately subjected to the scikit-learn k-means clustering algorithm \citep{Pedregosa2011}. A range of cluster configurations was evaluated during model development. Model performance was found to be relatively insensitive to the exact number of clusters beyond an inertia of 200, with diminishing returns observed as cluster granularity increased. These cluster assignments are represented as categorical variables in the LSTM framework. For example, the k-means clustering algorithm identified seven distinct LULC clusters among NYSM stations. Each station is therefore assigned a categorical value from 1 to 7, representing its LULC cluster assignment. This process allows the LSTM to incorporate geographic characteristics without introducing excessive noise or unnecessary complexity in the feature space.

\subsection{Data Curation}
\subsubsection{Target Mesonet Station \& Triangulate}
Our process for curating training data for an LSTM begins by identifying the mesonet station of interest. Once selected, we calculate the haversine distance to the nearest three mesonet stations to triangulate the data. Since LSTMs are trained on time series, this approach allows the LSTM to incorporate some spatial representation of how meteorological phenomena influence forecast error. Including information from the three closest stations improves model performance, additional stations provide negligible further improvement. This triangulation approach allows the LSTM to incorporate spatially distributed information while avoiding unnecessary model complexity associated with including a larger number of stations.

\subsubsection{Target NWP Model}
Given the HRRR’s fine grid spacing of 3 kilometers, the maximum distance between any HRRR grid point and a mesonet station is 2.12 kilometers. The LSTMs use HRRR grid points co-located with mesonet stations via a nearest-neighbor haversine distance. The median absolute difference in elevation between the co-located HRRR grid points and mesonet stations is typically between $\pm~30$ meters. While this reduces representativeness error relative to coarser-resolution models, some residual discrepancy may remain, particularly in regions of complex terrain.

\subsubsection{Forecast Hour \& Time Encoding}
Training is iterated recursively through the forecast hours. For example, for the HRRR, the training process begins with forecast hour 1, followed by forecast hour 2, and continues sequentially until reaching forecast hour 18. Mesonet observations and corresponding HRRR forecasts are collated based on valid hourly timestamps. To help the LSTM accurately capture the temporal variability of meteorological phenomena, we introduce a time encoding mechanism commonly used in ML \citep{lewinson2022three}. This involves applying a cyclic encoding scheme using sine and cosine transformations, enabling the LSTM to account for the influence of time of day and seasonality on forecast errors. While cyclic time encoding provides a continuous representation of the seasonal cycle, much of the seasonal and subseasonal variability is implicitly captured through the evolving meteorological state represented by the NWP and mesonet inputs.

\subsubsection{Calculate NWP Error}
The error associated with the parameter of interest is then identified, whether that be total hourly precipitation, 10-meter wind speed, or 2-meter (NYSM)/1.5-meter (OKSM) temperature. The error is found by subtracting the primary mesonet station’s observations from the NWP forecast, as seen in Equation \ref{eq:forecast_error}. 

\begin{align}
\mathrm{Forecast\ Error} &= \mathrm{NWP\ Forecast} - \mathrm{Mesonet\ Observation}
\label{eq:forecast_error}
\end{align}

\subsubsection{Train, Validation, and Test Data Split}
The LSTM is trained on data from the beginning of 2018 to the end of 2022 and validated on data from 2023. This time series is partitioned by time chronologically, with the validation set being the most recent split in the training data, to ensure that we do not involve training data from the future that may increase LSTM performance artificially \citep{Kapoor2023}. All LSTMs are then tested on data from 2024 to capture seasonal and sub-seasonal LSTM performance metrics.

\section{Machine Learning Model}
\subsection{Architecture}

A gamut of ML architectures were evaluated to identify an approach that balances generalizability and operational efficiency for the modeling paradigm discussed herein: LSTMs provided the most robust and consistent performance across variables and forecast lead times, effectively capturing temporal dependencies while maintaining stable training behavior. Simpler models, such as Random Forests (e.g., \citet{Gagne2017}), did not demonstrate sufficient predictive skill, as they lack an explicit mechanism for capturing temporal persistence and evolving error dynamics. More complex architectures designed to capture spatial and global dependencies, including ConvLSTM (e.g., \citet{Wang2022, zhang2024}), Vision Transformers (ViTs) (e.g., \citet{Kucuk2024}), and Vision Conformer models (e.g., \citet{saleem2024}), introduced substantial computational overhead without improving predictive performance, often producing noisy or weakly correlated outputs. Standard recurrent neural networks (RNNs, e.g., \citet{han2021}) exhibited partial skill but suffered from training instability and poor generalization due to vanishing and exploding gradient issues. Transformers are similar to LSTMs in their design to perform on sequential data \citep{vaswani2017}, yet, particularly important for our use case, LSTMs provide an inherent inductive bias toward temporal continuity and have demonstrated strong performance in settings with limited data and noisy geophysical signals. Given the moderate data volume, implicit temporal structure of forecast error, and need for operational efficiency, LSTMs represent a balanced and practical choice for this modeling paradigm.

\subsubsection{LSTM Encoder Architecture}

\begin{figure*}
    \centering
    \includegraphics[width=30pc]{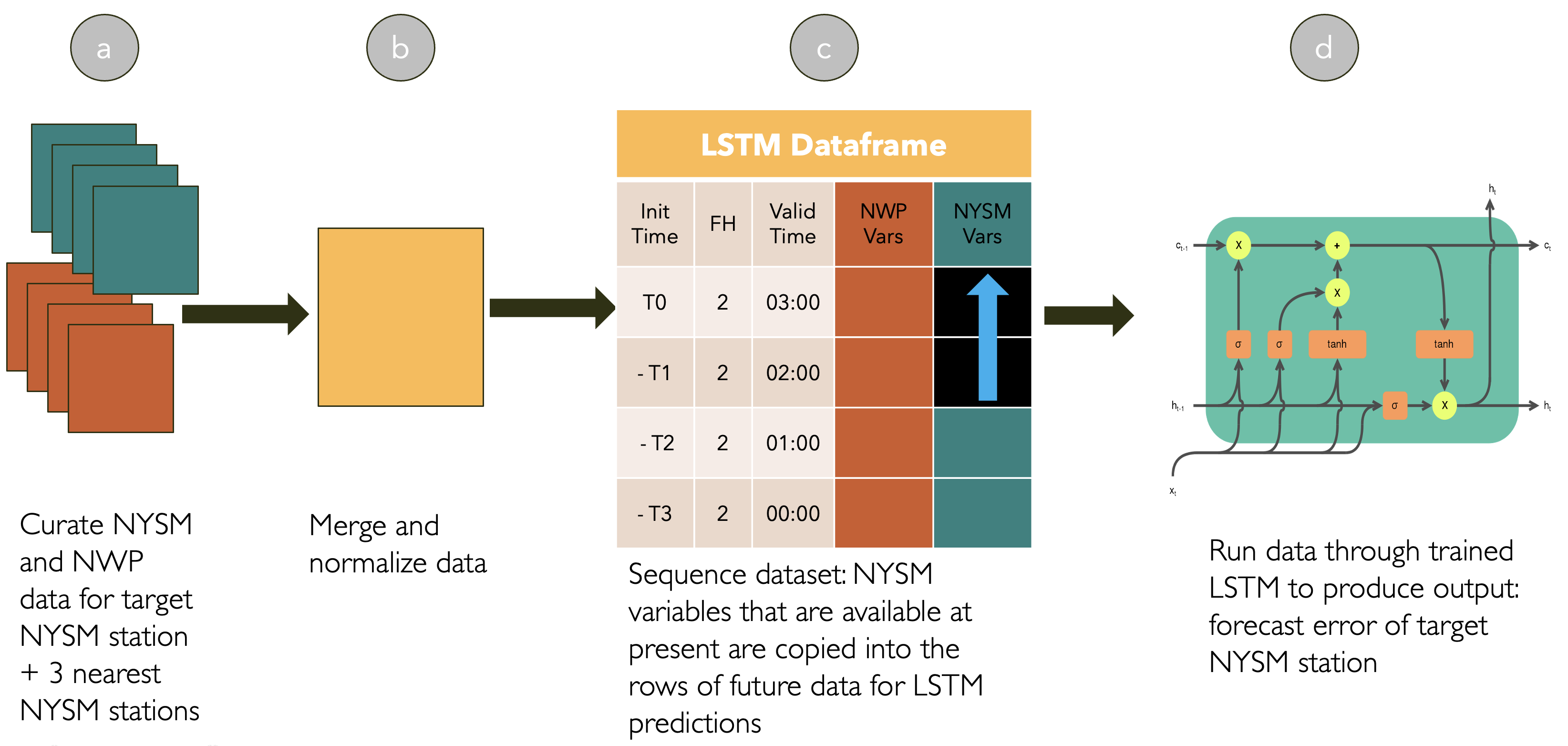}
    \caption{The diagram illustrates the persistence method applied to an LSTM for HRRR forecast error prediction, using the NYSM, and analogously for the OKSM.}\label{fig:persistence-encoder-method}
\end{figure*}

The LSTM was first introduced in 1997 \citep{Sepp1997} and builds upon the recurrent neural network (RNN) architecture but is modified to correct for the vanishing gradient problem from backpropagation of error \citep{Wang2021}. A detailed representation of an LSTM cell is provided in Fig.~\ref{fig:persistence-encoder-method}(d), and its gated operations are described in the documentation provided in \citet{PyTorch2024}, but at a high level, the LSTM can solve sequence prediction problems by adding the input gate, the forget gate, and the output gate to the memory unit in the feed-forward RNN \citep{Wang2021}. The extended memory unit determines which information to keep and forget based on operations at each of these gates \citep{Wang2021}. Due to the ability to remember information over longer time-scales, the LSTM network outperforms the RNN at capturing and generalizing long-term dependencies on the data \citep{Wang2021}. 

As described above, HRRR data is co-located with mesonet stations in space and time, merged, and then normalized using the standard z-score normalization algorithm (Fig. 1(a) and (b)) by batch. Additionally, batch-wise normalization reduces sensitivity to temporal and spatial variability in the data distribution, which is particularly important given the heterogeneity across stations and evolving atmospheric conditions. Each time series input to the LSTM encoder is specific to a given forecast hour. We apply a persistence method to align mesonet observations with future HRRR forecasts to preserve sequence integrity. As shown in Fig.~\ref{fig:persistence-encoder-method}(c), when the LSTM is used to predict forecast error, e.g., two hours ahead, there are naturally two missing rows corresponding to the unavailable mesonet observations at those future times. To maintain continuity in the input sequence, we persist (copy) the most recent mesonet observation into these missing future rows, ensuring the structure of the sequence remains consistent and therefore compatible with LSTM encoder operations. The resulting time series is then passed into the LSTM encoder Fig.~\ref{fig:persistence-encoder-method}(d). Note that other methods were tested (e.g., filling missing data with -999, NaN, masking) and found to be ineffective.

After an input time series passes through the gated LSTM operations (Fig.~\ref{fig:persistence-encoder-method}(d)), the final hidden state of the LSTM encoder is transferred to the decoder, as illustrated in Fig.~\ref{fig:entire-model}(a). The final hidden state effectively captures the encoded representation of the HRRR forecast and mesonet observations at the current time step.

\subsubsection{LSTM Decoder Architecture}

\begin{figure*}
    \centering
    \includegraphics[width=30pc]{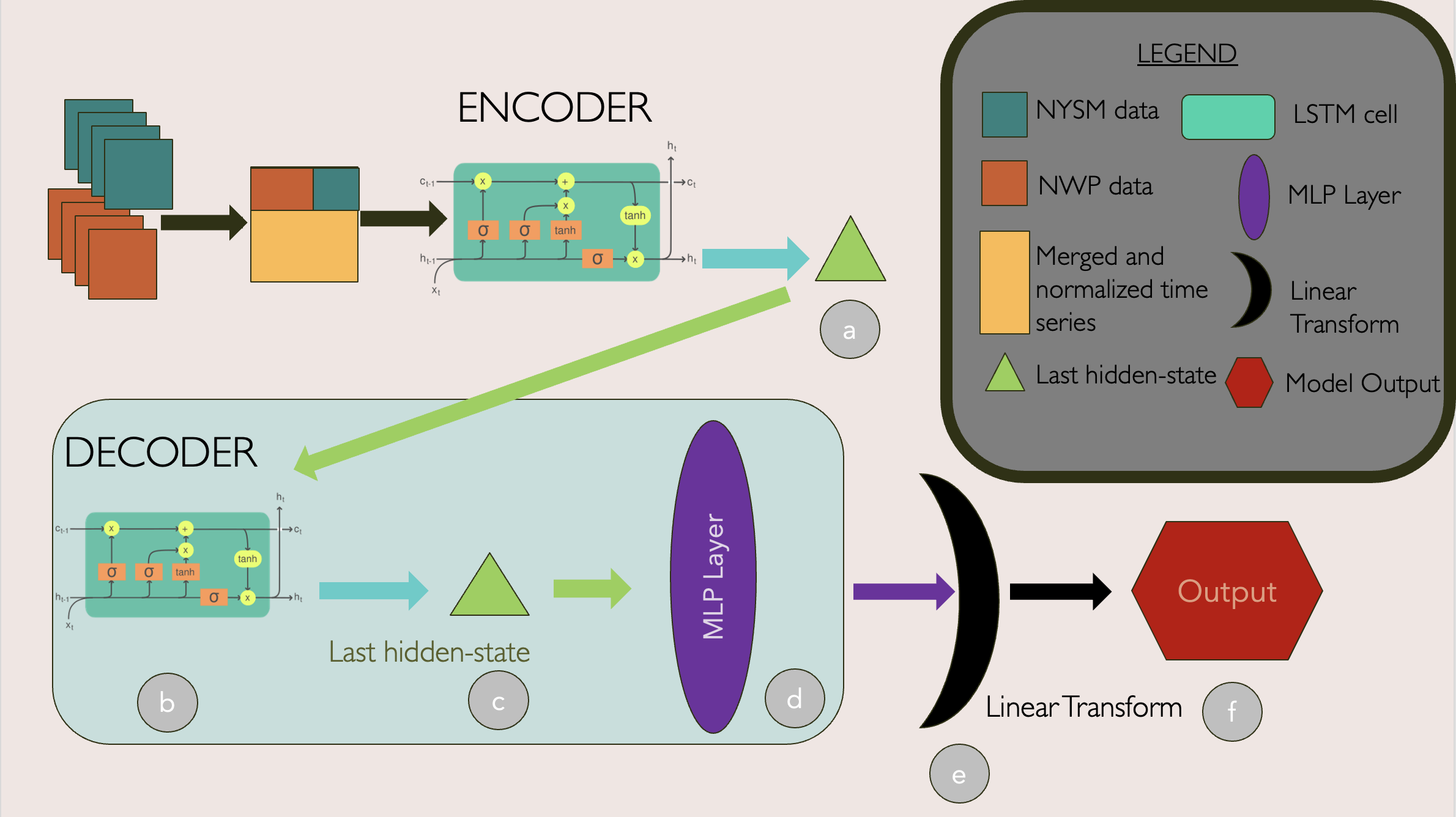}
    \caption{The diagram illustrates a high-level representation of the LSTM encoder-decoder workflow.}
    \label{fig:entire-model}
\end{figure*}

The flow described above and illustrated in Fig.~\ref{fig:persistence-encoder-method} is only the first component of the LSTM. Figure~\ref{fig:entire-model} illustrates the subsequent components. The LSTM decoder block (blue rectangle) begins with another LSTM cell (Fig.~\ref{fig:entire-model}(b)), which performs the same gated operations as described in \citet{PyTorch2024}. The last hidden state of the decoder (Fig.~\ref{fig:entire-model}(c), green triangle) is passed to a fully connected dense layer, or multi-layer perceptron (MLP, Fig.~\ref{fig:entire-model}(d), purple oval). The advantage of using an MLP output layer is that the hidden layers within the MLP contain learnable parameters that are updated while the LSTM is trained, making the MLP more effectively dynamic at capturing and modeling complex nonlinear relationships than a simple linear transformation of the last hidden state \citep{bishop2023}.

The decoder block (Fig.~\ref{fig:entire-model}, blue rectangle) is executed recursively, often referred to as “rolling out”, to predict forecast error across all forecast hours associated with the HRRR. The decoder cell executes this recursive process by accepting its own previously calculated hidden state and cell state as the input for the following calculation, or forecast hour. The decoder recursively updates \textit{n} number of times associated with the forecast hour targeted for output. 

\subsubsection{Linear Post Processing Function}

Lastly, we apply linear post-processing (black crescent, Fig.~\ref{fig:entire-model}(e)) to tailor the LSTM output to the individual NYSM station, forecast hour, and predictand of interest. The coefficients used for the linear post-processing calculations are determined using the validation fold of the data and are stored in a look-up table for testing and inference use. This linear transformation allows us to cost-effectively take a generalized LSTM output and introduce an effective bias term that further tailors the LSTM output to the target of interest. 

Refitting improves the explained variance ($R^2$) in over 80\% of cases across all variables, with the most pronounced gains for precipitation ($>95\%$), followed by temperature ($\sim83\%$) and wind ($\sim81\%$). The largest increases in $R^2$ are observed for precipitation ($\Delta R^2 \approx 0.81$), with meaningful improvements also seen for temperature ($\Delta R^2 \approx 0.32$) and wind ($\Delta R^2 \approx 0.24$). Together, these results demonstrate that the refitting process substantially enhances the model’s ability to capture variability and structure in forecast error, particularly in regimes characterized by high nonlinearity and intermittency.

\subsection{Model Training}
\subsubsection{Custom Loss Function}
The goal during model training is to minimize loss, or the quantifiable difference between the LSTM-predicted and target variables. The LSTM weights and parameters are updated using a custom loss function, as shown in Equation \ref{eq:outlier_loss}, designed to give greater weight to the correct prediction of outliers \citep{ebert-uphoff2021cira}. Equation \ref{eq:outlier_loss} enhances the overall LSTM performance by ensuring that outlier predictions are accounted for, something standard loss functions often avoid in favor of improving accuracy on more commonly expected patterns in the time series. Since the primary goal of the LSTM is to identify when the NWP model forecast output is incorrect, we prioritize accurate outlier predictions over mean-state points.

The sensitivity of the loss formulation to the weighting parameter $\alpha$ was also evaluated. Increasing $\alpha$ increases the penalty applied to large-error events and improves responsiveness to higher-magnitude forecast errors up to a threshold, though excessively large values can lead to overly conservative predictions and reduced sensitivity to extremes, while smaller values behave more similarly to standard loss formulations and under-emphasize outlier regimes.

\begin{equation}
\scalebox{1.}{$
    \text{OutlierFocusedLoss}(y_{\text{true}}, y_{\text{pred}}) = \frac{1}{n} \sum_{i=1}^{n} \left( \left( |y_{\text{true}, i} - y_{\text{pred}, i}| + 1 \right)^\alpha \times |y_{\text{true}, i} - y_{\text{pred}, i}| \right)
$},
\label{eq:outlier_loss}
\end{equation}
\noindent
where:
\begin{itemize}
    \item $y_{\text{true}, i}$ is the true value of the $i^{\text{th}}$ observation.
    \item $y_{\text{pred}, i}$ is the predicted value of the $i^{\text{th}}$ observation.
    \item $n$ is the total number of observations.
    \item $|y_{\text{true}, i} - y_{\text{pred}, i}|$ is the absolute error for the $i^{\text{th}}$ observation.
    \item $\alpha \in \mathbb{R}^+$ is a tunable hyperparameter that controls the sensitivity of the loss function to large errors.
    \item The term $\left(|y_{\text{true}, i} - y_{\text{pred}, i}| + 1\right)^\alpha$ amplifies the contribution of larger errors, encouraging the LSTM to focus on outliers.
\end{itemize}

\subsubsection{Hyperparameter Tuning}
Referencing Table~\ref{tab:hyperparameters}, hyperparameter tuning was performed using a two-stage procedure to balance efficiency and performance. An initial structured grid search was used to identify stable training regimes and constrain the hyperparameter space, exploring key optimization and architectural parameters (e.g., learning rate, batch size, weight decay, dropout, early-stopping patience, hidden size, number of layers, sequence length, and $\alpha$) under a consistent train–validation–test split. Bayesian optimization was then applied to refine performance within promising regions, using validation metrics (e.g., MAE or RMSE) as the objective, with additional diagnostics such as correlation and bias used to assess model behavior. All experiments were tracked using CometML to ensure reproducibility. Final hyperparameters (Table~\ref{tab:hyperparameters}) were selected based on validation performance, prioritizing both predictive skill and training stability, and a consistent configuration was applied across forecast lead times to maintain comparability and avoid overfitting.

\begin{table}[h!]
\centering
\resizebox{\textwidth}{!}{%
\begin{tabular}{ll}
\hline
\textbf{Hyperparameter} & \textbf{Value} \\
\hline
Batch Size              & 1000 \\
Learning Rate           & $5 \times 10^{-5}$ \\
Number of Layers        & 3 \\
Hidden Units            & 1728 \\
Sequence Length         & 30 \\
Regularization ($\lambda$) & 0.0 \\
Optimizer               & AdamW \\
Scheduler               & ReduceLROnPlateau (factor=0.1, patience=4) \\
Early Stopping          & Patience = 8 epochs \\
MLP Units (Decoder)     & 1500 \\
$\alpha$ (Loss Function) & 2.0 \\
\hline
\end{tabular}}
\caption{Hyperparameters for the LSTM model used in this study.}
\label{tab:hyperparameters}
\end{table}

\subsubsection{Model Output Evaluation}
Model performance was evaluated using aggregate metrics and targeted diagnostics. Mean absolute error (MAE), mean squared error (MSE), coefficient of determination ($R^2$), and Pearson correlation coefficient ($r$) quantified overall skill across stations and lead times. Additional diagnostics, including predicted-versus-observed scatter plots, time series analysis, and regime-based stratification (e.g., precipitation intensity, wind magnitude, temperature ranges), were used to assess bias, heteroscedasticity, and temporal structure. These analyses provide a comprehensive evaluation of model behavior and its dependence on physically meaningful variability.

\subsubsection{Training Stability}
Given the use of station-specific models and recursive decoding, training stability and overfitting control were carefully managed through complementary strategies. Early stopping (patience = 8) was employed as the primary regularization mechanism to prevent overfitting and limit error accumulation across forecast lead times. A dynamic learning rate scheduler (ReduceLROnPlateau; patience = 4) was used to stabilize optimization by reducing the learning rate during validation plateaus, enabling smoother convergence. Model capacity was controlled through hyperparameter tuning, with deeper or higher-capacity architectures yielding diminishing returns or reduced stability. Although weight decay was included in the search space, the optimal configuration corresponded to minimal regularization ($\lambda = 0.0$), with generalization instead governed by early stopping and architectural constraints.

\section{Results}
\label{Results}
LSTM performance\footnote{For clarity, throughout the \textbf{Results} section, ``prediction" is used to refer to LSTM prediction output, and ``forecast" is used to refer to HRRR forecast output.} is evaluated for three target variables across both the NYSM and OKSM domains: total hourly precipitation error, wind speed error, and temperature error. Independent models are trained for each variable and for each of the 244 stations in both networks. As shown in the Appendix, New York contains heterogeneous LULC and complex terrain, while Oklahoma is far more homogeneous with relatively flat, unobstructed topography -- with the exception of the eastern portion. The atmospheric regimes also differ: New York weather is driven largely by synoptic-scale variability with additional influences from continental air masses and coastal interactions along the Atlantic and Great Lakes, whereas Oklahoma is shaped primarily by convective processes along the dryline, together with synoptic and mesoscale patterns characteristic of the Southern Great Plains. These contrasting physical and dynamical environments provide a useful baseline for comparing LSTM skill. Table~\ref{tab:performance_summary} provides a consolidated summary of the results presented in this section, highlighting notable key differences in model performance across variables, regions, seasonality, and time of day.

\subsection{Precipitation Error}
Precipitation is one of the most consequential meteorological variables and remains a central challenge for accurate forecasting; it's notoriously difficult for NWP models to forecast due to pronounced space- and time-variability, especially in convective regimes. Precipitation also poses unique challenges for error prediction because it is a discontinuous, non-negative, accumulated quantity with skewed distributions and sharp spatial gradients that are difficult for physical and statistical models to capture. We focus our initial analysis on precipitation because of its critical role in operational meteorology and its substantially different climatological characteristics in New York and Oklahoma.

\subsubsection{New York State Mesonet}
\begin{figure*}[htbp]
    \centering
    \includegraphics[width=30pc]{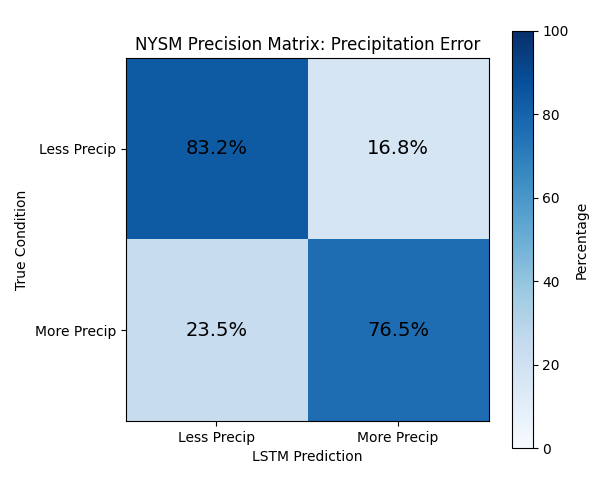}
    \caption{Confusion matrix summarizing the precision of LSTM predictions for precipitation points across the entire NYSM and forecast hours. Rows indicate the true condition, and columns indicate the LSTM’s prediction. More (less) precipitation translates to more (less) precipitation that occurred than was forecast by the HRRR.}
    \label{fig:confusion_matrix_nysm}
\end{figure*}

Using a standard ML definition of precision\footnote{Precision is defined as the proportion of predicted positive cases that are truly positive \citep{GoogleMLCrashCourseMetrics}.}, Fig.~\ref{fig:confusion_matrix_nysm} illustrates LSTM model performance in classifying the sign of HRRR precipitation error, with an overall combined precision of 79.85\%. Moreover, the LSTM is 6.7\% more precise in identifying wet bias (i.e., instances where the HRRR overpredicts precipitation) compared to dry bias (i.e., instances where the HRRR underpredicts precipitation).

\begin{figure*}[htbp]
    \centering
    \includegraphics[width=30pc]{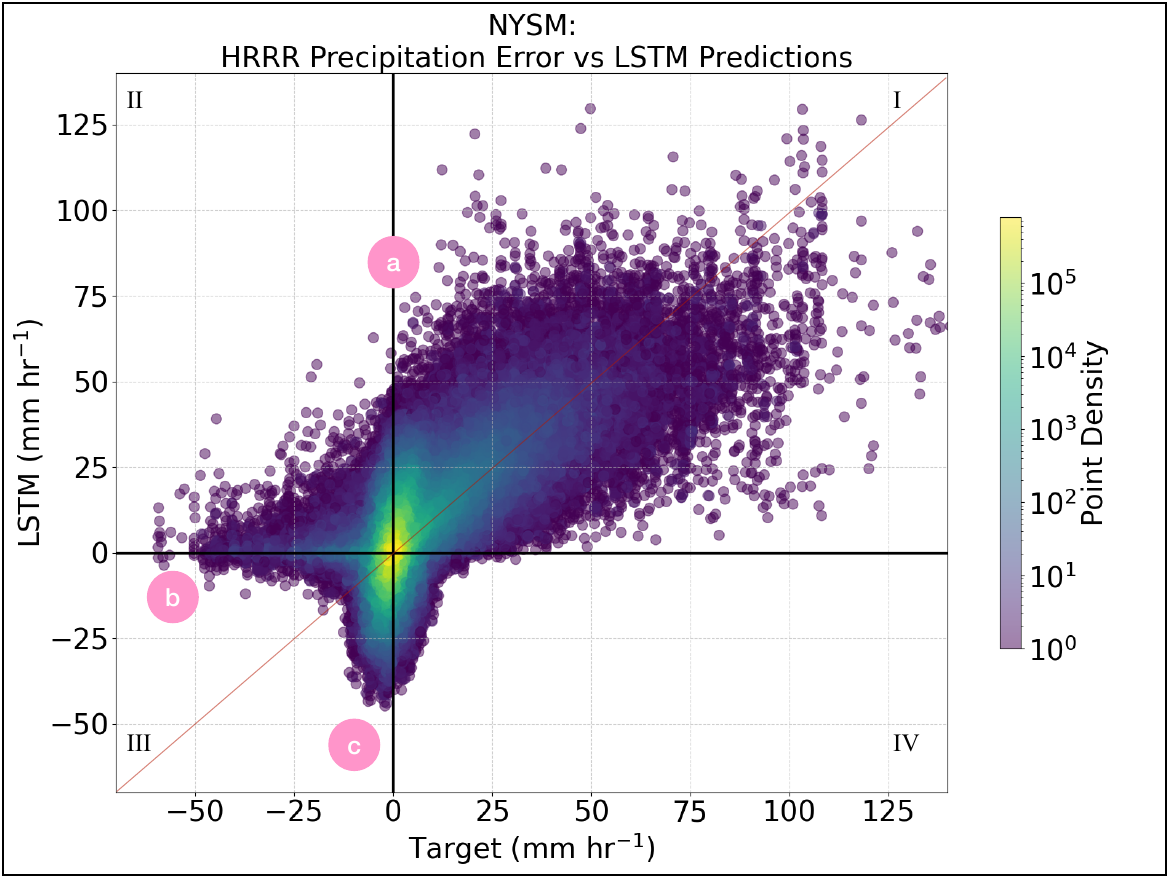}
    \caption{Scatterplot of the precipitation error across the NYSM network and all forecast hours, with the x-axis representing the true target error and the y-axis showing the corresponding LSTM-predicted error. The red diagonal line indicates the 1:1 line, where perfect predictions would lie.}
    \label{fig:scatterplots_nysm_tp}
\end{figure*}

Figure~\ref{fig:scatterplots_nysm_tp} compares true versus LSTM-predicted precipitation errors, with the red diagonal denoting the 1:1 line of perfect agreement. Within $\pm~5~\mathrm{mm~hr^{-1}}$, approximately $79\%$ of points fall on or near the 1:1 line. Further examination of the results reveals the asymmetric pattern first seen in Fig.~\ref{fig:confusion_matrix_nysm}: the LSTM captures positive precipitation errors (wet biases) well but systematically underestimates the magnitude of negative errors (dry biases). The strong covariance along the positive-error quadrant (Q1) further indicates that the LSTM effectively reproduces the magnitude of wet bias in HRRR forecasts. While some covariance is expected due to the physical relationship between precipitation magnitude and forecast error, the alignment along the 1:1 line indicates that the LSTM captures both the magnitude and sign of HRRR error with substantial fidelity. This asymmetry suggests that the model is not solely relying on precipitation magnitude as a proxy for error, but is instead capturing regime-dependent structure in forecast bias.

Referencing Fig.~\ref{fig:scatterplots_nysm_tp}(b), there are notable limitations to LSTM performance: approximately $20\%$ of negative-error cases cluster near the horizontal 0-line (i.e., $y = 0$), indicating that the LSTM predicts near-zero error when the true error is negative. While these negative-error events are often correctly identified in sign (see Fig.~\ref{fig:confusion_matrix_nysm}), their predicted magnitudes are substantially lower than the observed values (see Fig.~\ref{fig:scatterplots_nysm_tp}). This discrepancy suggests that, although the LSTM can identify instances where observed precipitation exceeds the HRRR forecast, it systematically underestimates the severity of these negative errors. Such behavior has important implications for operational forecasting, particularly in scenarios where underforecasting precipitation poses greater risk than overforecasting.

As shown in Fig.~\ref{fig:scatterplots_nysm_tp}(a), there is a concentration of points near the vertical 0-line (i.e., $x = 0$), indicating that the LSTM is highly sensitive to small precipitation errors ($<10~\mathrm{mm~hr^{-1}}$). In these cases, the model frequently predicts non-zero error (either overpredicting or underpredicting) even when the true HRRR error is minimal. Similarly, Fig.~\ref{fig:scatterplots_nysm_tp}(c) shows a concentration of points near the vertical 0-line, corresponding to LSTM false alarms -- cases where the model predicts non-zero error when little to no error is present. Most of these instances involve low-magnitude precipitation errors, suggesting that the LSTM is over-responsive to small errors across both positive and negative regimes. Importantly, this clustering of small-magnitude errors near zero likely reflects a combination of model behavior and physical limitations. From a modeling perspective, the LSTM may smooth predictions toward the mean in regimes with low signal-to-noise ratio, reducing sensitivity to subtle error structures. From a physical and observational perspective, light precipitation is inherently noisy, spatially intermittent, and difficult to resolve, making it challenging to distinguish between true signal and noise in both the HRRR forecasts and mesonet observations.

\begin{figure*}[htbp]
    \centering
    \includegraphics[width=30pc]{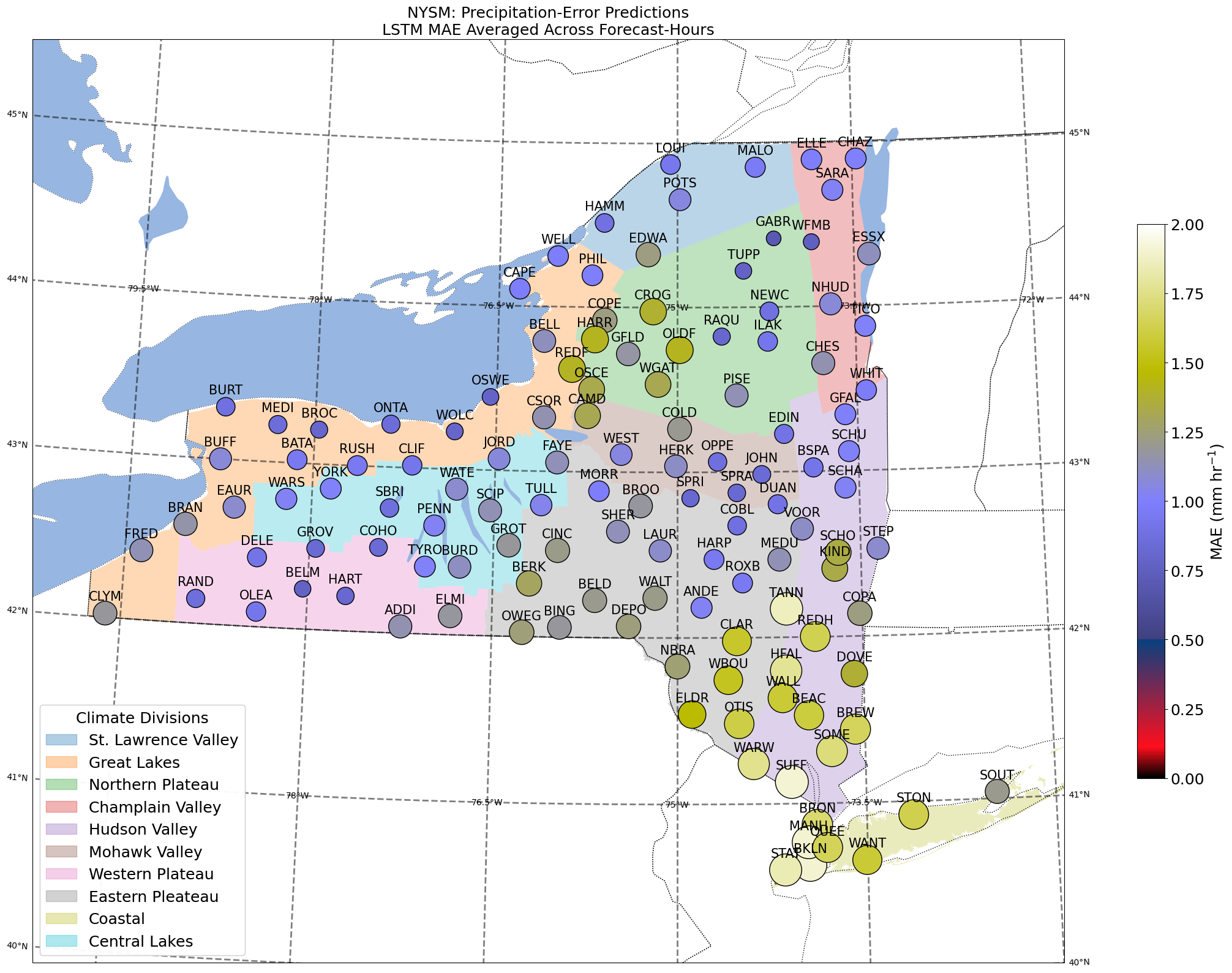}
    \caption{Average LSTM performance (MAE) for an NYSM station, averaged over all forecast lead times. The magnitude of the point is proportional to the MAE, where larger points translate to higher MAE. NCEI climate divisions \citep{NCEI2015} are displayed for reference. A shared color scale is used across domains to enable direct comparison of MAE magnitude; as a result, variability within the OKSM domain appears visually compressed relative to NYSM, and marker size is scaled to aid interpretation.}
    \label{fig:state_nysm_tp}
\end{figure*}

Figure~\ref{fig:state_nysm_tp} shows the MAE associated with the accuracy of LSTM predictions across the NYSM. There are two noticeable regions with elevated MAE. Most prominent are the Eastern Plateau, Hudson Valley, \& Coastal climate divisions. This area is defined by an average of $>1~\mathrm{mm~hr^{-1}}$ higher MAE as compared to the rest of the NYSM. The second region is Tug Hill, situated in the western portion of the Northern Plateau climate division. This area is defined by an average of $>0.5~\mathrm {mm~hr^{-1}}$ higher MAE as compared to the rest of the NYSM. These regions of elevated MAE also experience the highest amount of annual precipitation in the NYSM, as noted in \citet{bader2023methodology}. While this elevated precipitation frequency may contribute to the spatial error patterns observed, disentangling the influence of precipitation-driving dynamics \citep{campbell_steenburgh2017, swain} from the effects of simply receiving more precipitation is beyond the scope of this study. 

\begin{figure*}[htbp]
    \centering
    \includegraphics[width=30pc]{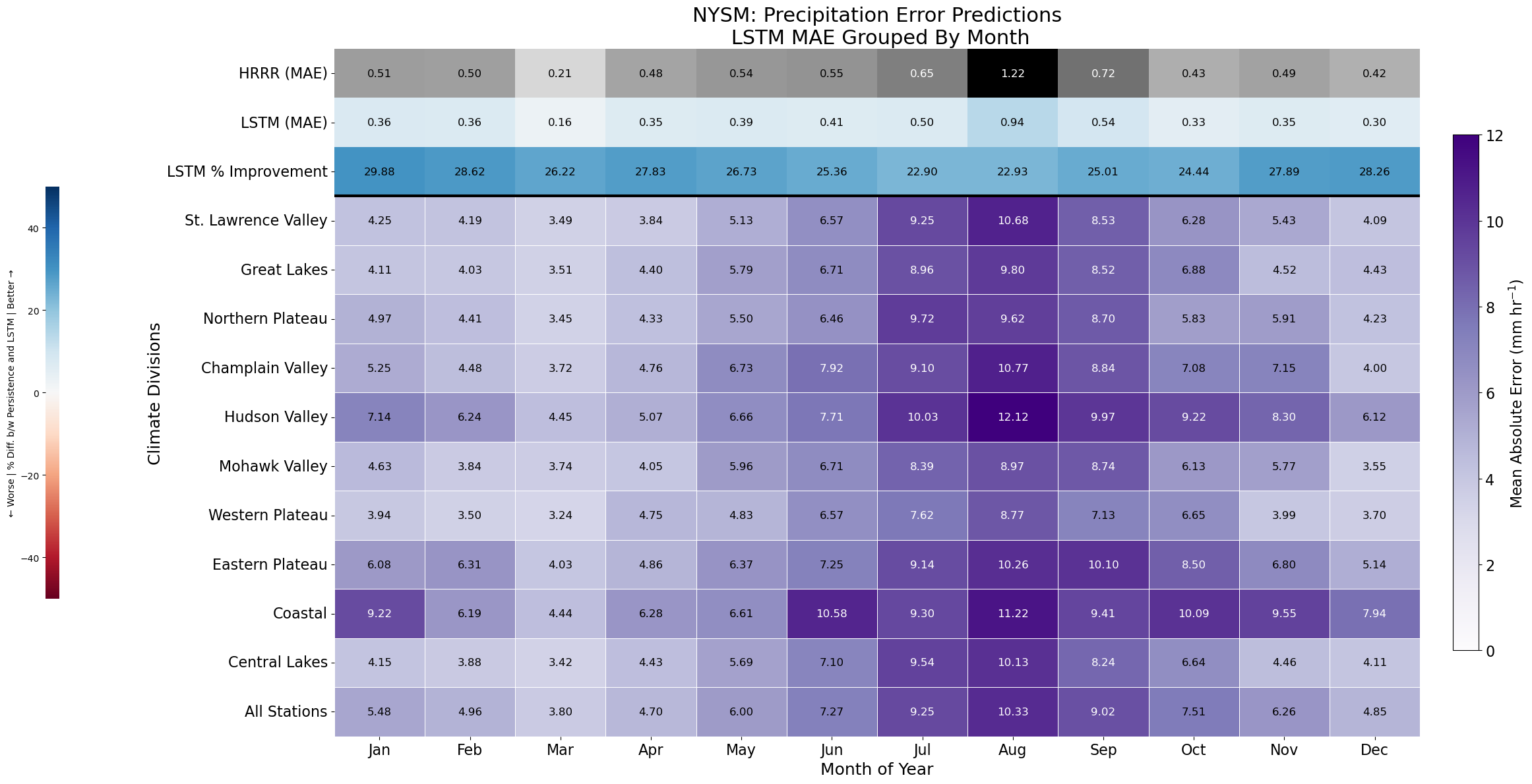}
    \caption{
    NYSM, MAE of LSTM precipitation-error predictions 
    in $\mathrm{mm\,hr^{-1}}$, grouped by month. Rows are arranged from top to bottom as follows:\\[3pt]
    1. HRRR MAE, \textbf{unfiltered}: grey shading proportional to the magnitude of the HRRR MAE average across all stations.\\
    2. LSTM MAE, \textbf{unfiltered}: average MAE across all stations. Blue shading indicates improvement relative to HRRR; red shading indicates degradation relative to HRRR. \\
    3. HRRR--LSTM MAE \% difference, \textbf{unfiltered}: shown using the left color bar to highlight where LSTM improves upon or underperforms HRRR.\\
    4. Climate-division panels*: one panel for each NCEI climate division MAE~\citep{NCEI2015}, enabling region-specific evaluation of LSTM performance (right color bar).\\
    5. All-stations aggregate*: average MAE across all stations (right color bar).\\[4pt]
    \textit{*Note: Panels are filtered to exclude zero-error LSTM predictions to better highlight model failure modes.}
    }
    \label{fig:time_of_year_nysm_tp}
\end{figure*}

Figure~\ref{fig:time_of_year_nysm_tp} presents monthly MAE values ($\mathrm{mm~hr^{-1}}$) for LSTM precipitation error predictions filtered for instances when the LSTM prediction error is non-zero to better highlight model failure modes. The results show strong seasonality: LSTM performance relative to the HRRR forecast decreases slightly during the convective season, and LSTM error is highest during the summer months (July-August), when the MAE magnitude exceeds twice the magnitude of the yearly minimum (approximately $4~\mathrm{mm~hr^{-1}}$), reaching an absolute maximum of $12.12~\mathrm{mm~hr^{-1}}$ in the Hudson Valley division in August. The Hudson Valley, Eastern Plateau, and Coastal climate divisions all display the most coherent secondary error maxima during the winter months (December–February), where MAE values increase by approximately $3~\mathrm{mm~hr^{-1}}$ relative to the yearly minima -- though this signature is present across divisions.  While aggregated metrics provide a useful summary of model performance, there is meaningful variability across individual stations, driven by local terrain, land–surface characteristics, and dominant atmospheric processes. From an operational perspective, this variability is critical, as it indicates that model performance and reliability are location-dependent and should be interpreted at the station level rather than solely through domain-averaged metrics. With that being said, these divisions with elevated errors throughout the year are also consistent with the regional MAE patterns shown in Fig.~\ref{fig:state_nysm_tp}.

\begin{figure*}
    \centering
    \includegraphics[width=30pc]{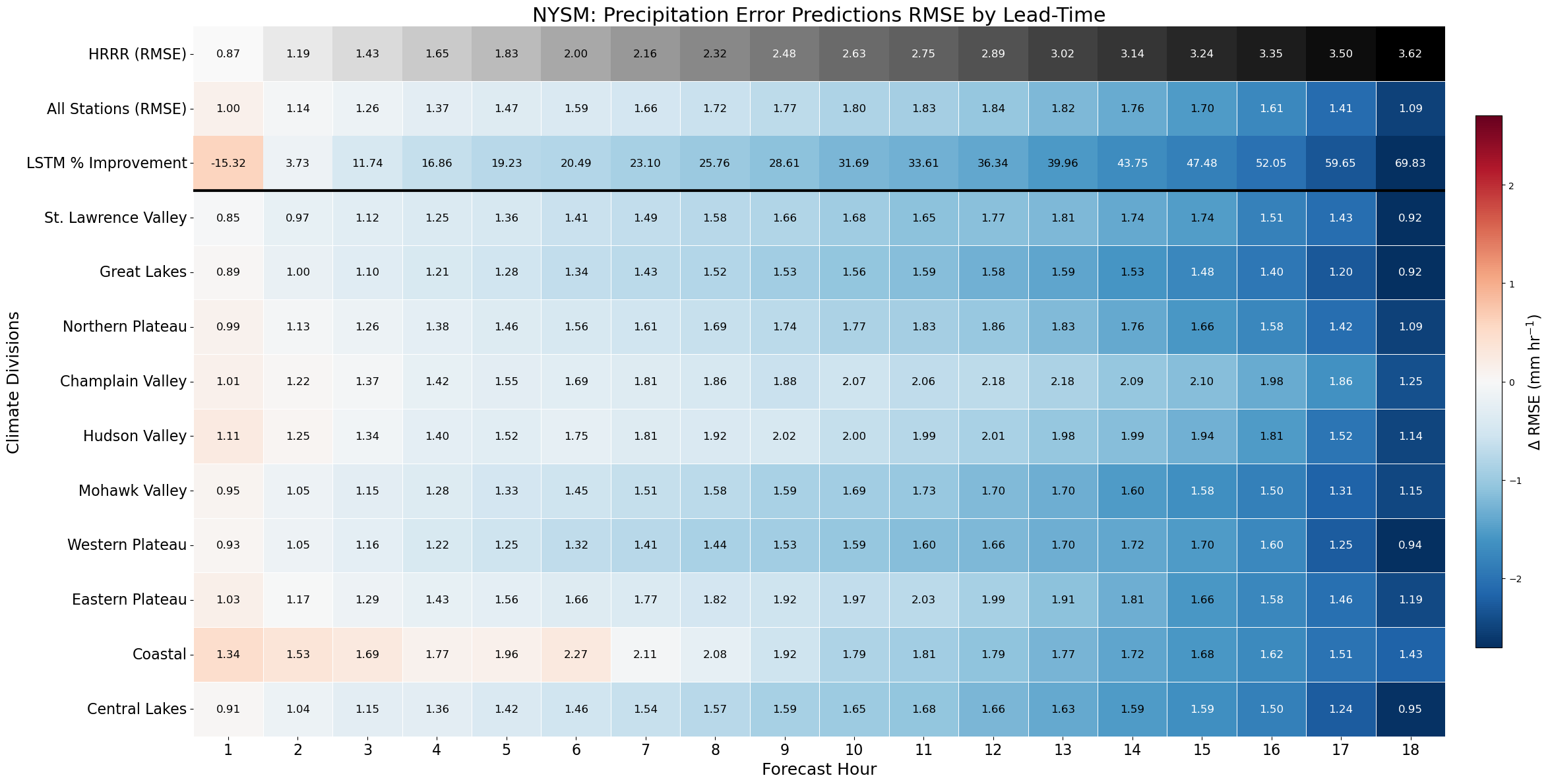}
    \caption{From top to bottom, panels show aggregate root-mean-square error (RMSE) in $\mathrm{mm~hr^{-1}}$ for the HRRR forecast, LSTM predictions, LSTM percent improvement relative to the HRRR (hue is proportional to \% improvement), and then each NYSM climate division. In the HRRR panel, RMSE magnitude is represented by grayscale shading. In the subsequent ML panels, colors denote RMSE differences relative to the HRRR forecast, with red shading indicating higher RMSE and blue shading indicating lower RMSE than HRRR.}
    \label{fig:nysm_tp_wb}
\end{figure*}

Figure~\ref{fig:nysm_tp_wb} illustrates the relative improvement of the LSTM model compared to the HRRR as a function of forecast lead time, with color shading inspired by \citet{WeatherBench2}. The top row shows a monotonic increase in HRRR root-mean-square error (RMSE) with lead time. The second row presents aggregate LSTM RMSE, with blue shading indicating improvement relative to HRRR, highlighting the model’s ability to systematically correct bias proportionally to lead time. The third row shows percent improvement relative to HRRR, while the remaining rows display RMSE by climate division across forecast lead times.

Figure~\ref{fig:nysm_tp_wb} enables analysis of LSTM behavior as a function of lead time without error filtering, highlighting general patterns rather than specific failure modes. Across the NYSM, LSTM RMSE increases with lead time from forecast hour (FH) 1–12, followed by a gradual decrease at longer lead times. In contrast, the relative LSTM improvement over the HRRR across the NYSM increases steadily with lead time across climate divisions. The Coastal climate division exhibits the slowest improvement relative to the HRRR with increasing lead time, with gains emerging at FH7 rather than FH3 as seen across most other divisions. This behavior is consistent with earlier findings in this section.

\subsubsection{NYSM Precipitation Error Discussion}

The LSTM false alarms in Fig.~\ref{fig:scatterplots_nysm_tp}(c) are strongest for observed light precipitation, which is inherently noisier and difficult to capture in Mesonet and HRRR data. The false alarms are concentrated in synoptic weather patterns, with $< 1 \%$ occurring in summer convective months. 

The LSTM’s overprediction bias (Fig.~\ref{fig:scatterplots_nysm_tp}(a)) is concentrated in summer convective months across the Eastern Plateau, Lower Hudson Valley, and Coastal divisions (Fig.~\ref{fig:time_of_year_nysm_tp}), accounting for $> 50 \%$ of all instances across divisions and time of year. These likely occur in environments where convective initiation is inhibited, such as when parcels fail to reach their level of free convection due to capping or insufficient instability, or the event passes just outside the observing station. Urban amplification and land–sea contrasts \citep{swain} in the discussed regions likely further contribute to the elevated MAE in Fig.~\ref{fig:state_nysm_tp}. 

Both vertical clusterings (Fig.~\ref{fig:scatterplots_nysm_tp}(a), (c)) also reflect the well-known “double-penalty” effect, wherein small timing or spatial errors in precipitation forecasts lead to disproportionate penalties in verification \citep{Gilleland_2009, lagerquist_2022, Bonavita2024}.

Cold-season precipitation patterns are modulated by lake-effect processes in the Great Lakes region, where orographic lifting enhances localized snowfall \citep{campbell_steenburgh2017}. These narrow snow bands, as well as the difficulty in predicting orographic enhancement to snowfall rates, as well as forcings linked to complex land-sea interactions, likely explain the secondary winter maxima in most climate divisions in Fig.~\ref{fig:time_of_year_nysm_tp} and the localized error structure across the western NYSM (Fig.~\ref{fig:state_nysm_tp}). 

This interpretation is supported quantitatively by the seasonal distribution of error events (Fig.~\ref{fig:time_of_year_nysm_tp}), with winter and early spring months (January–March) accounting for the largest fraction of erroneous predictions across multiple divisions (e.g., exceeding 20–30\% in several northern regions), consistent with cold-season processes such as lake-effect precipitation and synoptic forcing. In contrast, summer months exhibit a comparatively reduced frequency of these errors, despite higher convective activity, reflecting differences in precipitation structure and predictability across regimes. It is important to distinguish between intrinsic predictability differences in the atmospheric system and limitations of the modeling framework when interpreting these results. In many cases, elevated errors reflect a combination of both factors.

\subsubsection{Oklahoma State Mesonet}

\begin{figure*}[htbp]
    \centering
    \includegraphics[width=30pc]{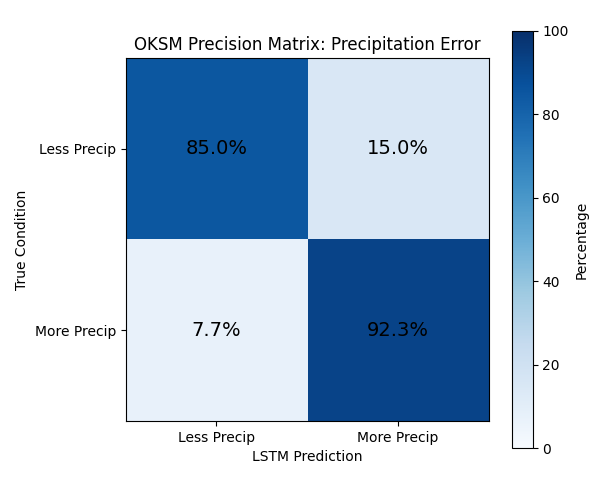}
    \caption{As in Fig.~\ref{fig:confusion_matrix_nysm}, but for the OKSM}
    \label{fig:confusion_matrix_oksm}
\end{figure*}

A confusion matrix for the OKSM domain is presented in Fig.~\ref{fig:confusion_matrix_oksm}. This figure illustrates the precision of the LSTM model in detecting HRRR precipitation forecast errors; The LSTM attains a combined precision of 88.65\%, representing an 8.8\% improvement over performance in the NYSM. The LSTM also demonstrates a 7.3\% higher precision in detecting dry bias points relative to wet bias points. Overall, these results indicate that the LSTM exhibits enhanced skill in detecting and predicting precipitation error within the OKSM domain.

\begin{figure*}[htbp]
    \centering
    \includegraphics[width=30pc]{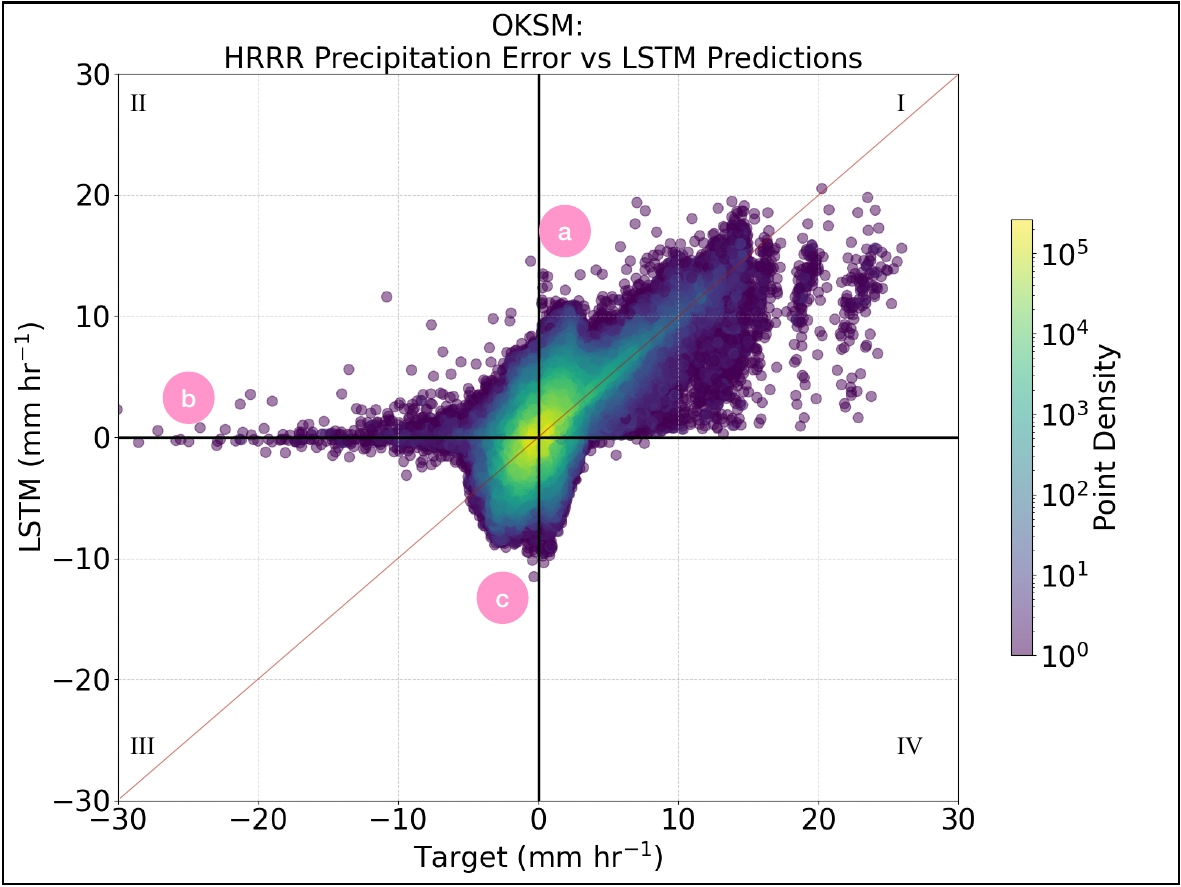}
    \caption{As in Fig.~\ref{fig:scatterplots_nysm_tp}, but for OKSM}
    \label{fig:scatterplots_oksm_tp}
\end{figure*}

Figure~\ref{fig:scatterplots_oksm_tp} compares true versus LSTM-predicted precipitation errors across the OKSM network and all forecast hours. The LSTM captures positive forecast errors reasonably well (quadrant I), with $99\%$ of targeted error points falling within $\pm~5~\mathrm{mm~hr^{-1}}$ of the 1:1 line. However, the LSTM struggles to represent the magnitude of negative errors and small magnitude errors. As in the NYSM (Fig.~\ref{fig:scatterplots_nysm_tp}), both positive (Fig.~\ref{fig:scatterplots_oksm_tp}(a)) and negative (Fig.~\ref{fig:scatterplots_oksm_tp}(c)) vertical clustering along the 0-line, as well as clustering along the negative horizontal 0-line (Fig.~\ref{fig:scatterplots_oksm_tp}(b)), are evident, reflecting systematic over- and underprediction of small magnitude error, and negative errors. The consistency of these “double penalty” signatures across both mesonets suggests that the discussed clusters are likely methodological, rather than being driven by geographical/dynamic forcings \citep{Gilleland_2009, lagerquist_2022, Bonavita2024}.

\begin{figure*}[htbp]
    \centering
    \includegraphics[width=30pc]{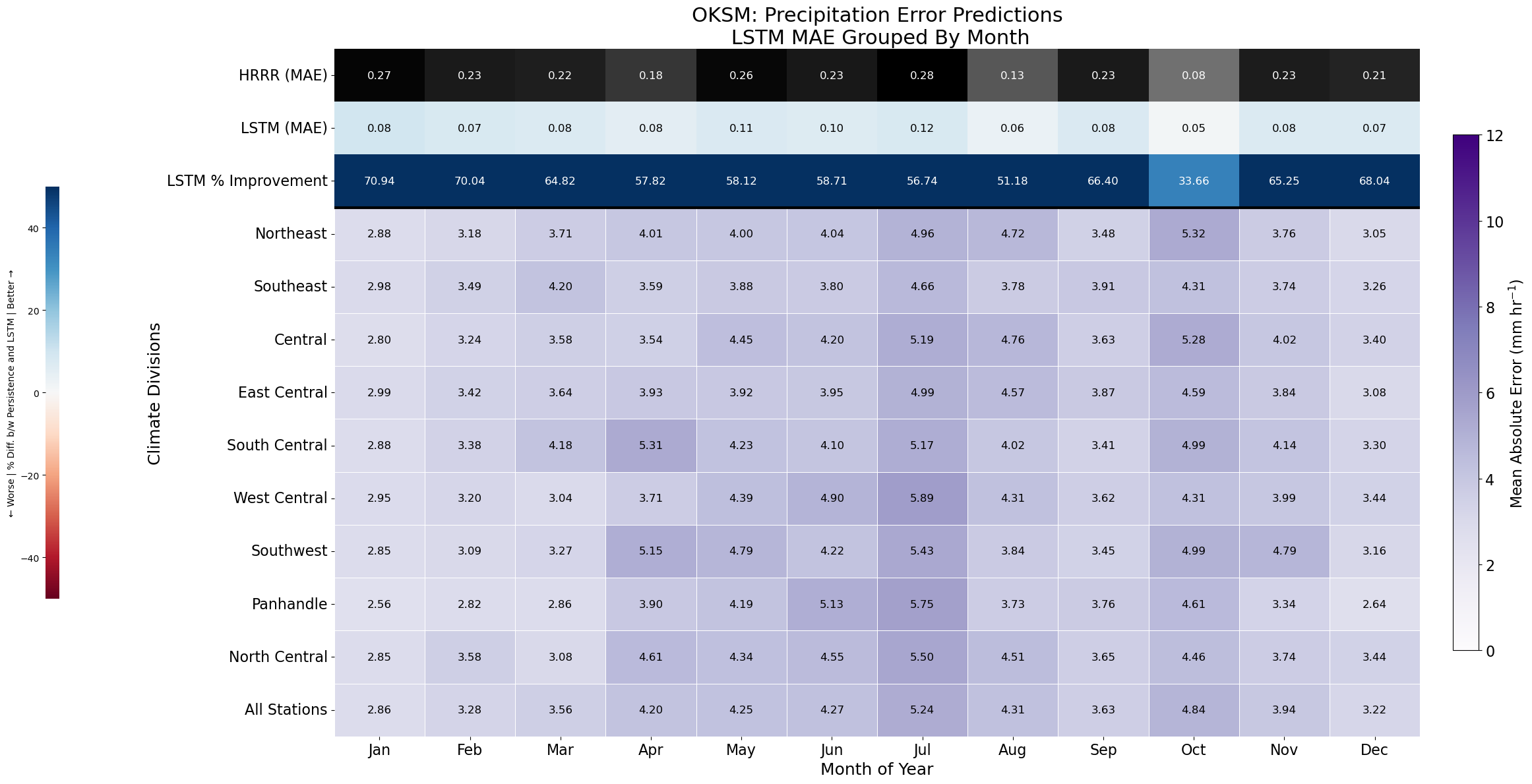}
    \caption{As in Fig.~\ref{fig:time_of_year_nysm_tp}, but for the OKSM.}
    \label{fig:time_of_year_oksm_tp}
\end{figure*}

Figure~\ref{fig:time_of_year_oksm_tp} shows the monthly MAE of LSTM predictions in $\mathrm{mm~hr^{-1}}$ for precipitation error prediction filtered for LSTM predictions with nonzero error. Oklahoma, which experiences convective weather during much of the year, exhibits seasonal peaks in LSTM error during periods of heightened convective activity (Fig.~\ref{fig:time_of_year_oksm_tp}), though to a lesser magnitude than we see in the NYSM. Summer months (May–August) generally show a slight degradation in LSTM improvement compared to the HRRR baseline, as well as higher LSTM prediction error compared to winter, with an average increase in MAE compared to the division minima of approximately $2~\mathrm{mm~hr^{-1}}$ and July as a relative error maxima across all climate divisions (approximately $5~\mathrm{mm~hr^{-1}}$). October and March/April also demonstrate a relative maxima in error, with an average increase in MAE compared to the division minima of approximately $2~\mathrm{mm~hr^{-1}}$, most notably within the Northeast, Central, South Central, and Southwest climate divisions. 

\begin{figure*}[htbp]
    \centering
    \includegraphics[width=30pc]{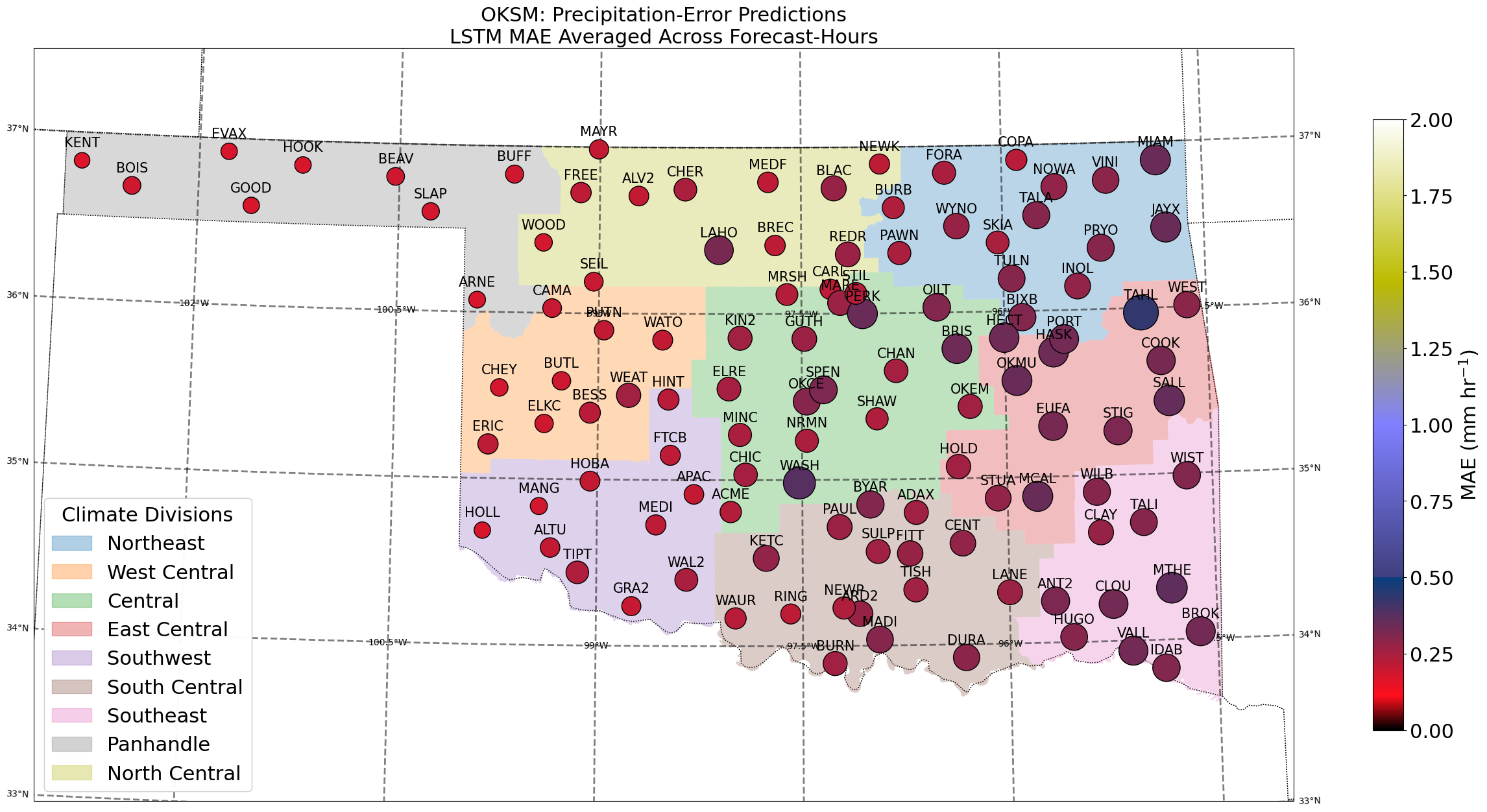}
    \caption{As in Fig.~\ref{fig:state_nysm_tp}, but for the OKSM.}
    \label{fig:state_oksm_tp}
\end{figure*}

Figure~\ref{fig:state_oksm_tp} shows the LSTM MAE for each OKSM station, averaged over all forecast hours. The spatial distribution of LSTM error forms a northeast–southwest gradient across the state, with elevated errors concentrated in the Central, East Central, South Central, Northeast, and Southeast divisions. Given the relatively uniform geography of the OKSM domain, spatial variance in MAE remains minimal, with these higher-error regions exhibiting only a modest increase of about $0.25~\mathrm{mm~hr^{-1}}$.

\begin{figure*}[htbp]
    \centering
    \includegraphics[width=30pc]{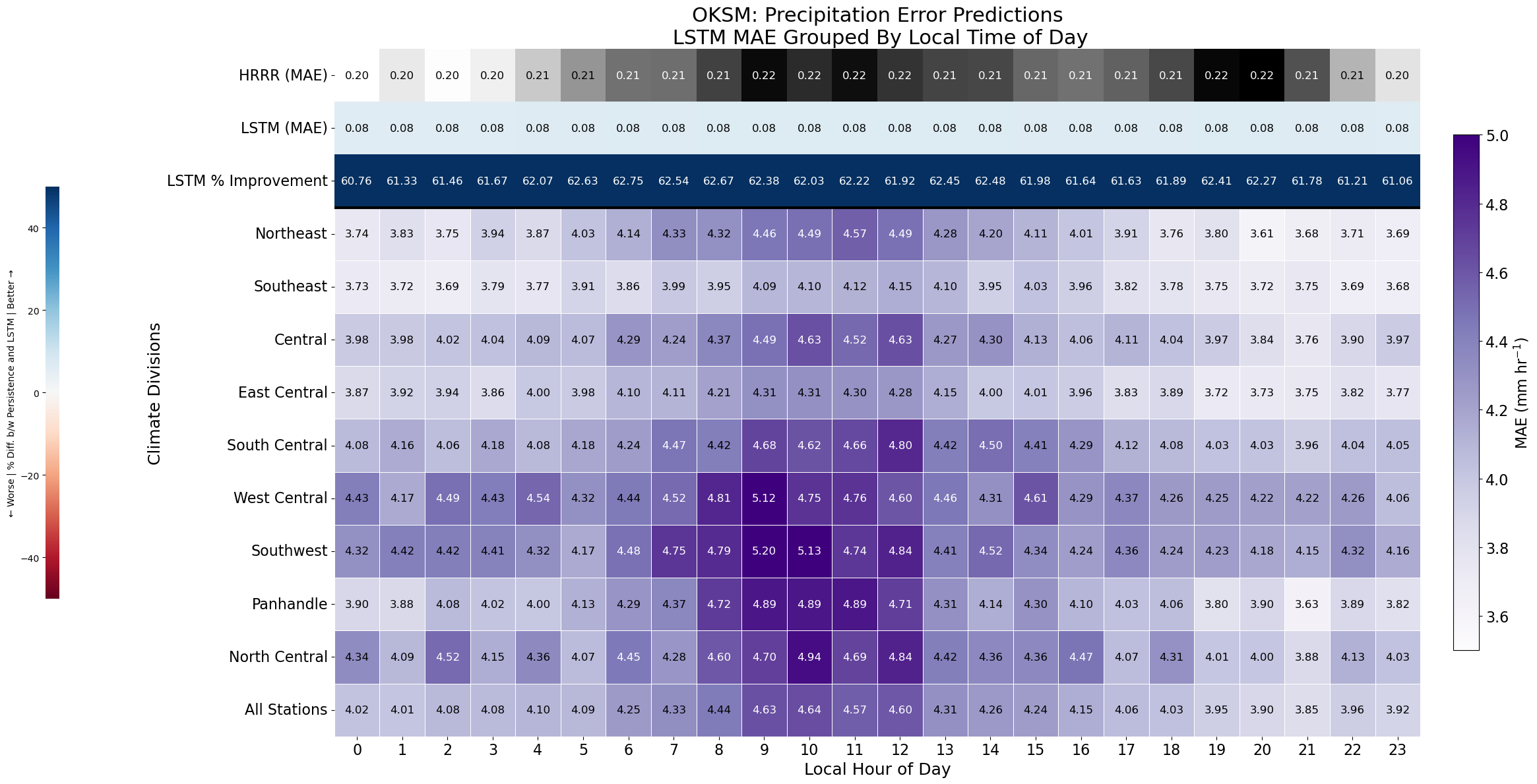}
    \caption{OKSM, MAE of LSTM predictions in $\mathrm{mm~hr^{-1}}$ for precipitation error, grouped by local time of day. Panels are arranged from top to bottom with the same layout and color conventions as Fig.~\ref{fig:time_of_year_nysm_tp}.}
    \label{fig:time_of_day_oksm_tp}
\end{figure*}

Figure~\ref{fig:time_of_day_oksm_tp} shows MAE of LSTM predictions in $\mathrm{mm~hr^{-1}}$ for precipitation error predictions, grouped by time of day; again, this data is filtered for LSTM predictions with non-zero error to better highlight model failure modes. Unlike the NYSM (not shown), the OKSM exhibits a discernible diurnal error signature in precipitation error predictions. Specifically, LSTM error peaks during the morning hours (0900 to 1200), with an average increase in MAE compared to the division minima of approximately $1~\mathrm{mm~hr^{-1}}$. West Central, Southwest, and North Central have coherent secondary maxima in the early morning hours (0000 to 0400) before sunrise, with an average increase in MAE compared to the division minima of approximately $0.25~\mathrm{mm~hr^{-1}}$. The early morning error maxima shown here are consistent with mesoscale convective processes and boundary layer transitions, discussed further in the following section. Importantly, this diurnal signature is not confined to a single region, but is observed consistently across multiple climate divisions, including the West Central, Southwest, North Central, and Panhandle regions. The recurrence of this pattern across geographically distinct areas suggests that it reflects a coherent, domain-wide behavior rather than a localized anomaly.

\begin{figure*}
    \centering
    \includegraphics[width=30pc]{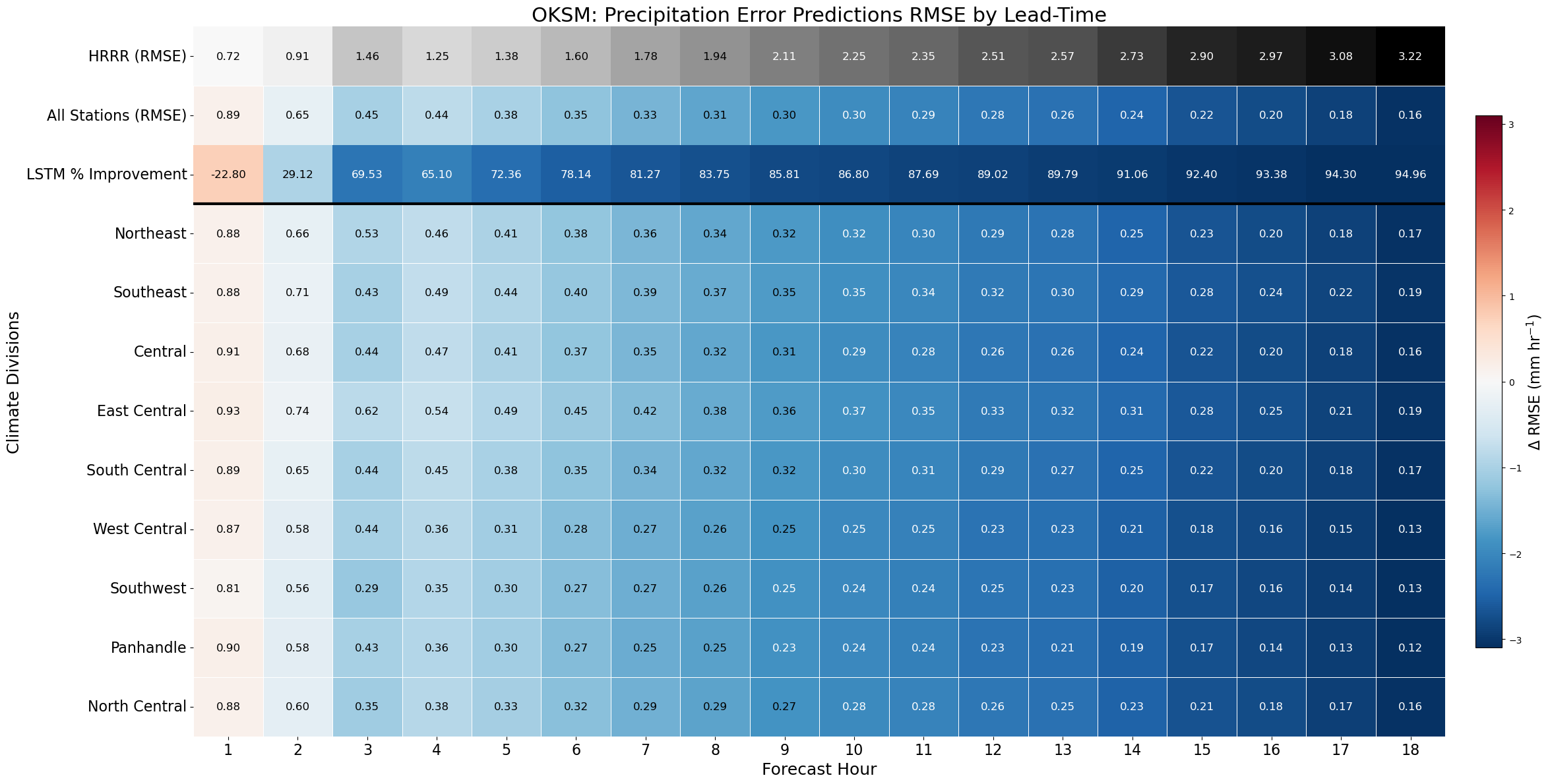}
    \caption{As in Fig.~\ref{fig:nysm_tp_wb}, but for the OKSM.}
    \label{fig:oksm_tp_wb}
\end{figure*}

Figure~\ref{fig:oksm_tp_wb} illustrates the relative improvement of the LSTM model compared to the HRRR as a function of forecast lead time. As in the NYSM, the OKSM shows a monotonic increase in HRRR RMSE with lead time. The second row presents aggregate LSTM RMSE, with blue shading indicating improvement relative to HRRR, highlighting the model’s ability to systematically correct bias proportionally to lead time. The third row shows percent improvement relative to HRRR, while the remaining rows display RMSE by climate division across forecast lead times.

Figure~\ref{fig:oksm_tp_wb} enables analysis of LSTM behavior as a function of lead time without error filtering, highlighting general patterns rather than specific failure modes. Similarly to the NYSM, across the OKSM, LSTM RMSE increases with lead time from FH 1–12, followed by a gradual decrease at longer lead times. In contrast, relative LSTM improvement over the HRRR increases steadily with lead time across climate divisions. The OKSM shows substantially greater improvement than the NYSM across most lead times (excluding FH 1–2), often yielding nearly double the percent improvement relative to the HRRR. As shown in this section, error exhibits minimal variation across climate divisions in the OKSM.

\subsubsection{OKSM Precipitation Error Discussion}

The LSTM appears to struggle with precipitation linked to frontal–dryline interactions, where small positional shifts can drastically change convective outcomes \citep{dryline1, dryline2}. Elevated errors align with climatological dryline zones across central Oklahoma (Fig.~\ref{fig:state_oksm_tp}), possibly reflecting regions of inherently low predictability due to strong sensitivity to small positional shifts in convection, while also highlighting environments that are challenging for the LSTM to represent \citep{hoch2005}. The patterns in Fig.~\ref{fig:time_of_year_oksm_tp} (third row) might reflect dryline-induced convection, which is most volatile during the spring and fall. The LSTM shows the least improvement over the HRRR in October and exhibits performance degradation in late spring as well. In contrast, arid Panhandle and West Central divisions show larger errors primarily in summer (Fig.~\ref{fig:time_of_year_oksm_tp}), coinciding with peak convective activity \citep{ocs_oklahoma_climate}.

Early morning error maxima (Fig.~\ref{fig:time_of_day_oksm_tp}) in the West Central, Southwest, Panhandle, and North Central divisions seem to correspond to convective initiation by atmospheric bores and mesoscale outflows \citep{Haghi2017, Haghi2019, Lin2021}. The associated maxima are also likely linked to the spin-up of the planetary boundary layer (PBL), a key driver of convective initiation and amplification across the Southern Plains \citep{hane2003, Hane2008_MorningMCS_PartII, zheng2019}.

These results, most evident in Fig.~\ref{fig:time_of_year_nysm_tp} \& \ref{fig:time_of_year_oksm_tp} (third row), suggest the LSTM performs best under synoptic-scale precipitation regimes, which are generally more predictable, but struggles in mesoscale or convective contexts characterized by high spatial and temporal variability. These challenges reflect both intrinsic limits to predictability and the model’s reduced ability to represent rapidly evolving, vertically driven processes. The lack of observational vertical information (e.g., shear, instability, moisture advection) represents a limitation of the modeling framework, which likely reduces its ability to capture these processes. At the same time, these regimes are also intrinsically more difficult to predict due to their high spatial and temporal variability. Overall, differences in model performance across regions and regimes should be interpreted as arising from both intrinsic atmospheric predictability and the ability of the model to represent the governing physical processes.

\subsection{Wind Error}

Wind magnitude was selected as another primary predictand due to its critical operational importance for the energy sector (e.g., wind power forecasting), as well as its direct societal impacts, including transportation safety, infrastructure resilience, and wildfire risk. Near-surface wind exhibits pronounced temporal variability and is governed by physical processes that differ substantially from precipitation. Wind magnitude reflects a complex interplay among pressure-gradient forces, surface drag, turbulent momentum transport, and thermally driven circulations within the PBL, with additional modulation by geography and mesoscale forcing.

Despite these complexities, forecast error and bias in wind magnitude are known to be strongly correlated with the wind speed itself \citep{Gaudet2024, Seto2025, Collins2024, Fovell2025}. As a result, the LSTM proves to be better able to learn and anticipate recurring forecast error patterns in wind, enabling more effective correction of systematic biases.

\subsubsection{New York State Mesonet}
\label{NYSM Wind Error}

\begin{figure*}
    \centering
    \includegraphics[width=30pc]{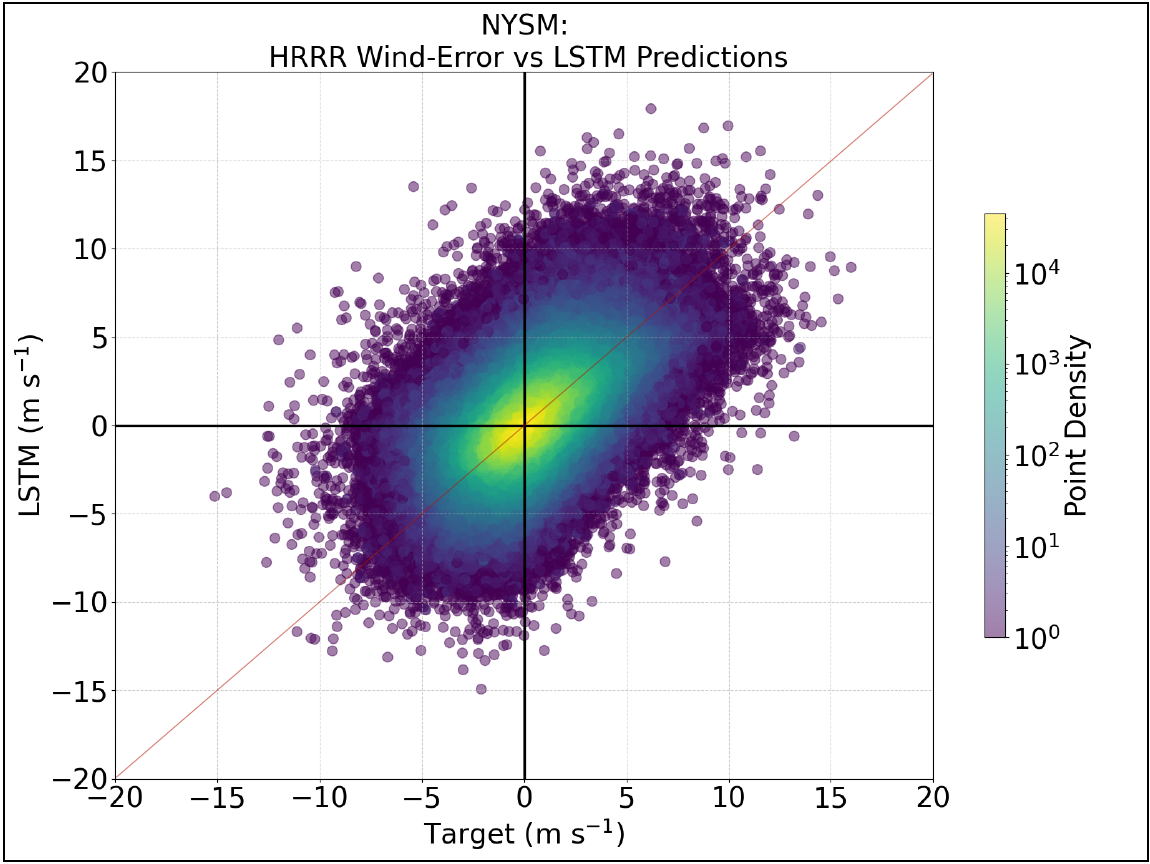}
    \caption{Scatterplot of the wind error across the NYSM network and all forecast hours, with the x-axis representing the true target error in $\mathrm{m~s^{-1}}$ and the y-axis showing the corresponding LSTM-predicted error in $\mathrm{m~s^{-1}}$. The red diagonal line indicates the 1:1 line, where perfect predictions would lie.}
    \label{fig:scatterplots_nysm_wind}
\end{figure*}

Figure~\ref{fig:scatterplots_nysm_wind} compares true and LSTM-predicted wind errors (in $\mathrm{m~s^{-1}}$). The LSTM effectively identifies and predicts both overforecast and underforecast wind, with $92\%$ of targeted error points falling within $\pm~2~\mathrm{m~s^{-1}}$ of the 1:1 line, accurately detecting the occurrence of wind error and its magnitude.

\begin{figure*}
    \centering
    \includegraphics[width=30pc]{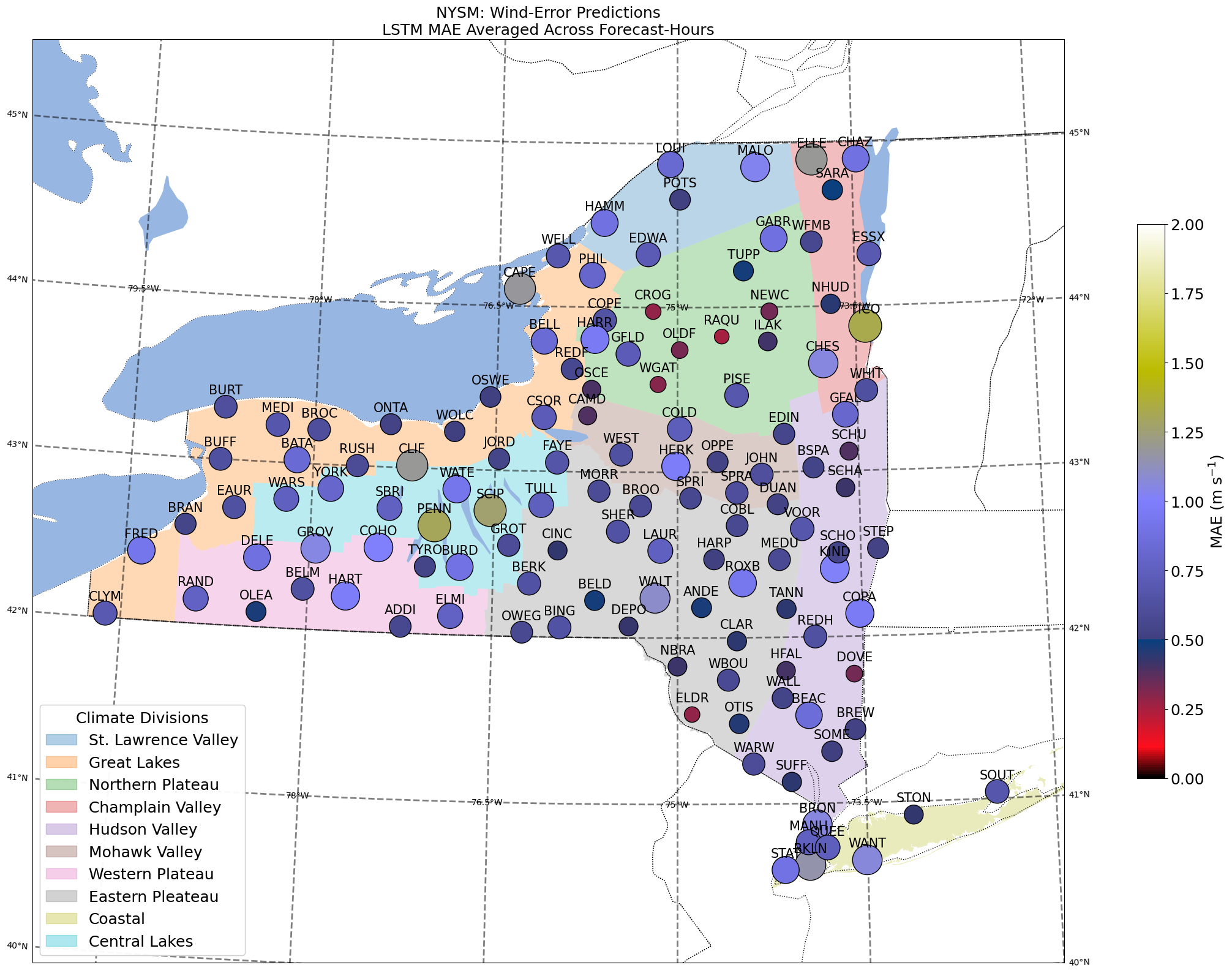}
    \caption{Average LSTM performance (MAE) for an NYSM station, averaged over all forecast lead times. The magnitude of the point is proportional to the MAE, where larger points translate to higher MAE. NCEI climate divisions \citep{NCEI2015} are displayed for reference. A shared color scale is used across domains to enable direct comparison of MAE magnitude; as a result, variability within the OKSM domain appears visually compressed relative to NYSM, and marker size is scaled to aid interpretation.}
    \label{fig:nysm_state_mae_wind}
\end{figure*}

Figure~\ref{fig:nysm_state_mae_wind} shows the LSTM performance (MAE, $\mathrm{m~s^{-1}}$) for each NYSM station, averaged over all forecast hours. LSTM error exhibits a slight negative correlation with station elevation (correlation: $-0.128$, p-score: $0.15$; see Fig.~\ref{fig:elevation_nysm}). The lowest errors occur in the Northern Plateau, with secondary minima in the more topographically complex Eastern Plateau and Taconic Mountains (Hudson Valley). These regions show a reduced MAE of $\sim1~\mathrm{m~s^{-1}}$ compared to the rest of the domain. 

\begin{figure*}[htbp]
    \centering
    \includegraphics[width=30pc]{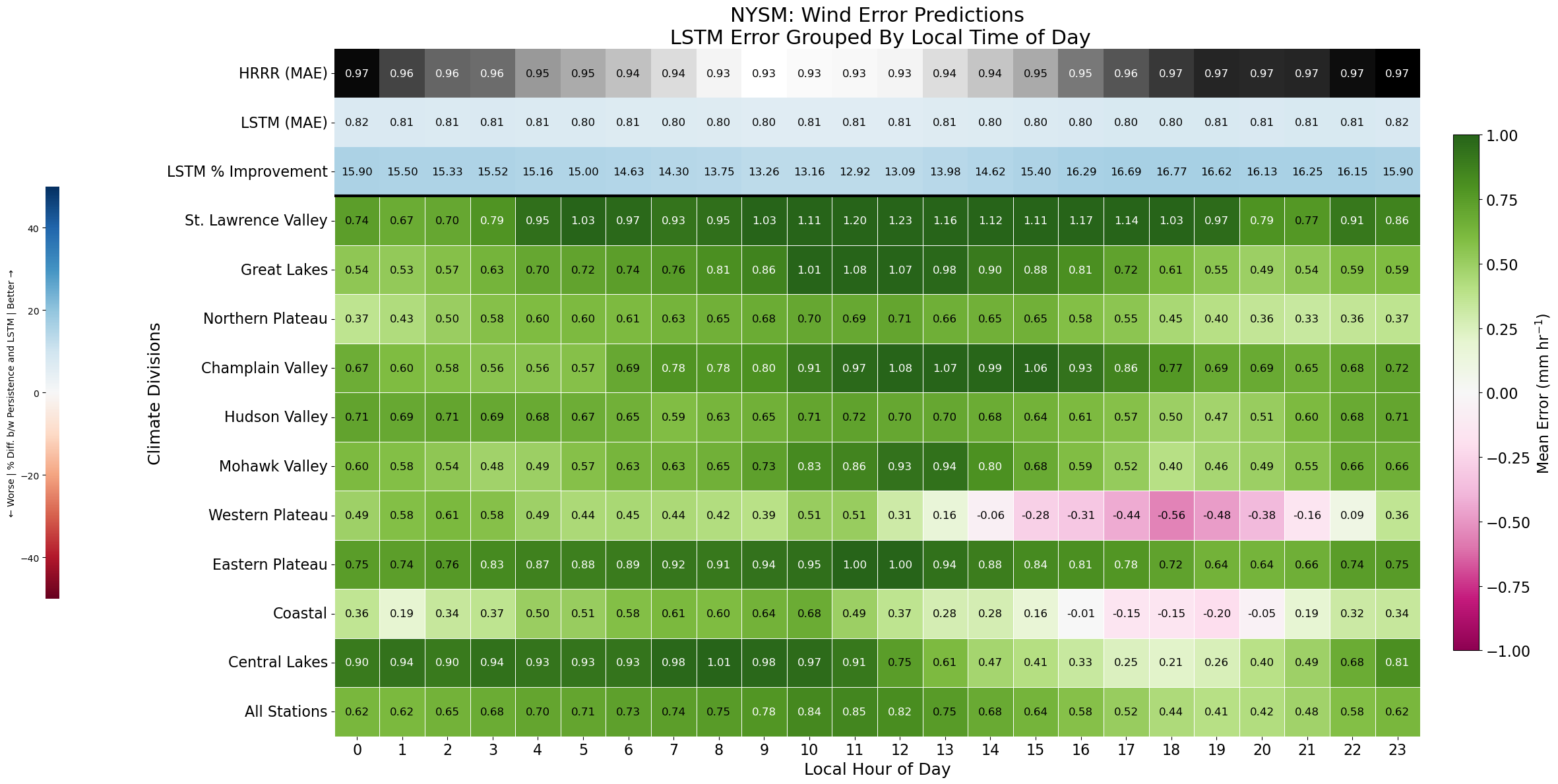}
    \caption{NYSM, mean error of LSTM predictions for wind error in $\mathrm{m~s^{-1}}$, grouped by local time of day. Panels are arranged from top to bottom with the same layout and color conventions as Fig.~\ref{fig:time_of_year_nysm_tp}. The top three rows provide a direct comparison between HRRR and LSTM diurnal error structure, highlighting where the LSTM improves upon or underperforms the HRRR baseline.}
    \label{fig:time_of_day_nysm_wind}
\end{figure*}

Figure~\ref{fig:time_of_day_nysm_wind} shows the mean LSTM error (in $\mathrm{m~s^{-1}}$) grouped by time of day, filtered to highlight model failure modes. Distinct diurnal patterns emerge across climate divisions, with prediction skill generally decreasing around solar noon, with an average error increase of about $0.5~\mathrm{m~s^{-1}}$ relative to the division minima. LSTMs perform best shortly after sunset (1800–2100), marked by the highest percent improvement over the HRRR, and lowest relative errors across all divisions. This pattern is most pronounced in the northern divisions (Champlain Valley, Northern Plateau, St. Lawrence Valley, Great Lakes), which also exhibit a modest reduction in LSTM errors before sunrise (0300–0500). The high-elevation Eastern Plateau shows a similar structure, and the Great Lakes division displays error characteristics that closely resemble those of the Northern and Eastern Plateau regions.

Figure~\ref{fig:time_of_day_nysm_wind} also reveals that the Mohawk and Hudson Valley climate divisions exhibit similar error signatures. As discussed, LSTM errors peak primarily at solar noon, with an average increase in error compared to the division minima of approximately $0.4~\mathrm{m~s^{-1}}$, and maintain a relative error minima before sunrise and after sunset; however, the Hudson and Mohawk Valleys exhibit secondary error maxima around midnight, with an average increase in error compared to the division minima of approximately $0.2~\mathrm{m~s^{-1}}$.

Finally, the Western Plateau exhibits error characteristics broadly similar to those of the Coastal division, despite major contrasts in geography and local dynamics. Both divisions show pronounced afternoon error and are the only regions with sustained underprediction, with average relative error increases of about $0.4~\mathrm{m~s^{-1}}$ from the division minima. This timing contrasts with most other divisions, which generally improve during the early evening hours. The Coastal division also shows modest nocturnal improvement (1900–0300), with an average error decrease of roughly $0.4~\mathrm{m~s^{-1}}$, and increased prediction accuracy in the early afternoon (1400–1600). The Central Lakes division, by contrast, maintains nearly uniform performance throughout the diurnal cycle, with only a slight late-afternoon improvement (1600–1900) corresponding to an average error decrease of about $0.75~\mathrm{m~s^{-1}}$.

\begin{figure*}
    \centering
    \includegraphics[width=36pc]{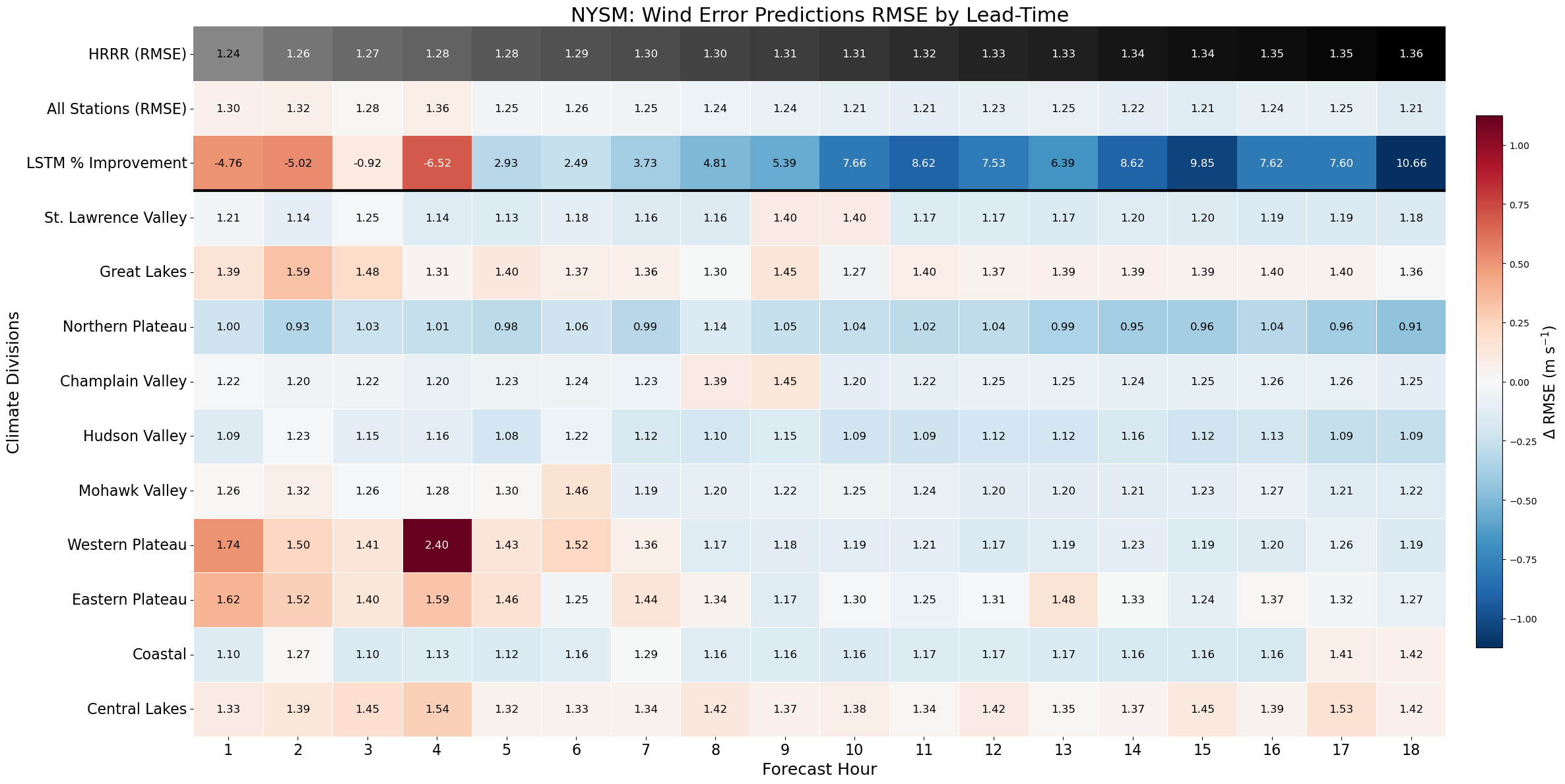}
    \caption{The same color conventions as in Fig.~\ref{fig:nysm_tp_wb}, but in $\mathrm{m~s^{-1}}$.}
    \label{fig:nysm_wind_wb}
\end{figure*}

Figure~\ref{fig:nysm_wind_wb} illustrates the relative improvement of the LSTM model compared to the HRRR as a function of forecast lead time. The NYSM shows a monotonic increase in HRRR RMSE with lead time associated with near-surface wind forecasts. The second row presents aggregate LSTM RMSE, with blue shading indicating improvement relative to HRRR, highlighting the model’s ability to correct bias. The third row shows percent improvement relative to HRRR, while the remaining rows display RMSE by climate division across forecast lead times.

Despite substantial gains in MAE and diurnal analysis, RMSE evaluated by forecast lead time reveals a more nuanced view of LSTM wind error predictions across the NYSM. As with precipitation, aggregate wind error improvement increases with lead time, though with subtle variation. However, this improvement varies considerably across divisions. The Great Lakes and Central Lakes divisions stand out as consistently underperforming the HRRR across divisions, never achieving improvement over the baseline. The Mohawk Valley, Western Plateau, and Eastern Plateau also exhibit delayed improvement, not surpassing the HRRR until approximately FH7, FH8, and FH9, respectively. The St. Lawrence Valley, Champlain Valley, and Coastal climate divisions exhibit intermittent underperformance relative to the HRRR at mid to longer lead times, typically limited to one or two forecast hours.

\subsubsection{NYSM Wind Error Discussion}
As with precipitation, it is important to distinguish between intrinsic predictability differences in the atmospheric system and limitations of the modeling framework when interpreting these results. As shown in Fig.~\ref{fig:time_of_day_nysm_wind}, LSTM error is lowest prior to morning PBL spin-up and following afternoon mix-out, with enhanced improvement relative to the HRRR during the early evening hours immediately following sunset. This pattern reflects increased LSTM skill under stable PBL conditions and during well-mixed periods, which are generally more predictable, as well as improved alignment between the model inputs and dominant physical processes.

Northern and upland climate divisions exhibit the most coherent diurnal error cycles, consistent with the dominance of katabatic flows in complex terrain \citep{ZardiWhiteman2013}. Elevated nocturnal errors within valley regions likely arise from turbulent, channelized flows interacting with a stratified PBL, which represent both intrinsically complex flow regimes and environments that are challenging for the LSTM to represent using surface-based predictors alone \citep{Ricardo2006, Card2023_MHC}.

The Coastal division has marked underprediction in the afternoon during peak PBL mixing, and exhibits modest nighttime improvement and enhanced midday accuracy, likely associated with the erratic timing and inland penetration of sea-breeze circulations, which introduce both intrinsic variability and modeling challenges due to their mesoscale and transient nature \citep{McCabeFreedman2023, MakWalsh1976}.
 
\subsubsection{Oklahoma State Mesonet}

\begin{figure*}
    \centering
    \includegraphics[width=30pc]{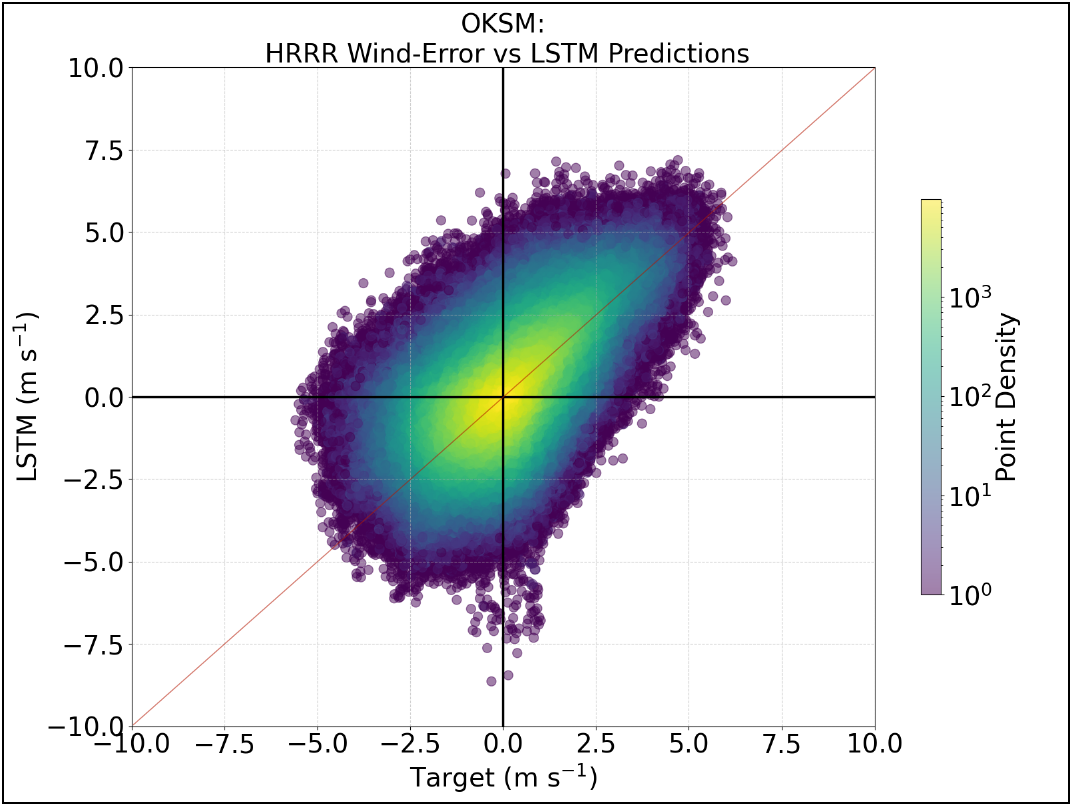}
    \caption{As in Fig.~\ref{fig:scatterplots_nysm_wind}, but for the OKSM.}
    \label{fig:scatterplots_oksm_wind}
\end{figure*}

As shown in Fig.~\ref{fig:scatterplots_oksm_wind}, the scatterplot compares true against LSTM-predicted wind errors, with $95\%$ of targeted error points falling within $\pm~2~\mathrm{m~s^{-1}}$ of the 1:1 line. Figure \ref{fig:scatterplots_oksm_wind} highlights a slight asymmetry in prediction skill: the LSTM is more adept at predicting positive forecast errors (i.e., identifying HRRR overforecasts), as these values align more closely with the 1:1 line. In contrast, negative errors (underforecast) are less accurately predicted, suggesting that the LSTM may be biased or less sensitive to the conditions that lead to underforecast wind, specifically in the context of unimpeded, relatively flat topography.

\begin{figure*}
    \centering
    \includegraphics[width=30pc]{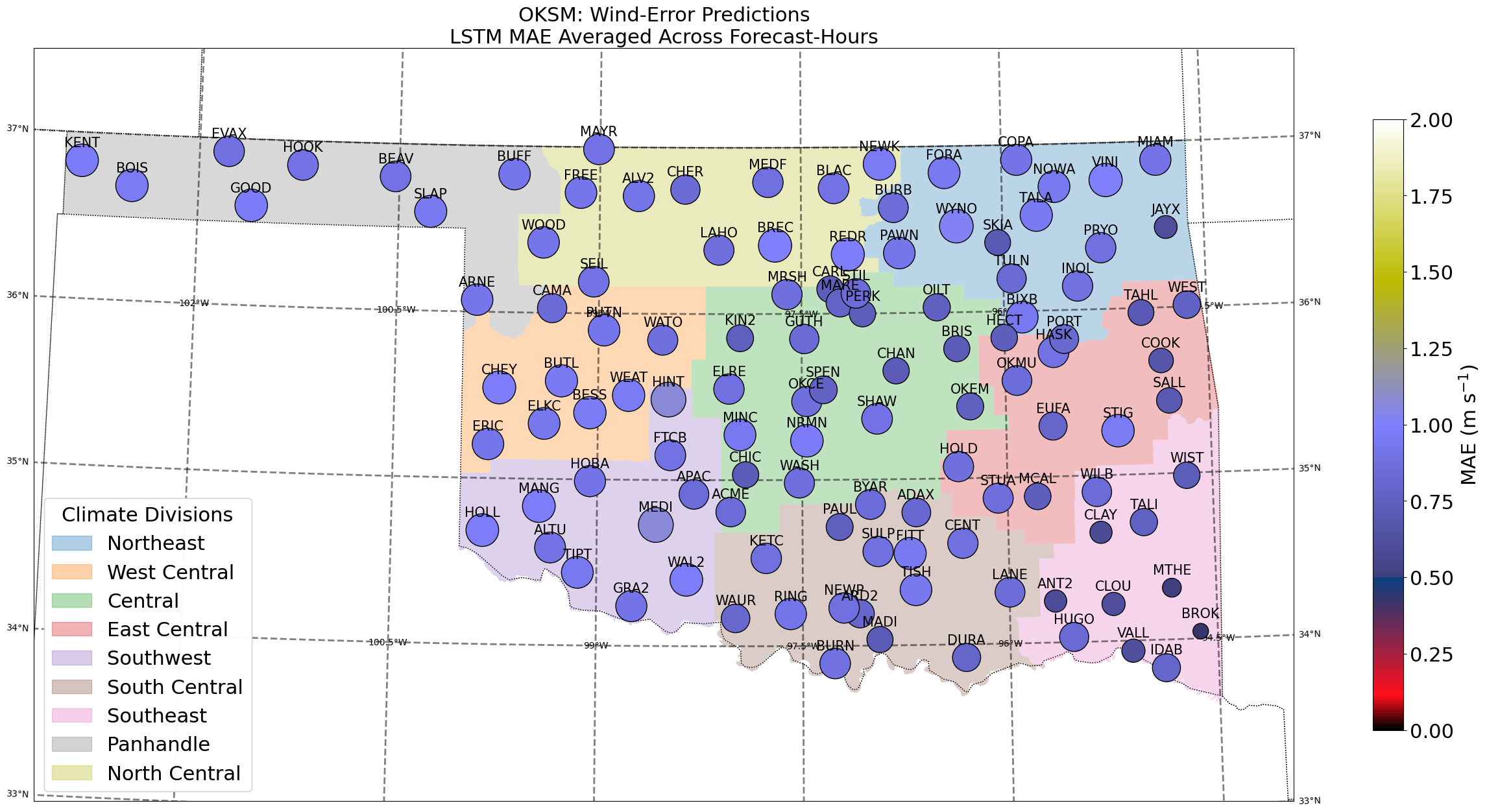}
    \caption{As in Fig.~\ref{fig:nysm_state_mae_wind}, but for the OKSM.}
    \label{fig:state_mae_oksm_wind}
\end{figure*}

Figure~\ref{fig:state_mae_oksm_wind} shows the average LSTM performance (MAE, $\mathrm{m~s^{-1}}$) for an OKSM station. The Southeast division exhibits the lowest errors, with an average improvement in model performance of approximately $0.35~\mathrm{m~s^{-1}}$ relative to the domain mean. In contrast, the Panhandle, North Central, Southwest, and West Central divisions display the highest MAE values, corresponding to an average performance degradation of about $0.5~\mathrm{m~s^{-1}}$, compared to the domain mean. While these spatial differences in MAE are consistent across divisions, they remain relatively subtle in magnitude (on the order of $\sim0.25$--$0.5~\mathrm{m~s^{-1}}$) and should be interpreted with consideration of observational uncertainty in near-surface wind measurements. Such uncertainty can arise from instrument precision, siting effects, and unresolved microscale variability, which may contribute to or obscure small spatial gradients in error.

\begin{figure*}[htbp]
    \centering
    \includegraphics[width=30pc]{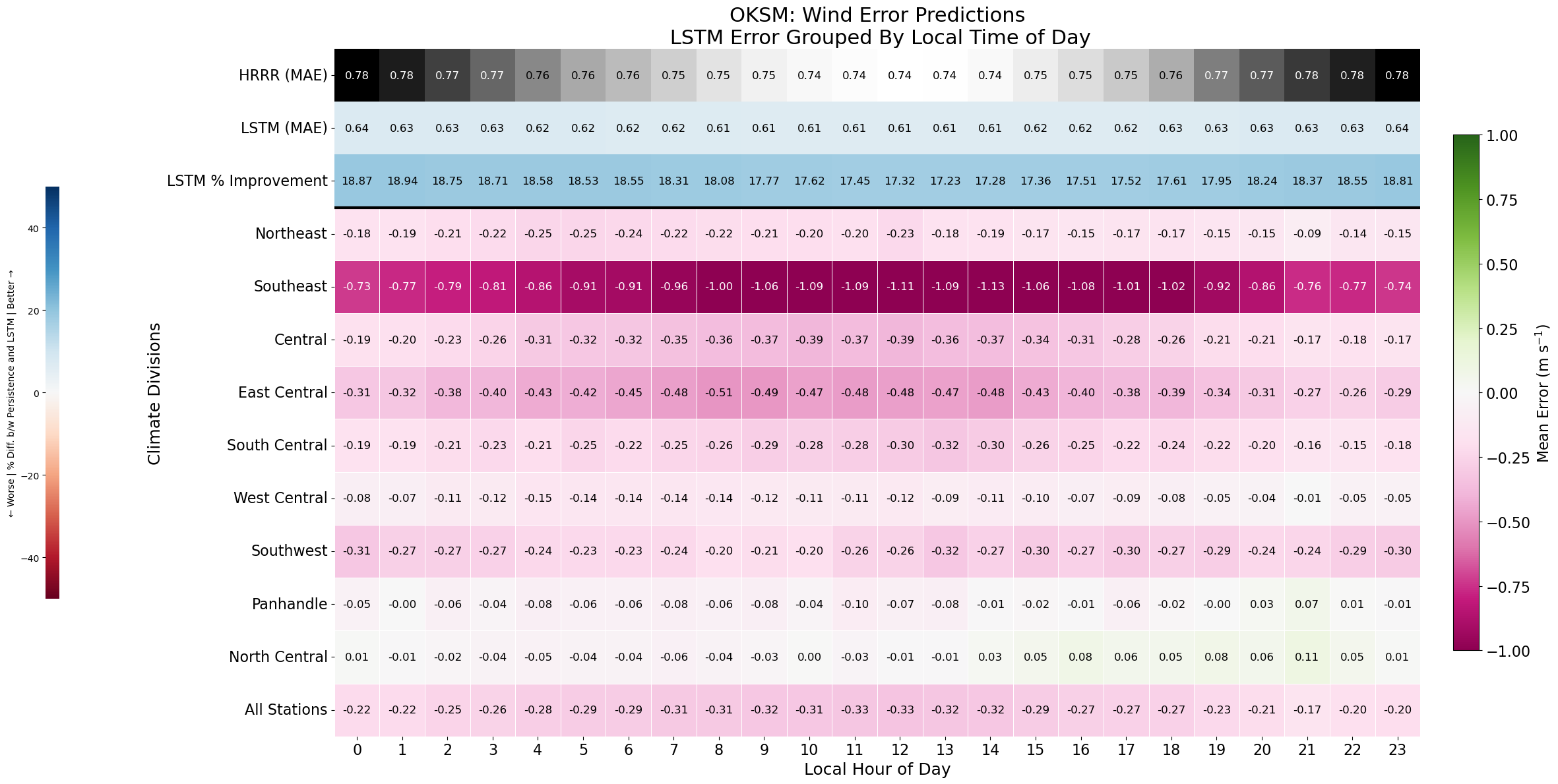}
    \caption{As in Fig.~\ref{fig:time_of_day_nysm_wind}, but for the OKSM.}
    \label{fig:time_of_day_oksm_wind}
\end{figure*}

Figure~\ref{fig:time_of_day_oksm_wind} presents the mean LSTM error in $\mathrm{m~s^{-1}}$, grouped by time of day, filtered to highlight model failure modes. Error signatures vary notably across climate divisions; the Central, Southeast, East Central, and South Central divisions exhibit distinct diurnal cycles, with peak errors occurring near solar noon and an average increase of approximately $0.25~\mathrm{m~s^{-1}}$ relative to the nighttime division minima (2000–0000), when overall model performance improves. The Southeast division exhibits the most consistent and pronounced underprediction across the OKSM domain, with peak error near solar noon and an average underprediction magnitude of approximately $1~\mathrm{m~s^{-1}}$.

Referencing Fig.~\ref{fig:time_of_day_oksm_wind}, the Northeast and West Central divisions exhibit similar diurnal error patterns, with the lowest model skill occurring near sunrise (0400–0600). During this period, the average degradation in model performance is approximately $0.15~\mathrm{m~s^{-1}}$ relative to the division minima, after which skill gradually improves toward midnight.

The North Central and Southwest climate divisions exhibit the least distinct diurnal error patterns (Fig.~\ref{fig:time_of_day_oksm_wind}). The Southwest division maintains relatively stable model skill throughout the day, aside from a negligible improvement during the early morning hours (0800–1000). A weak late-evening overprediction signature is observed in the Panhandle, the only region displaying a pattern comparable to that of the North Central division. As shown in Fig.~\ref{fig:time_of_day_oksm_wind}, the Panhandle otherwise maintains stable performance. These regions also exhibit elevated MAE relative to the broader OKSM domain (Fig.~\ref{fig:state_mae_oksm_wind}), suggesting that although the overall magnitude of LSTM error is larger, it does not manifest as a coherent diurnal signal when evaluated using the filtered true mean.

\begin{figure*}
    \centering
    \includegraphics[width=30pc]{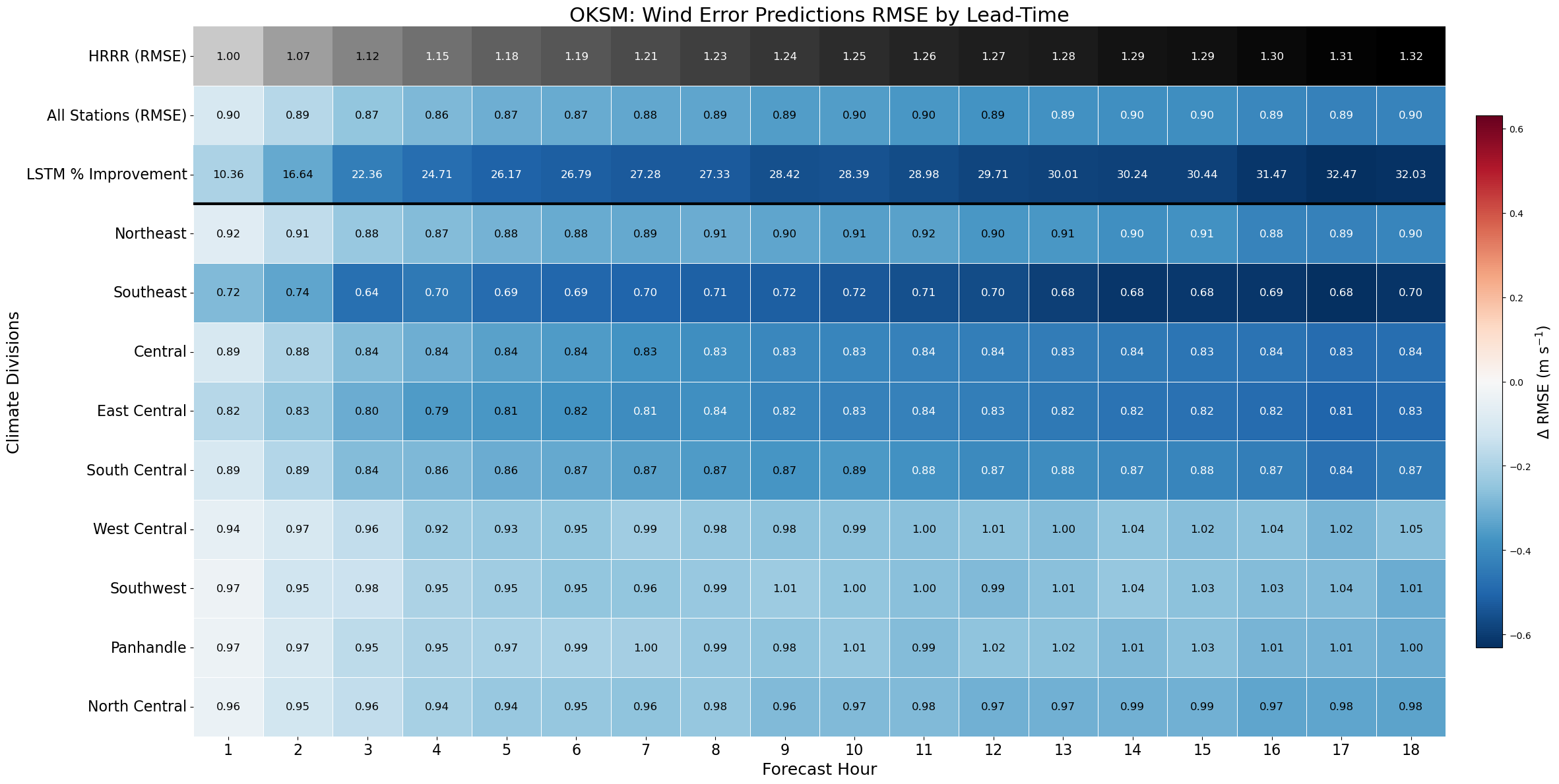}
    \caption{As in Fig.~\ref{fig:nysm_wind_wb}, but for the OKSM.}
    \label{fig:oksm_wind_wb}
\end{figure*}

Figure~\ref{fig:oksm_wind_wb} enables analysis of LSTM behavior as a function of lead time without error filtering, highlighting general patterns rather than specific failure modes. Across the OKSM, LSTM RMSE increases with lead time from FH 1–12, followed by a gradual decrease at longer lead times. In contrast, relative improvement over the HRRR increases steadily with lead time across climate divisions, with the OKSM showing substantially greater gains than the NYSM -- often nearly doubling the percent improvement. RMSE exhibits minimal variation across climate divisions and lead times in the OKSM, representing a marked contrast to the more variable behavior observed in the NYSM. Consistent with earlier analysis, a subtle spatial divide is evident, with the northeastern quadrant of the OKSM showing slightly less improvement relative to the HRRR than the southwestern quadrant.

\subsubsection{OKSM Wind Error Discussion}

Diurnal error patterns in the OKSM domain reflect strong coupling between PBL evolution and local mesoscale processes (Fig.~\ref{fig:time_of_day_oksm_wind}). LSTMs here underpredict error, unlike in the NYSM, which may reflect both greater intrinsic predictability associated with simpler terrain and more uniform PBL structure, as well as improved compatibility between the model inputs and the dominant physical processes. The humid Southeast division shows the lowest MAE (Fig.~\ref{fig:state_mae_oksm_wind}), as moisture-driven stability and complex terrain likely dampen energy transport, creating conditions that are both more predictable and more amenable to representation by the LSTM \citep{Dewani2023}.

Northeast and West Central divisions exhibit morning degradation (0400–0600, Fig.~\ref{fig:time_of_day_oksm_wind}), consistent with atmospheric bores/mesoscale outflows, which reduce predictability, which reduce intrinsic predictability, particularly during PBL spin-up, and also represent processes that are difficult for the LSTM to capture given its reliance on surface-based inputs \citep{Haghi2017, HaghiDurran2021}. Conversely, the North Central, Southwest, and Panhandle divisions show weak diurnal structure (Fig.~\ref{fig:time_of_day_oksm_wind}) but larger overall MAE (Fig.~\ref{fig:state_mae_oksm_wind}). This likely reflects the region’s flat terrain and more predictable PBL evolution; however, the elevated MAE suggests that while the large-scale behavior is more predictable, the model may struggle to capture smaller-scale variability or subtle error structures in these environments \citep{Demoz2002, Couvreux2009}. LSTM skill improves under stable or mixed conditions, but transitional PBL regimes and convective variability remain limiting factors, reflecting both reduced intrinsic predictability and limitations in the model’s ability to represent rapidly evolving boundary layer processes.

\subsection{Temperature Error}

Temperature is included as a primary predictand due to its relevance for thermodynamics, its broad societal (e.g., heat stress, morbidity) and energy-sector impacts, and its substantial inter-annual variability relative to the other target variables. Furthermore, temperature is governed by radiative and surface–atmosphere exchange processes that differ appreciably from those driving wind and precipitation, providing an opportunity to assess the capacity of LSTMs to generalize across diverse atmospheric dynamics.

Despite temperature being a continuous variable in both space and time, it is the least accurately predicted variable across all three predictors when evaluated by MAE, and LSTM percent improvement relative to the HRRR. Nevertheless, LSTM performance remains reasonably accurate, potentially aided by the systematic bias in the HRRR, which transitions from a cold bias at the lowest forecast temperatures to a warm bias at the highest \citep{Gaudet2024, James2022}. 

\subsubsection{New York State Mesonet}

\begin{figure*}
    \centering
    \includegraphics[width=30pc]{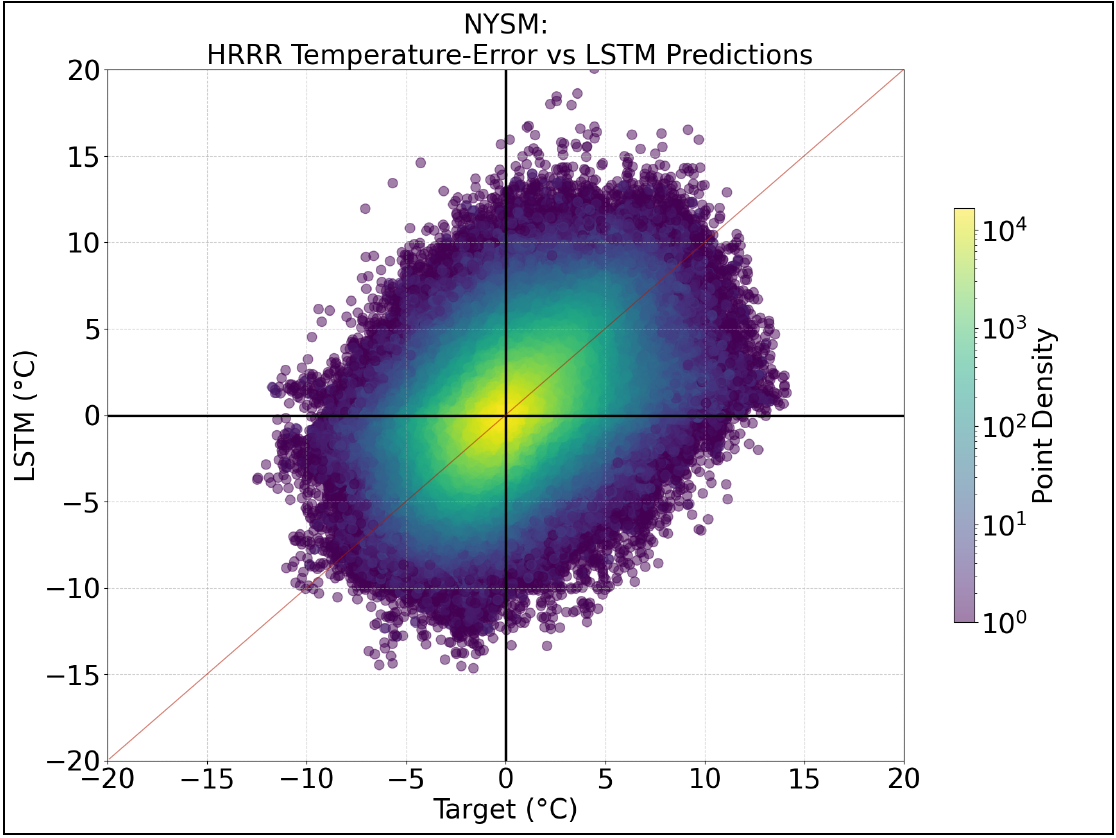}
    \caption{Scatterplot of the temperature error across the NYSM network and all forecast hours, with the x-axis representing the true target error in $^\circ\mathrm{C}$ and the y-axis showing the corresponding LSTM-predicted error in $^\circ\mathrm{C}$. The red diagonal line indicates the 1:1 line, where perfect predictions would lie.}
    \label{fig:scatterplots_nysm_temp}
\end{figure*}

Figure~\ref{fig:scatterplots_nysm_temp} shows a scatterplot of the temperature error across the NYSM, where $74\%$ of targeted error points fall within $\pm~2^\circ\mathrm{C}$ of the 1:1 line. Temperature error data displays greater variance, with more scatter away from the diagonal, suggesting that the LSTM’s prediction confidence is less consistent for temperature compared to wind and precipitation. 

\begin{figure*}
    \centering
    \includegraphics[width=30pc]{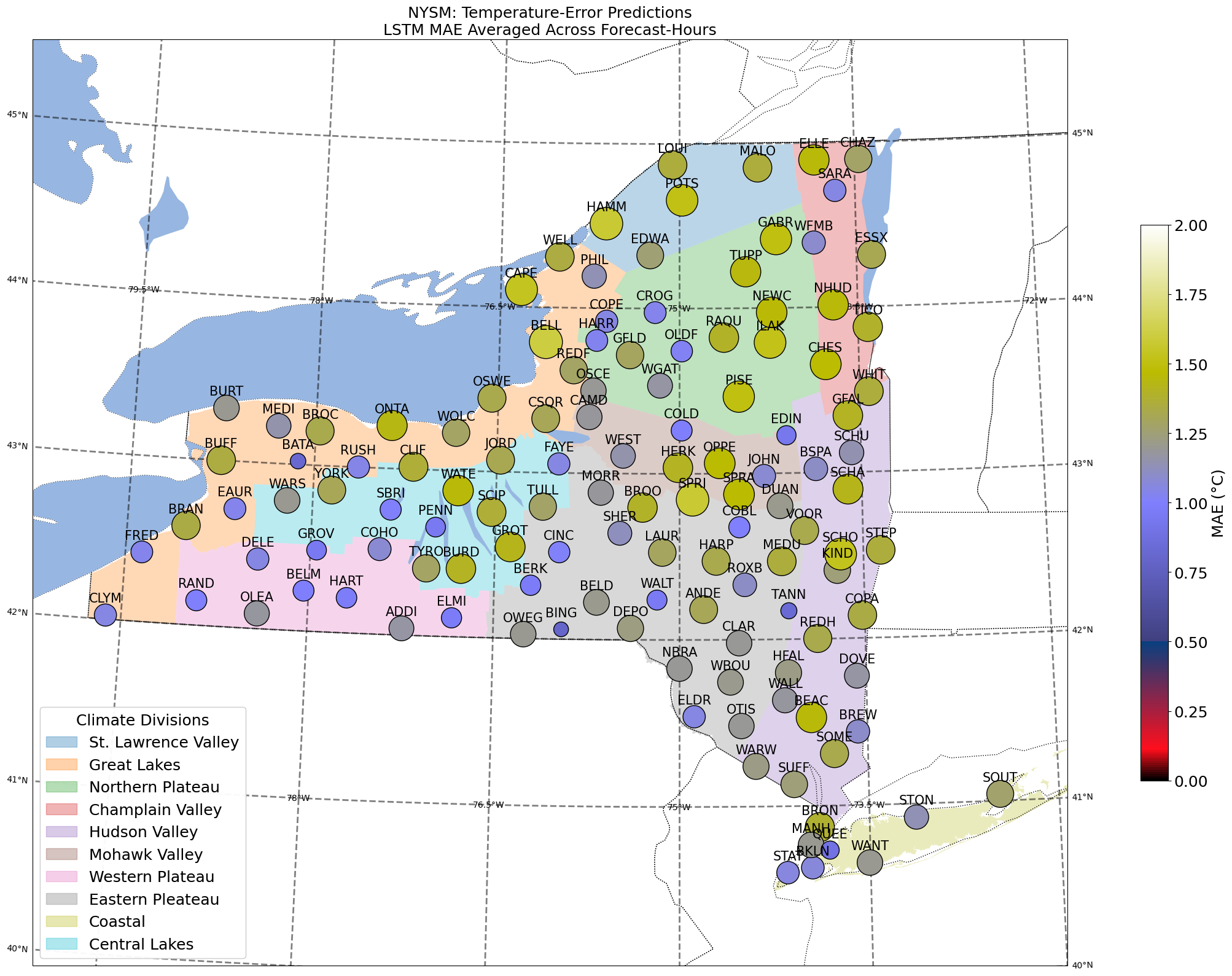}
    \caption{Average LSTM performance (MAE) for an NYSM station, averaged over all forecast lead times. The magnitude of the point is proportional to the MAE, where larger points translate to higher MAE. NCEI climate divisions \citep{NCEI2015} are displayed for reference. A shared color scale is used across domains to enable direct comparison of MAE magnitude; as a result, variability within the OKSM domain appears visually compressed relative to NYSM, and marker size is scaled to aid interpretation.}
    \label{fig:state_nysm_temp}
\end{figure*}

Figure~\ref{fig:state_nysm_temp} shows the average LSTM performance (MAE, $^\circ\mathrm{C}$) for an NYSM station. A subtle spatial pattern in LSTM performance is evident in Fig.~\ref{fig:state_nysm_temp}, showing a weak latitudinal gradient with slightly better temperature error predictions at more southerly NYSM stations (correlation: $0.329$, p-score: $0.00$). This trend is most pronounced in the Mohawk Valley division, marking the onset of a modest north–south gradient in LSTM accuracy, most clearly expressed along the 75\textdegree{}W to 74\textdegree{}W meridian. 

\begin{figure*}[htbp]
    \centering
    \includegraphics[width=30pc]{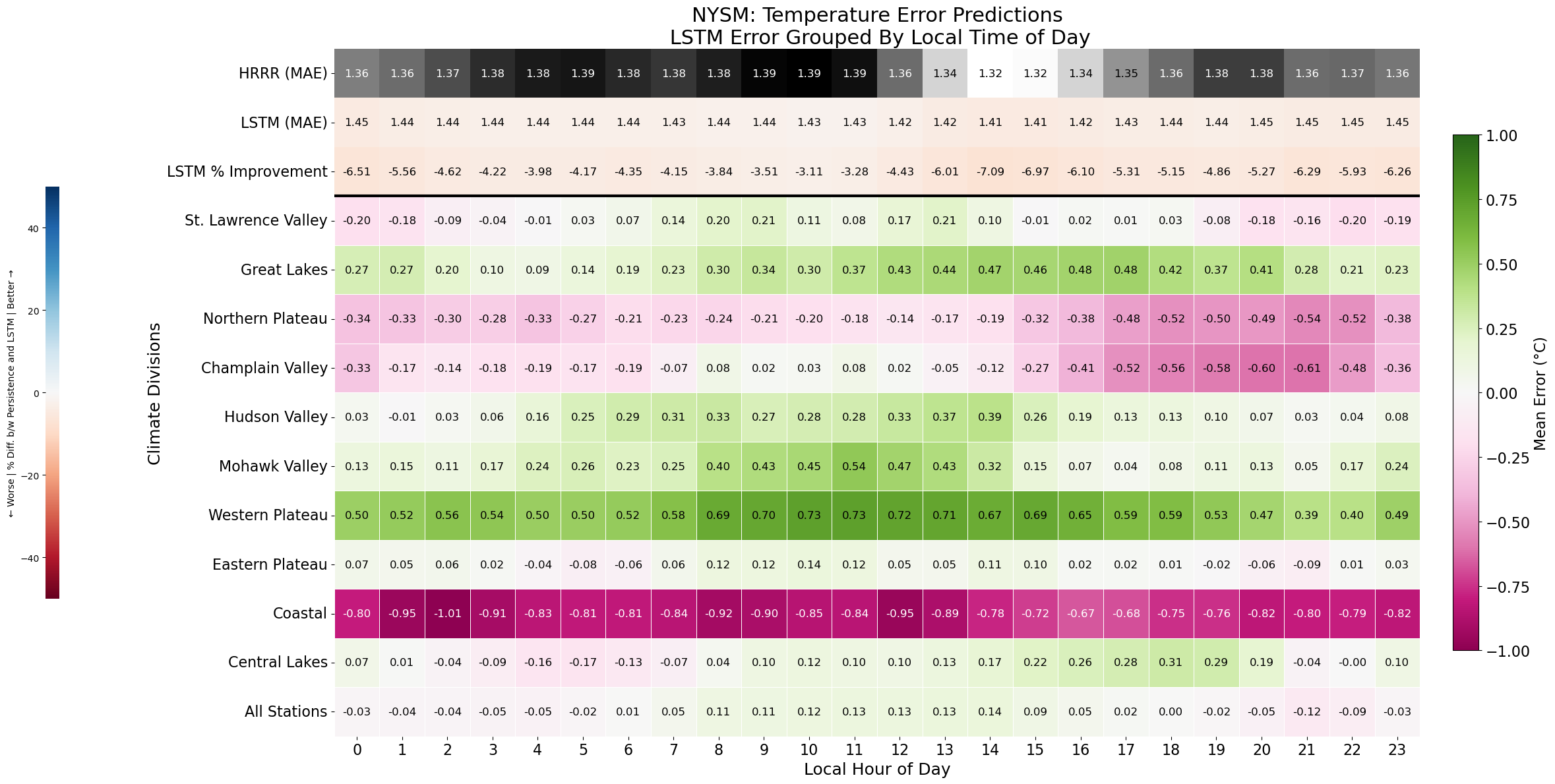}
    \caption{NYSM, mean error of LSTM predictions for temperature error in $^\circ\mathrm{C}$, grouped by local time of day. Panels are arranged from top to bottom with the same layout and color conventions as Fig.~\ref{fig:time_of_year_nysm_tp}. The top three rows provide a direct comparison between HRRR and LSTM diurnal error structure, highlighting where the LSTM improves upon or underperforms the HRRR baseline.}
    \label{fig:time_of_day_nysm_temp}
\end{figure*}

Figure~\ref{fig:time_of_day_nysm_temp} shows the filtered mean error of LSTM predictions for temperature error in $^\circ\mathrm{C}$, grouped by time of day. Notably, temperature is the only predictand for which the LSTM predictions do not outperform HRRR forecasts; however, we include temperature as a diagnostic case rather than a primary success metric, as its comparatively weaker performance highlights known challenges in near-surface temperature error prediction. In this context, the analysis provides insight into the limits of the current modeling framework and the role of boundary layer processes in constraining predictive skill. In most NYSM climate divisions, LSTM error maintains a relative maxima around solar noon, with an average increase in error of approximately $0.15^\circ\mathrm{C}$ compared to the division minima -- particularly the Great Lakes, Western Plateau, Mohawk Valley, and Hudson Valley.

Referencing Fig.~\ref{fig:time_of_day_nysm_temp}, the Eastern Plateau, St. Lawrence Valley, Central Lakes, Champlain Valley, and Northern Plateau divisions each deviate from the broader diurnal error patterns observed elsewhere. The Eastern Plateau and St. Lawrence Valley exhibit improved LSTM performance during the morning (0300–0500) and early evening (1600–1900). Conversely, these divisions exhibit degraded performance at night (2000–0200), corresponding to an average error increase of approximately $0.1^\circ\mathrm{C}$ relative to the division minima. In contrast, the Champlain Valley and Northern Plateau show reduced accuracy during the early nocturnal period (1500–2200), with an average error increase of about $0.5^\circ\mathrm{C}$ relative to the division minima.

The Coastal climate division diverges from other regions, showing a pronounced underprediction of temperature error (Fig.~\ref{fig:time_of_day_nysm_temp}), with an average increase of approximately $0.75^\circ\mathrm{C}$ relative to other divisions. Its performance remains relatively consistent throughout the day, with a slight improvement in the late afternoon (1500–1700). The Central Lakes division also exhibits a distinct diurnal pattern, with slight underprediction in the early morning (0300–0700; $\sim$0.15$^\circ\mathrm{C}$ above the division minima) and modest overprediction in the late afternoon and early evening (1500–2000; $\sim$0.25$^\circ\mathrm{C}$ above the division minima).

\begin{figure*}
    \centering
    \includegraphics[width=30pc]{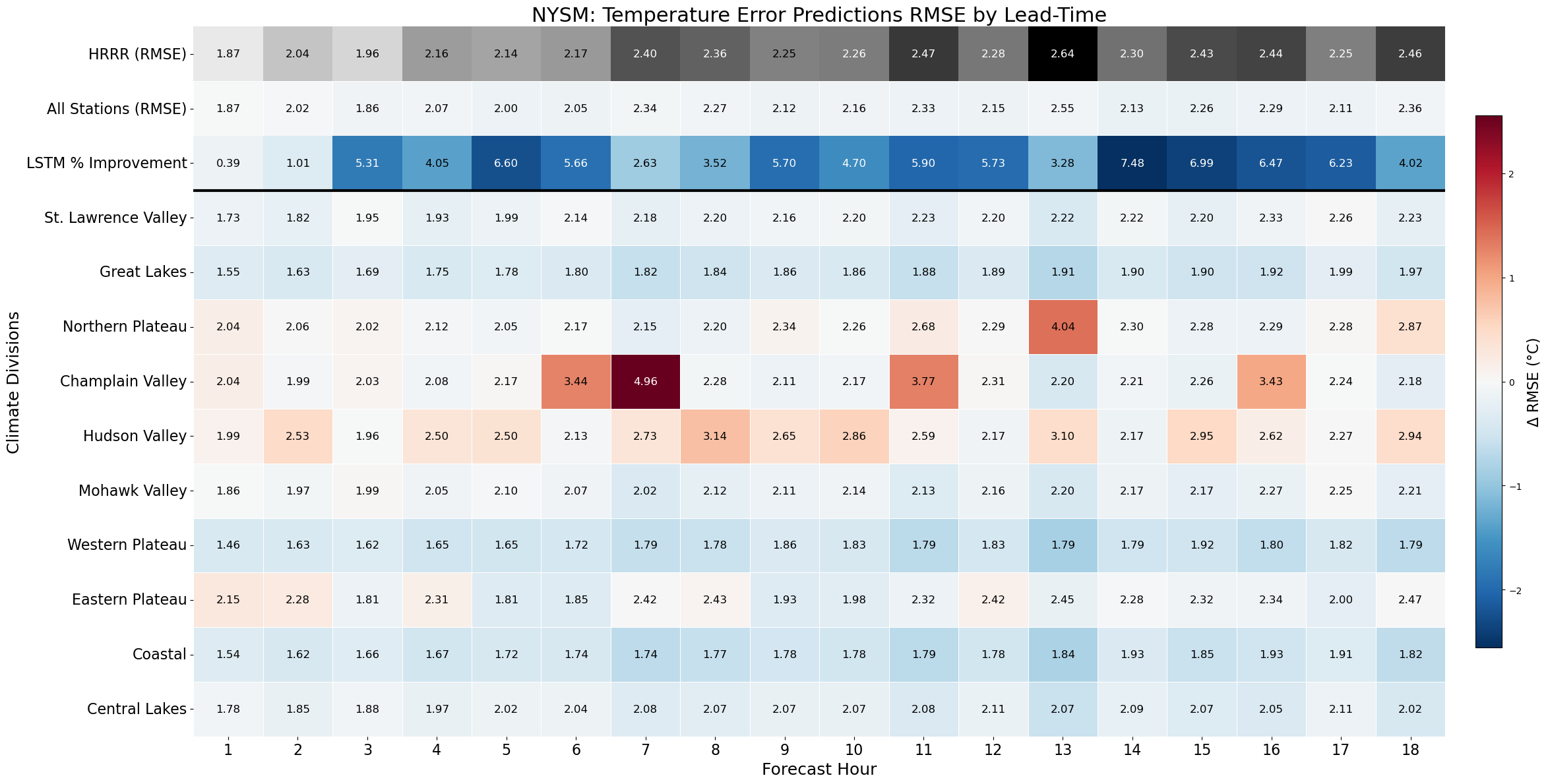}
    \caption{The same color conventions as in Fig.~\ref{fig:nysm_tp_wb}, but in $^\circ\mathrm{C}$.}
    \label{fig:nysm_t2m_wb}
\end{figure*}

Referencing Fig.~\ref{fig:nysm_t2m_wb}, despite slight degradation relative to the HRRR when evaluated using MAE and diurnal analysis, RMSE by forecast lead time provides a more nuanced view of LSTM temperature error predictions across the NYSM. As with the other predictors, aggregate temperature error improvement increases with lead time, though with greater variability than observed for precipitation and wind. The Northern Plateau, Champlain Valley, Hudson Valley, and Eastern Plateau exhibit considerable variability across lead times, including brief periods where LSTM RMSE exceeds that of the HRRR, indicating short-lived degradations in performance. These occurrences are typically limited to one or a few consecutive forecast hours. The Champlain Valley and Northern Plateau show the largest RMSE excursions, exceeding 4°C for isolated forecast hours, while the Hudson Valley stands out for having the most frequent instances of underperformance relative to the HRRR.

\subsubsection{NYSM Temperature Error Discussion}

Referencing Fig.~\ref{fig:time_of_day_nysm_temp}, error magnitudes that generally peak midday likely reflect enhanced PBL overturning, corresponding to peak solar irradiance. Increased daytime turbulence and mixing introduce variability in near-surface temperature, reducing intrinsic predictability during this period, while also creating conditions that are more difficult for the LSTM to represent using surface-based predictors alone. Several northern and upland climate divisions deviate from this general pattern. This inverted diurnal behavior is likely influenced by temperature inversions over the Central Lakes \citep{Laird2009FingerLakes} and Champlain Valley \citep{tardy2000lake}, which alter nocturnal PBL structure, reducing predictability and limiting the LSTM’s ability to represent temperature error given its reliance on surface-based inputs. Similar processes may also explain the nocturnal degradation observed in the St. Lawrence Valley \citep{Carrera2009}.

Referencing Fig.~\ref{fig:time_of_day_nysm_temp}, the Coastal climate division diverges further from these inland patterns. This behavior is likely shaped by land–sea interactions and urban amplification effects that modify latent and sensible heat fluxes as well as vertical and horizontal mixing \citep{McCabeFreedman2023, swain}. The combined effects of thermal inertia from the ocean and the urban heat island dampen diurnal variability, producing a smoother and more consistent error signal. However, these same factors make temperature‐error prediction more difficult, as both the coastal and the urban environment act as substantial, spatially and temporally complex heat reservoirs.

While spatial and temporal patterns vary in strength across regions, they likely reflect underlying atmospheric processes related to PBL depth and vertical mixing. Northern stations, more often influenced by continental air masses, tend to experience shallower PBLs and reduced mixing \citep{Zhang2020PBL, Seidel2012PBL}. Complex terrain in the northern part of the state further introduces orographic blocking, cold-air damming, and inversion formation \citep{ZardiWhiteman2013}, contributing to the subtle north–south gradient in LSTM performance, which likely reflects both differences in intrinsic predictability and the model’s ability to capture these processes. In contrast, southern divisions are more often affected by warmer, maritime air, leading to deeper PBLs and enhanced vertical transport \citep{Zhang2020PBL, Seidel2012PBL}.

\subsubsection{Oklahoma State Mesonet}

\begin{figure*}
    \centering
    \includegraphics[width=30pc]{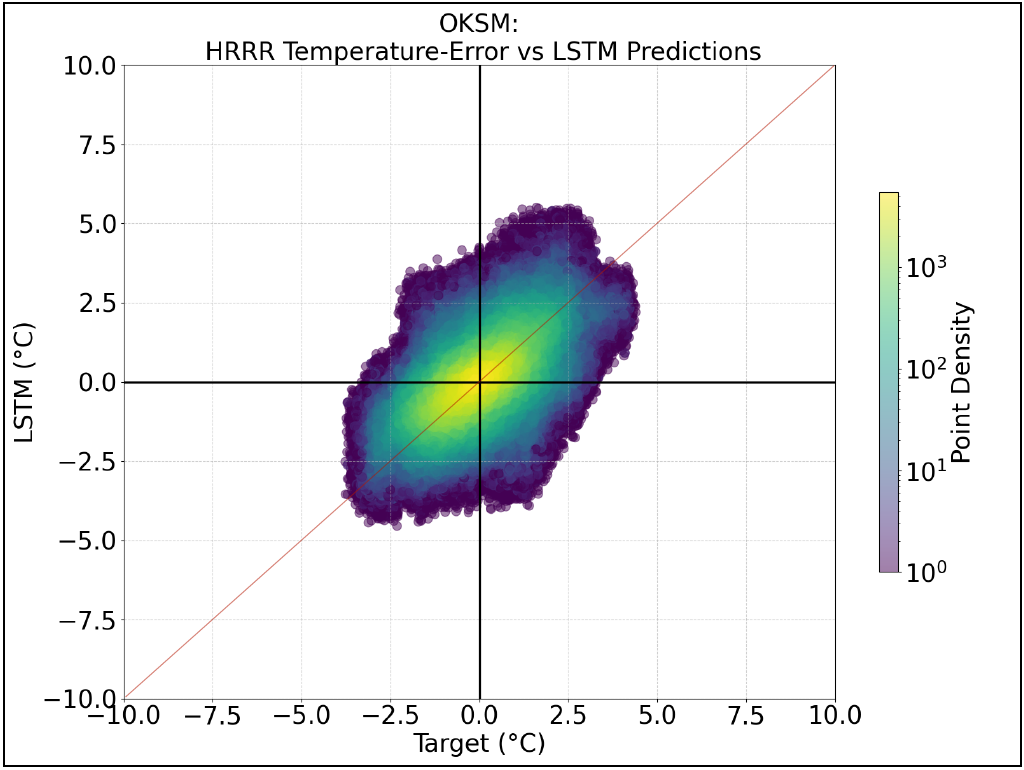}
    \caption{As in Fig.~\ref{fig:scatterplots_nysm_temp}, but for the OKSM.}
    \label{fig:scatterplots_oksm_temp}
\end{figure*}

Figure~\ref{fig:scatterplots_oksm_temp} shows a scatterplot of the temperature error across the OKSM, where $98\%$ of targeted error points fall within $\pm~2^\circ\mathrm{C}$ of the 1:1 line. LSTM performance generally exhibits a smaller magnitude of error as compared to the NYSM, and its ability to capture both overpredictions and underpredictions is the most symmetrical of the three predictors examined for the OKSM domain.

\begin{figure*}
    \centering
    \includegraphics[width=30pc]{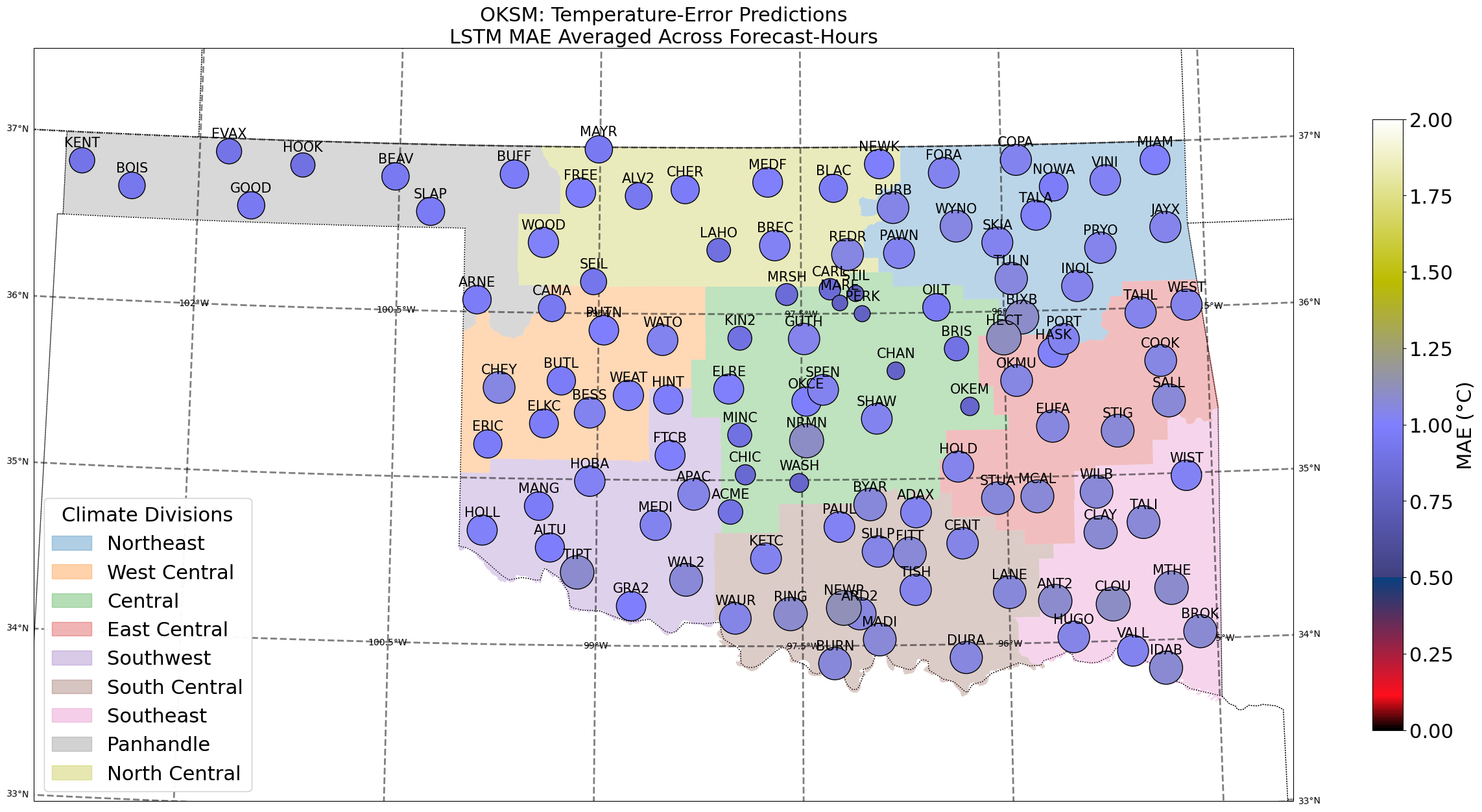}
    \caption{As in Fig.~\ref{fig:state_nysm_temp}, but for the OKSM.}
    \label{fig:state_oksm_temp}
\end{figure*}

Figure~\ref{fig:state_oksm_temp} shows OKSM MAE ($^\circ\mathrm{C}$). Across the domain, a slight improvement in LSTM skill is evident from southeast to northwest, where average errors decrease by about $0.3^\circ\mathrm{C}$. Notably, several stations in the Central division show the lowest errors statewide, but the overall variance across the domain remains minimal.

\begin{figure*}[htbp]
    \centering
    \includegraphics[width=30pc]{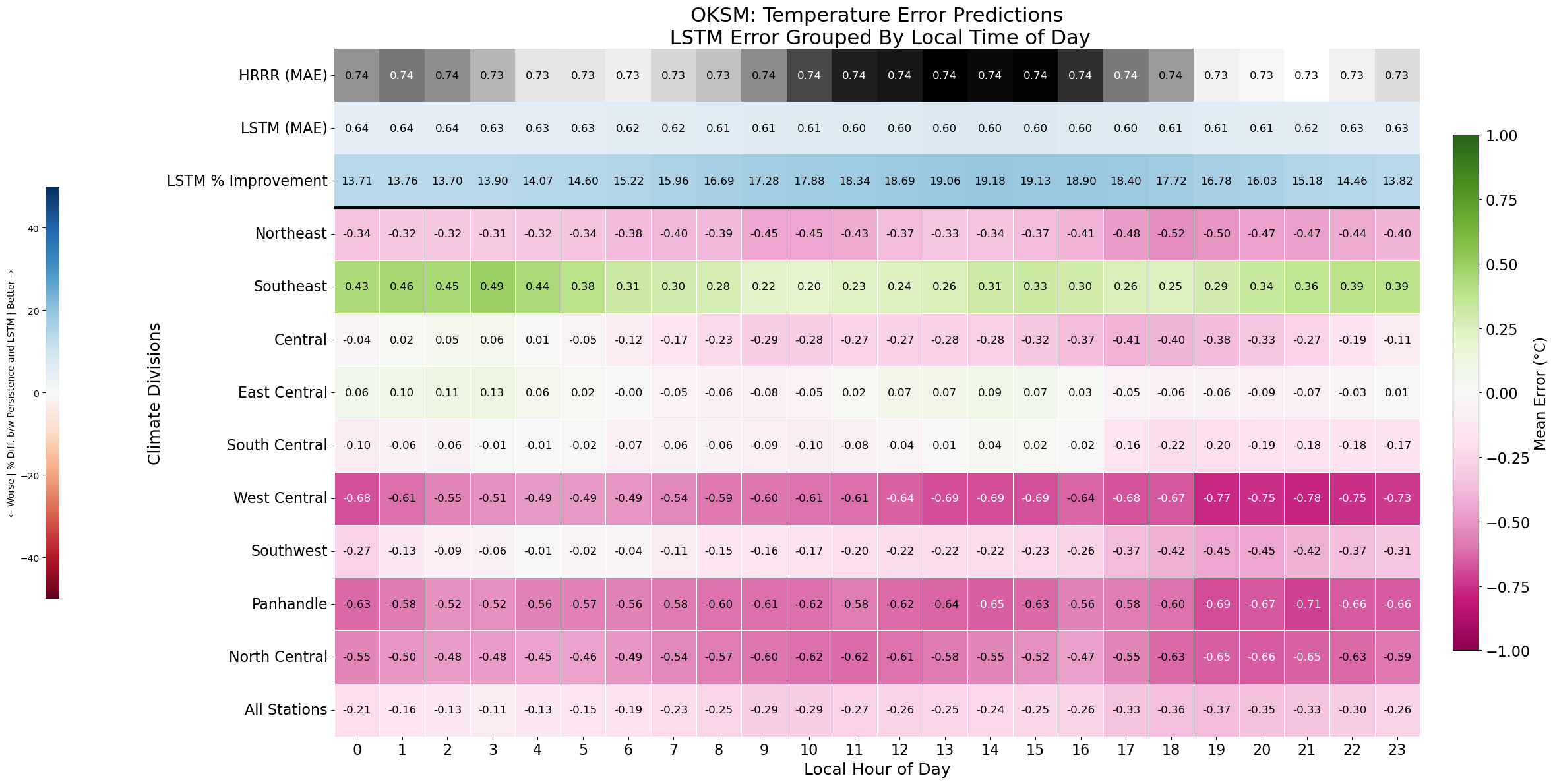}
    \caption{As in Fig.~\ref{fig:time_of_day_nysm_temp}, but for the OKSM.}
    \label{fig:time_of_day_oksm_temp}
\end{figure*}

Figure~\ref{fig:time_of_day_oksm_temp} shows the filtered mean error of LSTM predictions in $^\circ\mathrm{C}$, grouped by time of day. In contrast to the NYSM, the OKSM temperature LSTM error prediction is a demonstrable improvement over the HRRR forecasts. Moreover, the diurnal error pattern for the LSTM performance is less conclusive than in the NYSM and manifests as unique regional error signatures. Similar to wind error, temperature error prediction  skill tends to decrease around solar noon ($\sim0.1^\circ\mathrm{C}$, relative to the division minima).

Additionally, in the West Central and Central climate divisions, an absolute maximum in LSTM error occurs during the transition from daytime to nighttime conditions (1600 to 2200), where average error increases by about $0.3^\circ\mathrm{C}$, relative to the division minima. Conversely, an absolute minimum in LSTM error is often observed shortly before sunrise (0300 to 0500). For the Northeast, Panhandle, and North Central climate divisions, the temporal signatures are similar to those previously described, but errors remain relatively more stable, and the LSTM markedly underpredicts forecast error.

The Southeast climate division stands out for its consistent overprediction by the LSTM, with performance degradation during the late evening and early morning hours (0000–0400), when average errors increase by about $0.2^\circ\mathrm{C}$ relative to the division minima. The East Central, South Central, Central, and Southwest divisions show the least coherent diurnal patterns (Fig.~\ref{fig:time_of_day_oksm_temp}) but exhibit the highest overall MAE values (Fig.~\ref{fig:state_oksm_temp}).

\begin{figure*}
    \centering
    \includegraphics[width=30pc]{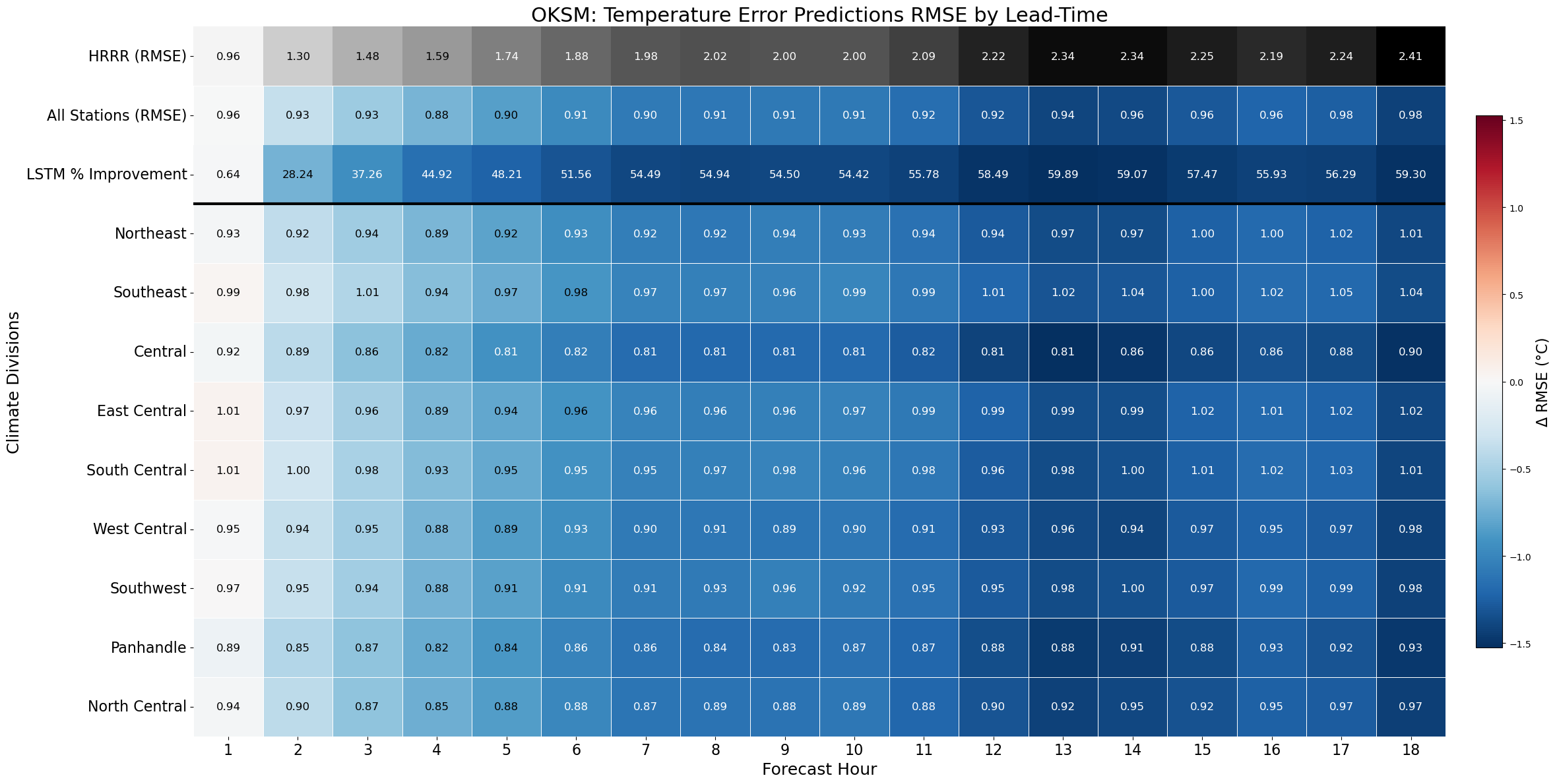}
    \caption{As in Fig.~\ref{fig:nysm_t2m_wb} but for the OKSM.}
    \label{fig:oksm_t2m_wb}
\end{figure*}

Figure~\ref{fig:oksm_t2m_wb} provides a lead-time perspective on LSTM temperature error behavior without error filtering, emphasizing overall trends. Across the OKSM, LSTM RMSE generally increases with lead time, with only modest fluctuations. Meanwhile, improvement relative to the HRRR strengthens progressively across climate divisions, with the OKSM again outperforming the NYSM, often by a factor of two in percent improvement. Overall, RMSE remains relatively uniform across both climate divisions and lead times in the OKSM, in contrast to the more variable and heterogeneous patterns observed in the NYSM.

\subsubsection{OKSM Temperature Error Discussion}

LSTM error generally peaks midday (Fig.~\ref{fig:time_of_day_oksm_temp}), likely reflecting increased solar irradiance and PBL overturning, which reduce intrinsic predictability and introduce additional complexity for the LSTM to capture. In contrast, LSTM performance improvements relative to the HRRR are greatest from midday through late afternoon, before degrading after sunset and worsening into the nocturnal period, with slight recovery prior to sunrise. This coincides with the lowest LSTM prediction errors, which typically occur shortly before the morning spin-up of the convective PBL (Fig.~\ref{fig:time_of_day_oksm_temp}), when the atmosphere is most stable and well-stratified.  

Evening error maxima related to underprediction in the West Central and Central divisions coincide with PBL collapse and low-level jet onset \citep{tinney2017llj, song2005llj}, intensified by dryline-induced gradients, which represent both intrinsically complex and rapidly evolving processes and conditions that are challenging for the LSTM to represent. The Southeast division stands out for its consistent overprediction by the LSTM (Fig.~\ref{fig:time_of_day_oksm_temp}), particularly during the late evening and early morning hours. This degradation aligns with temperature inversions common to the Ouachita region and strong warm, moist advection from the Gulf of Mexico\footnote{Following Executive Order $14172$, ``Restoring Names That Honor American Greatness'' ($90 FR 8629, Jan.\ 20, 2025$), U.S.\ government publications and regulations have been updated to refer to the area traditionally called the Gulf of Mexico as the ``Gulf of America'', with the U.S.\ Board on Geographic Names and federal agencies implementing this name change in official federal databases and regulatory text.} \citep{Rowden2018, Abuduwali}, mirroring but contrasting in sign with NYSM nocturnal bias in valleys.

Divisions with transient dryline influence show weak diurnal coherence (Fig.~\ref{fig:time_of_day_oksm_temp}) and highest MAE (Fig.~\ref{fig:state_oksm_temp}), while persistently humid or dry regimes exhibit more structured error cycles, highlighting the role of stable PBL regimes in both enhancing intrinsic predictability and improving LSTM prediction skill. Overall, variations in temperature error prediction across regions and regimes reflect a combination of intrinsic atmospheric predictability and the model’s ability to represent the governing physical processes.

\begin{table*}[ht]
\centering
\small
\setlength{\tabcolsep}{4pt}
\resizebox{\textwidth}{!}{%
\begin{tabular}{l l l c c c p{4cm}}
\hline
\textbf{Variable} & \textbf{Region / Division} & \textbf{Time} & \textbf{HRRR MAE} & \textbf{LSTM MAE} & \textbf{\% Improvement} & \textbf{Summary} \\
\hline

Precipitation & NYSM (All) & Annual & 0.56 & 0.42 & $+25\%$ & 
Wet bias captured; dry bias underestimated; small-error overprediction \\

Precipitation & NYSM (All) & Summer & 0.81 & 0.62 & $+23\%$ & 
Convection-driven variability \\

Precipitation & OKSM (All) & Annual & 0.21 & 0.08 & $+62\%$ & 
Higher skill compared to NYSM; Similar bias signatures; small-error overprediction \\

Precipitation & OKSM (All) & Summer & 0.21 & 0.09 & $+57\%$ & 
Convection-driven variability \\

Precipitation & OKSM (All) & Fall/Spring & 0.13 & 0.07 & $+46\%$ & 
Dryline-driven variability \\

Precipitation & OKSM (Western Half) & Early Morning & 0.21 & 0.06 & $+71\%$ & 
Outflow and bore variability \\

\hline

Wind Speed & NYSM (All) & Annual & 0.95 & 0.81 & $+15\%$ & 
Strong skill; terrain and PBL-driven variability \\

Wind Speed & NYSM (All) & Solar Noon & 0.93 & 0.81 & $+13\%$ & 
PBL-driven variability \\

Wind Speed & NYSM (Northern Plateau) & Annual & 0.94 & 0.69 & $+27\%$ & 
Highest skill \\

Wind Speed & OKSM (All) & Annual & 0.76 & 0.69 & $+9\%$ & 
Highest skill; spatially uniform performance \\

Wind Speed & OKSM (All) & Solar Noon & 0.74 & 0.61 & $+18\%$ & 
PBL-driven stability \\

Wind Speed & OKSM (Southeast) & Annual & 0.76 & 0.52 & $+31\%$ & 
Highest skill \\

Wind Speed & OKSM (Panhandle) & Annual & 0.75 & 0.76 & $-2\%$ & 
Lowest skill \\

\hline

Temperature & NYSM (All) & Annual & 1.37 & 1.43 & $-5\%$ & 
Diurnal bias captured; variance smoothing \\

Temperature & NYSM (All) & Solar Noon & 1.36 & 1.42 & $-4\%$ & 
PBL-driven stability \\

Temperature & NYSM (Coastal) & Annual & 1.42 & 1.29 & $+9\%$ & 
PBL depth improvement \\

Temperature & NYSM (Northern Plateau) & Annual & 1.44 & 1.49 & $-3\%$ & 
PBL depth degradation \\

Temperature & OKSM (All) & Annual & 0.74 & 0.62 & $+16\%$ & 
Uniform performance; improved symmetry \\

Temperature & OKSM (All) & Solar Noon & 0.74 & 0.60 & $+19\%$ & 
PBL-driven stability \\

\hline
\end{tabular}}
\caption{Summary of LSTM performance relative to HRRR across variables, regions, and temporal regimes. MAE is reported in native units for each variable. Percent improvement is computed relative to the HRRR baseline.}
\label{tab:performance_summary}
\end{table*}

\section{Summary and Conclusions}

LSTMs were trained using the NYSM \& OKSM networks to predict forecast error of three target variables in the HRRR: precipitation error, wind error, and temperature error. The LSTMs were trained on data from 2018 to 2023 and tested on data from 2024. Independent LSTMs were trained specifically to a mesonet station and target variable, but are generalizable across forecast lead times. To better capture rare but high-impact events, we incorporated an outlier-focused loss function that prioritizes extreme errors in the training process.

LSTM performance was assessed primarily using MAE and mean error, with results further analyzed by geography, time of day, time of year, lead time, and associated improvement to HRRR forecasts. This multi-faceted evaluation provides a comprehensive understanding of LSTM forecast error prediction skill in the HRRR domain. Error signatures across predictors appear to deteriorate along mesoscale boundaries influenced by topography and latent processes, particularly during periods of peak or complex PBL activity, suggesting a potential physical mechanism underlying LSTM limitations, namely that training on surface-level features without vertically resolved information may limit the model’s ability to capture processes occurring above the surface. However, further work is needed to confirm this hypothesis.

Performance for LSTM precipitation error prediction appears to be negatively affected during warm-season convective events dominated by vertical motion and instability, with topography and storm frequency seemingly exerting secondary effects. In the OKSM, an early-morning error signature is evident, especially across the northwestern divisions, most representative of the Great Plains. Precipitation error predictions also exhibit an asymmetry: the LSTM accurately captures the magnitude of positive errors (wet-bias) but underestimates negative magnitudes (dry-bias), though it correctly identifies most negative-error points. It should also be noted that the LSTM is oversensitive to small magnitude precipitation errors -- consistently overpredicting small magnitude targets. 

The LSTM tends to slightly overpredict wind errors for the NYSM domain and slightly underpredict for the OKSM domain. Wind error maintains the most covariance and is mostly consistent in LSTM performance across over- and underpredictions in the NYSM domain. Notably, the OKSM domain is slightly less confident in predicting negative wind error. A key outcome of the domain comparison is that topography and associated latent energy characteristics may create conditions for consistent LSTM failure modes, tied to diurnal dependencies. Despite these temporally localized degradations, divisions with more complex terrain and higher humidity generally exhibit lower overall MAE values.

LSTM temperature error predictions exhibit strong regional dependence and vary across temporal scales, leading to markedly different implications for performance between the two domains. We posit that divisions with more stable and predictable PBL dynamics tend to exhibit more coherent diurnal error patterns and lower overall MAE values. In contrast, climate divisions influenced by mesoscale boundaries or complex topography generally show higher MAE but less coherent diurnal signatures, reflecting greater variability in local atmospheric processes. While temperature error prediction is reasonably accurate, the LSTM output is smoother and less variable than the target. OKSM exhibits the highest covariance and symmetry in this predictor, whereas the NYSM shows less confident performance. 

The consistent gradient in topography, LULC, and moisture produces an emergent, yet subtle, spatial error gradient across all predictors in the OKSM domain. In contrast, the more heterogeneous LULC and complex topography of the NYSM domain produce spatial error patterns that appear less coherent and more diverse. The LSTM performs better in the OKSM across all three predictors, which may be attributed to its more homogeneous terrain, simpler topography, and the higher baseline accuracy of the HRRR in this region. While these interpretations are physically consistent with known atmospheric processes, they are inferred from model behavior and diagnostic analysis, and further investigation is required to explicitly validate these mechanisms. 

These results suggest that the proposed framework is most operationally effective when applied in environments with predictable error structure and may require additional contextual information to improve performance in more complex regimes. Future work will focus on incorporating vertically resolved atmospheric information into the modeling framework, particularly through the integration of data from the NYSM profiler network. This addition is expected to improve the representation of boundary layer processes, including stability, mixing, and convective development, which are not fully captured by surface-based predictors and are associated with observed performance limitations.

From an operational perspective, several key findings emerge. First, the LSTM provides the most consistent improvement for wind error prediction across both domains, demonstrating strong covariance and stability across forecast lead times, making it the most reliable variable for real-time application. Second, precipitation error prediction shows meaningful skill in identifying high-impact events, particularly wet-bias regimes, though performance degrades during convective conditions and exhibits sensitivity to small-magnitude errors. Third, temperature error prediction is more regionally dependent and less consistently improves upon HRRR performance, particularly in the NYSM domain, where complex terrain and boundary layer variability introduce additional uncertainty.

The LSTM predicted error fields provide station-specific, real-time guidance on the expected magnitude and direction of HRRR forecast bias, enabling forecasters to adjust deterministic output accordingly. This framework can support more informed decision-making by identifying when and where forecast confidence should be reduced, particularly in regimes associated with known model deficiencies. Overall, the relative accuracy of these results underscores the potential for targeted ML approaches to substantially enhance forecast error prediction in high-resolution NWP systems, such as the HRRR. This application offers forecasters a reliable means of assessing forecast uncertainty at the point of use, and can be applied to other high-resolution NWP systems of interest at any mesonet.

\appendix
\begin{figure}
    \centering
    \includegraphics[width=\textwidth]{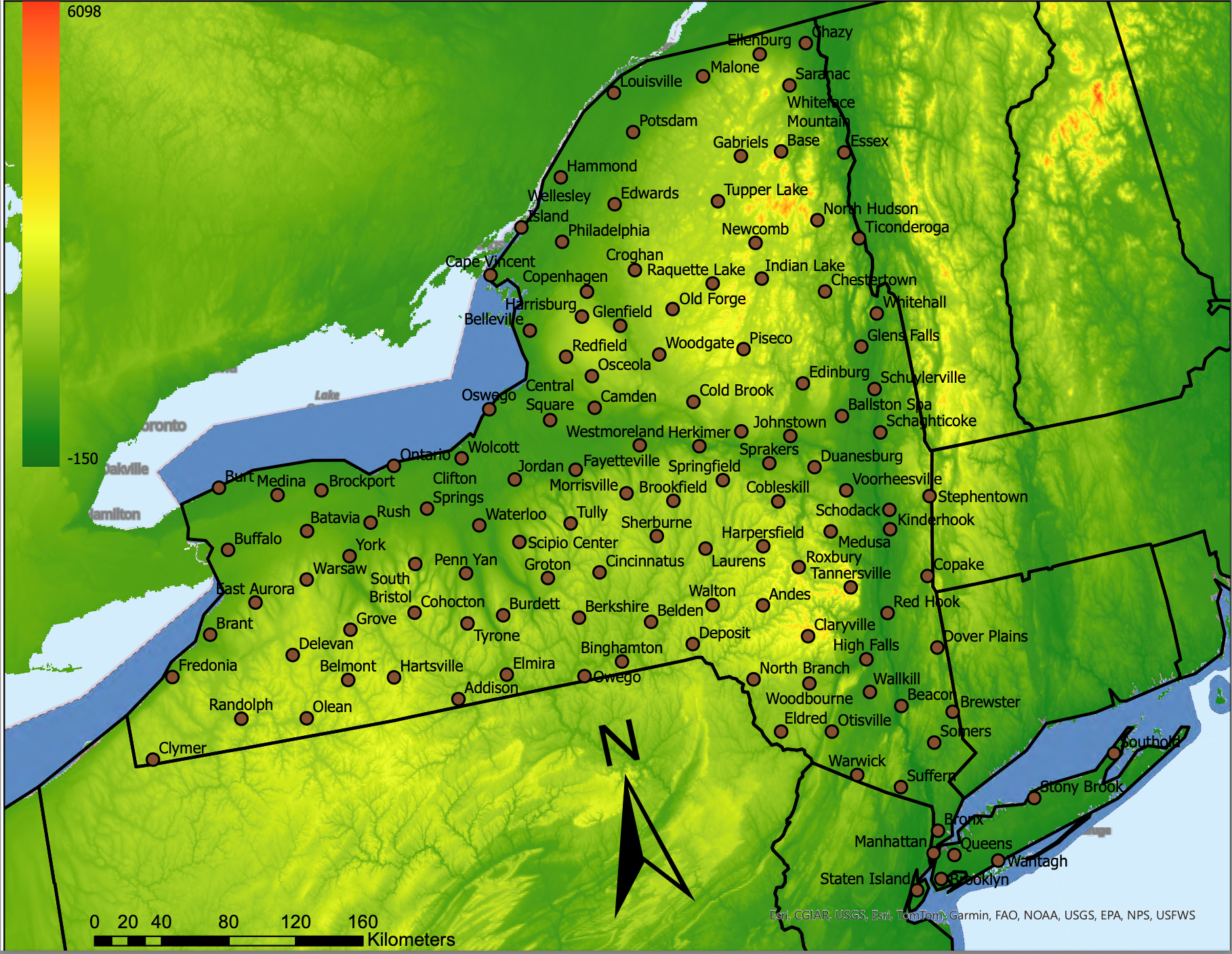}
    \caption{NYSM Network overlaid on an Elevation Map in meters, using \citet{EarthResources1997}.}
    \label{fig:elevation_nysm}
\end{figure}

\begin{figure}
    \centering
    \includegraphics[width=\textwidth]{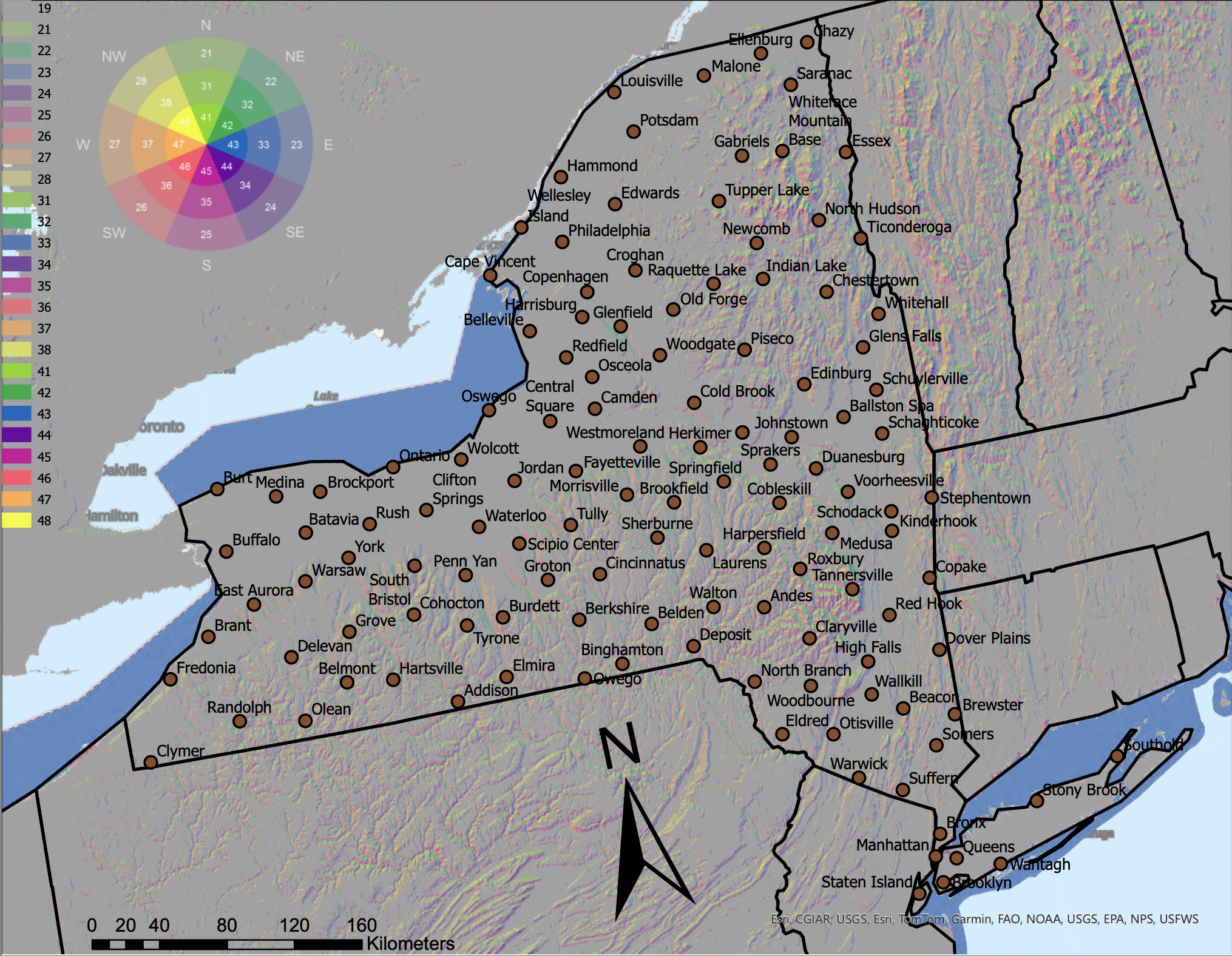}
    \caption{NYSM Network overlaid on an Aspect/Slope Map, using \citet{EarthResources1997} and Aspect/Slope Analysis in arcGIS.}
    \label{fig:aspect_slope_nysm}
\end{figure}

\begin{figure}
    \centering
    \includegraphics[width=\textwidth]{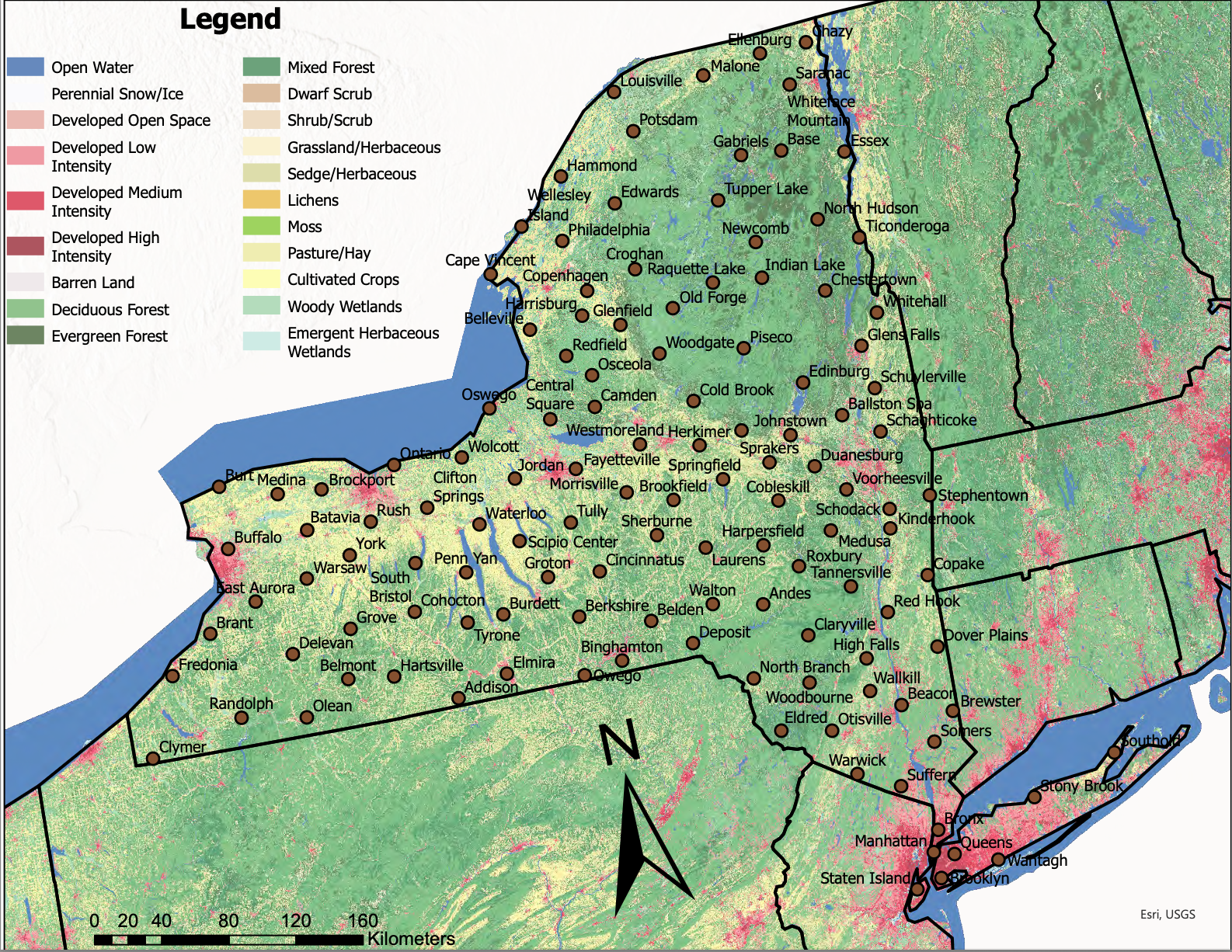}
    \caption{NYSM Network overland on the National Land-cover Database Map, using \citet{Dewitz2021} \& \citet{nlcd2023}.}
    \label{fig:nlcd_nysm}
\end{figure}

\begin{figure}
    \centering
    \includegraphics[width=\textwidth]{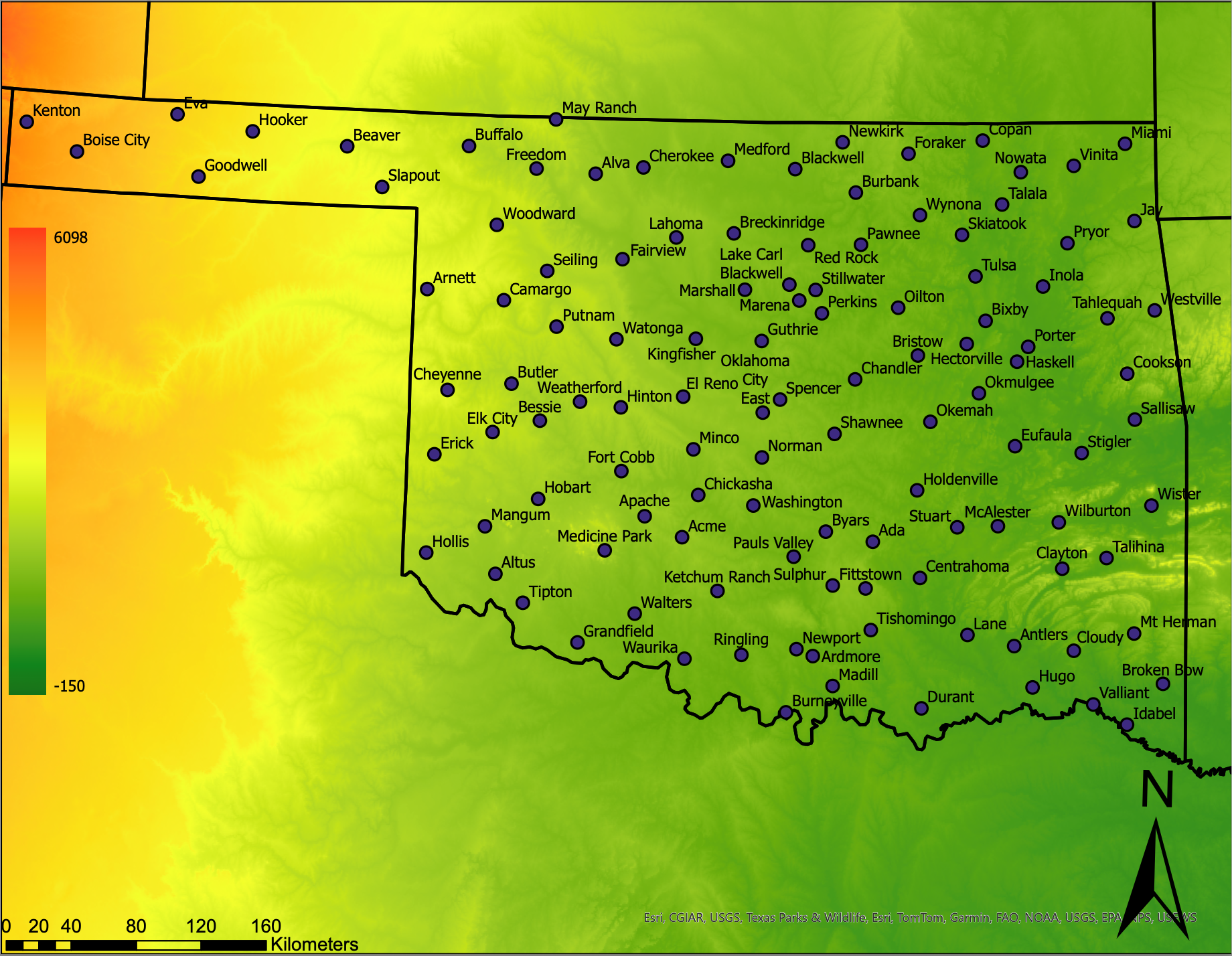}
    \caption{As in Fig.~\ref{fig:elevation_nysm}, but for the OKSM.}
    \label{fig:elevation_oksm}
\end{figure}

\begin{figure}
    \centering
    \includegraphics[width=\textwidth]{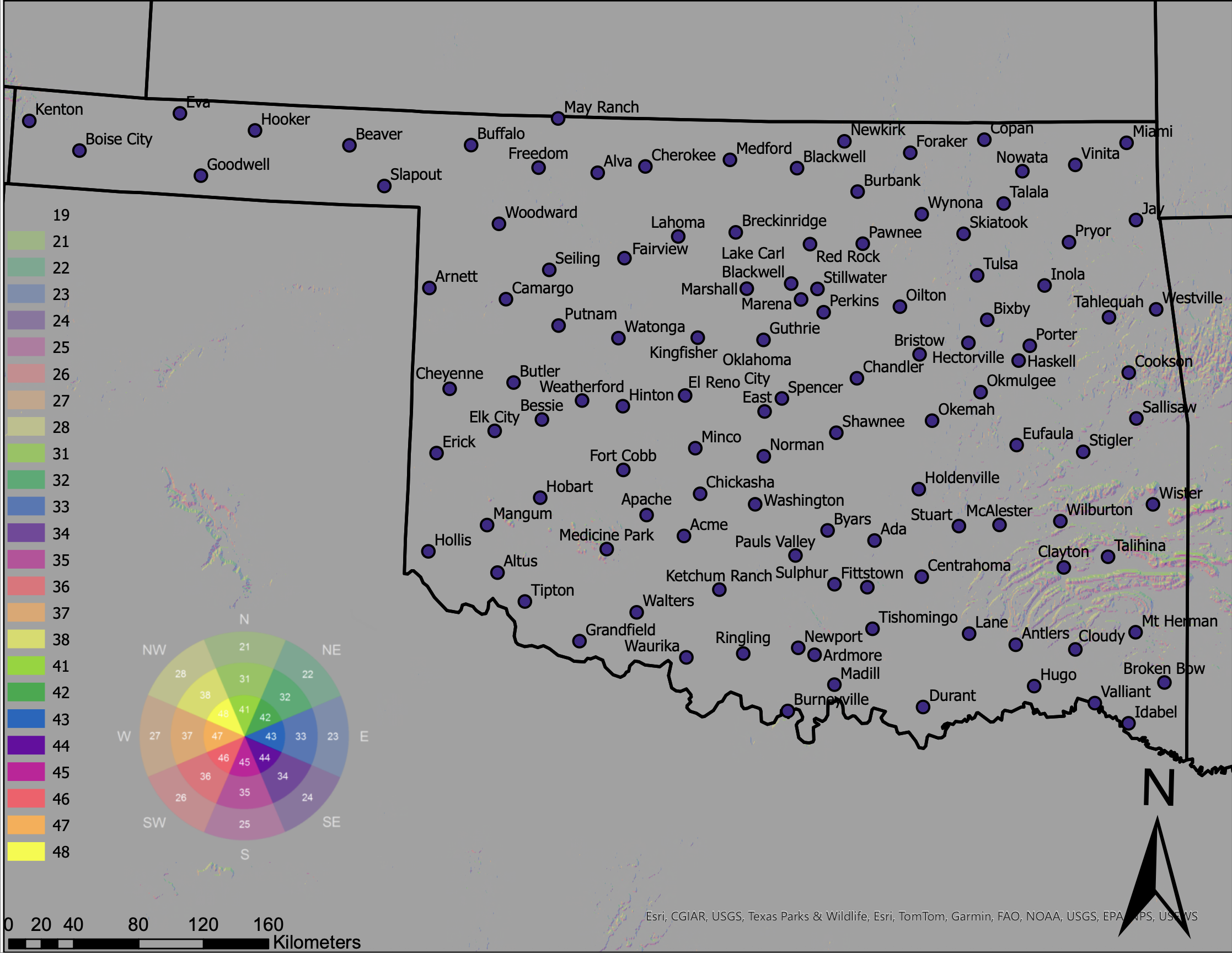}
    \caption{As in Fig.~\ref{fig:aspect_slope_nysm}, but for the OKSM.}
    \label{fig:aspect_slope_oksm}
\end{figure}

\begin{figure}
    \centering
    \includegraphics[width=\textwidth]{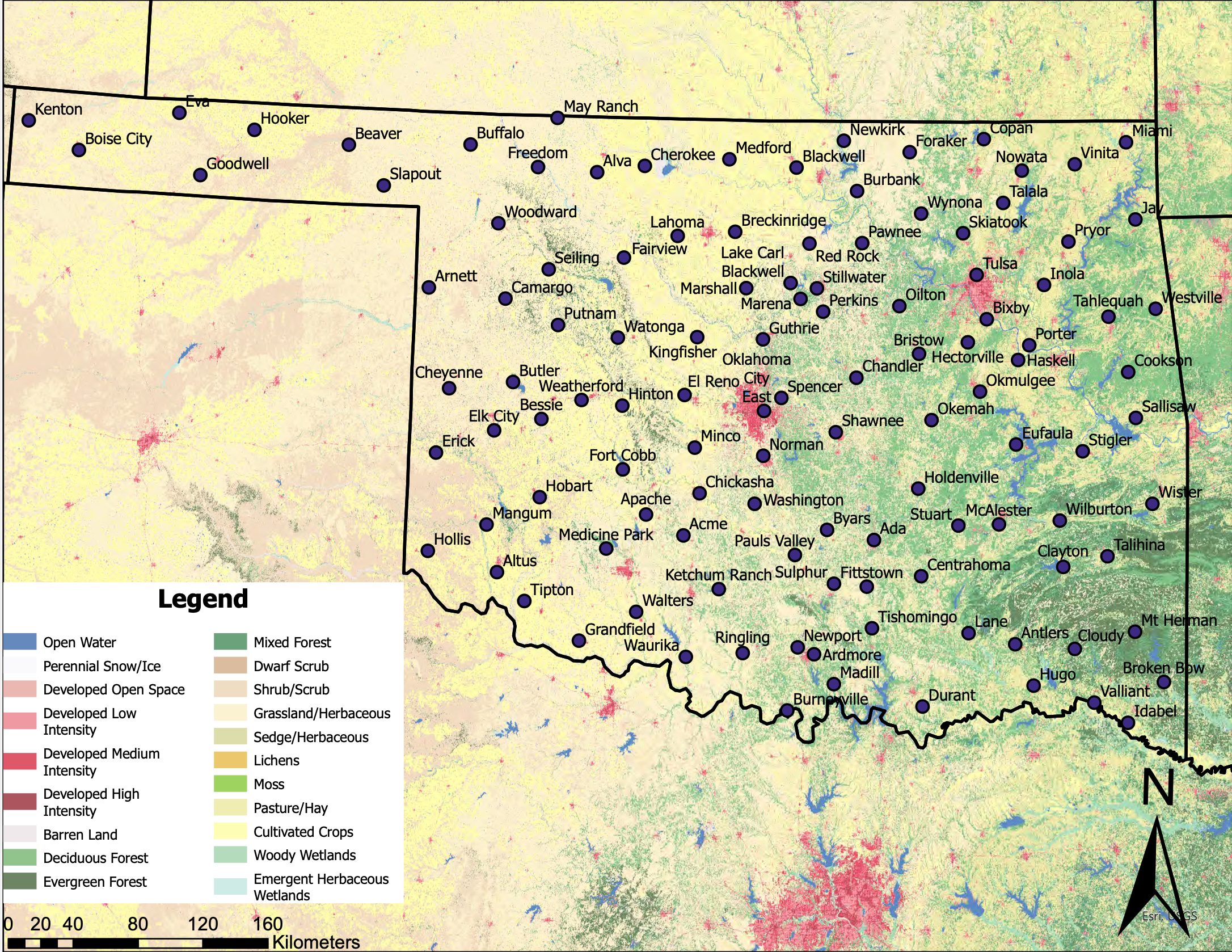}
    \caption{As in Fig.~\ref{fig:nlcd_nysm}, but for the OKSM.}
    \label{fig:nlcd_oksm}
\end{figure}

\section*{Acknowledgments}
This material is based upon work supported by the U.S. National Science Foundation under Grant No. RISE-2019758. \\ This research is made possible by the New York State (NYS) Mesonet. Original funding for the NYS Mesonet buildup was provided by the Federal Emergency Management Agency (FEMA) under grant FEMA-4085-DR-NY. Continued operation and maintenance of the NYSM are supported by the National Mesonet Program, University at Albany, and a combination of federal, private, and other grants. \\ Oklahoma Mesonet data are provided courtesy of the Oklahoma Mesonet, which is jointly operated by Oklahoma State University and the University of Oklahoma. Continued funding for maintenance of the network is provided by the taxpayers of Oklahoma.

\section*{Data Statement}
HRRR data were accessed from the University of Utah MesoWest HRRR archive \citep{blaylock2017cloud} and from Amazon Web Services. \\ NYSM data can be requested through the New York State Mesonet website: \url{http://nysmesonet.org}. \\ OKSM data can be requested through the Oklahoma State Mesonet website: \url{https://www.mesonet.org} \\ The code used for data preprocessing, model training, and figure generation is publicly available at \url{https://github.com/shmaronshmevans/inference_ai2es_forecast_err}.

%%% BIBLIOGRAPHY %%%

\end{document}